\documentclass[twocolumn]{aastex6}

\def \hbeta{H$\beta$}
\def \ebmv{E(B-V)}
\def \halpha{H$\alpha$}
\def \Msol{{\rm M}_{\odot}}
\def \Lsol{{\rm L}_{\odot}}

\def \logm{\log(M/\Msol)}
\def \logLIR{\log({\rm L}_{{\rm FIR}}/\Lsol)}
\def \lya{Ly$\alpha$}
\def \ewha{{\rm EW}({\rm H}\alpha)}
\def \ewlya{{\rm EW}({\rm Ly}\alpha)}
\def \h2{{\rm H_{2}}}
\def \oabund{12+\log({\rm O/H})}

\def \hbeta{H$\beta$~}
\def \halpha{H$\alpha$~}

\def \Cii{[C{\scriptsize ~II}]}

\def \oi{[O{\scriptsize ~I}]}
\def \oii{[O{\scriptsize ~II}]}
\def \oiii{[O{\scriptsize ~III}]}

\def \nii{[N{\tiny\,II}]}

\def \Ga{\textit{GALEX0959+0151}}
\def \Gb{\textit{GALEX1000+0157}}
\def \Gc{\textit{GALEX1000+0201}}
\def \Gaa{GALEX0959+0151}
\def \Gbb{GALEX1000+0157}
\def \Gcc{GALEX1000+0201}
\def \CO{CO$(1-0)$}

\def \LIR{L_{{\rm FIR}}}
\def \LUV{L_{{\rm UV}}}

\def \LCII{L_{{\rm [CII]}}}
\def \IRXB{IRX$-\beta$}
\def \dn4000{D_{{\rm n}}(4000) }

\defcitealias{CAPAK15}{C15}
\defcitealias{CHARLOTFALL00}{CF}

\begin{document}


\title{Are High Redshift Galaxies hot? - Temperature of $z>5$ Galaxies and Implications on their Dust Properties}


\author{
Andreas L. Faisst$^{1}$,
Peter L. Capak$^{1}$,
Lin Yan$^{2}$,
Riccardo Pavesi$^{3}$,
Dominik A. Riechers$^{3}$,
Ivana Bari\v{s}i\'{c}$^{4}$,
Kevin C. Cooke$^{5}$,
Jeyhan S. Kartaltepe$^{5}$,
Daniel C. Masters$^{1}$
}





\affil{$^1$Infrared Processing and Analysis Center, California Institute of Technology, Pasadena, CA 91125, USA}
\affil{$^2$Cahill Center for Astronomy and Astrophysics, California Institute of Technology, Pasadena, CA 91125, USA}
\affil{$^3$Department of Astronomy, Cornell University, Space Sciences Building, Ithaca, NY 14853, USA}
\affil{$^4$Max-Planck Institut f\"ur Astronomie, K\"onigstuhl 17, D-69117, Heidelberg, Germany}
\affil{$^5$School of Physics and Astronomy, Rochester Institute of Technology, 84 Lomb Memorial Drive, Rochester, NY 14623, USA}


\email{afaisst@ipac.caltech.edu; Twitter: @astrofaisst}

\begin{abstract}
	Recent studies have found a significant evolution and scatter in the \IRXB~relation at $z>4$, suggesting different dust properties of these galaxies.
	The total far-infrared (FIR) luminosity is key for this analysis but poorly constrained in normal (main-sequence) star-forming $z>5$ galaxies where often only one single FIR point is available.
	To better inform estimates of the FIR luminosity, we construct a sample of local galaxies and three low-redshift analogs of $z>5$ systems.
	The trends in this sample suggest that normal high-redshift galaxies have a warmer infrared (IR) SED compared to average $z<4$ galaxies that are used as prior in these studies. The blue-shifted peak and mid-IR excess emission could be explained by a combination of a larger fraction of the metal-poor inter-stellar medium (ISM) being optically thin to ultra-violet (UV) light and a stronger UV radiation field due to high star formation densities.
	Assuming a maximally warm IR SED suggests $0.6\,{\rm dex}$ increased total FIR luminosities, which removes some tension between dust attenuation models and observations of the \IRXB~relation at $z>5$.
	Despite this, some galaxies still fall below the minimum \IRXB~relation derived with standard dust cloud models. We propose that radiation pressure in these highly star-forming galaxies causes a spatial offset between dust clouds and young star-forming regions within the lifetime of O/B stars.  These offsets change the radiation balance and create viewing-angle effects that can change UV colors at fixed IRX.  We provide a modified model that can explain the location of these galaxies on the \IRXB~diagram.

\end{abstract}

\keywords{galaxies: ISM --- dust, extinction --- galaxies: formation --- galaxies: evolution --- galaxies: ISM --- galaxies: high-redshift}



\section{Introduction}
	Modern galaxy surveys have provided large and robust samples of $z>5$ galaxies that span a wide range of properties. These rich datasets are beginning to give us insights into the physical properties of early galaxies only one billion year after the Big Bang.
	In particular, the specific star formation rate (sSFR; the rate of mass build-up) is found to increase rapidly by more than a factor of 50 with increasing redshift at $z<2$ \citep[e.g.,][]{ELBAZ07,DADDI07a,NOESKE07,KARIM11,ILBERT15} and is expected to keep rising continuously out to $z\sim6$ \citep{SCHAERER13,STARK13,DEBARROS14,GONZALEZ14,TASCA15,JIANG16,FAISST16a,MARMOLQUERALTO16}.  While the average low-redshift star-forming galaxy is represented by a smooth bulge/disk profile, galaxies in the early Universe are irregular with turbulent clumps in the ultra-violet (UV) light representing regions of on-going vigorous star formation \citep[e.g.,][]{ELMEGREEN09,FORSTERSCHREIBER11,HEMMATI14,TACCHELLA15,MASON16}.  Furthermore, the amount of dust-obscured star formation increases significantly between $0<z<4$ where it can be measured directly and appears to begin declining at $z>4$ based on UV colors, indicating a change in the inter-stellar medium (ISM) properties of galaxies in the early Universe \citep[e.g.,][]{BOUWENS09,BOUWENS12,FELDMANN15}. However, the far-infrared (FIR) properties of \textit{normal}\footnote{As opposed to sub-millimeter galaxies (SMGs) and dusty star-forming galaxies (DSFGs) for which rich data sets in the FIR are available out to high redshifts.} main-sequence $z>5$ galaxies have been difficult to study even with the \textit{Atacama Large Millimeter Array} (ALMA), mostly providing only one single data point at these wavelengths.

The \IRXB~diagram \citep{MEURER95,MEURER99} is one of the few probes available to-date to study the ISM properties for large samples of normal $z>5$ galaxies within reasonable observation times on the facilities that are currently available. 
This diagram connects the ratio of total FIR to UV luminosity ($\LIR/\LUV \equiv {\rm IRX}$)\footnote{The UV luminosity ($\LUV$) is defined as the monochromatic luminosity measured at rest-frame $1600\,{\rm \AA}$. The total FIR luminosity ($\LIR$, sometimes also called ``TIR'') is integrated at $3-1100\,{\rm \mu m}$.} with the UV continuum spectral slope ($\beta$)\footnote{The UV continuum slope (defined as $f_{\lambda} \propto \lambda^\beta$) is derived for $1600 < \lambda < 2600$ \citep[e.g.,][]{MEURER99}.}. While the former is a proxy of the total dust mass, the latter depends on the column density of dust along the line-of-sight to the observer that is attenuating the UV light of stars and hence creating a red UV color.
	The relation is therefore sensitive to a range of ISM properties including dust geometries, dust-to-gas ratios, dust grain properties, and the spatial distribution of dust.  Furthermore, the scatter and trends in this diagram are strongly related to evolutionary trends in the ISM. 
	It is further to note here, that the UV slope $\beta$ also depends on other galaxy properties such as star-formation history (SFH), age of the stellar population, and metallicity. However, in comparison to dust, these effects are shown to contribute little to $\beta$ in the case of young galaxies in the early Universe \citep[][]{BOUWENS12,SCOVILLE15}.

Studies of low-redshift galaxies have shown that galaxies of different types populate different regions on the \IRXB~diagram. Specifically, young metal-poor galaxies similar to the Small Magellanic cloud (SMC) show a flatter relation between IRX and $\beta$ compared to local starburst galaxies, which occupy regions of higher IRX at similar $\beta$ \citep[e.g.,][]{BUAT05,SIANA09,BUAT10,HOWELL10,REDDY10,TAKEUCHI10,OVERZIER11,BOQUIEN12,REDDY12,BATTISTI17}.
	The differences between these populations of galaxies on the \IRXB~diagram can be attributed to differences in the shape of the internal dust attenuation curves \citep[e.g.,][]{GORDON00,BURGARELLA05,BUAT12,BATTISTI17}, different star-formation histories \citep[e.g.,][]{KONG04,BOQUIEN09,MUNOZMATEOS09}, and/or different dust geometries \citep[][]{SEIBERT05,CORTESE06,BOQUIEN09,MUNOZMATEOS09}.
	Studies at $z<3$ suggest \IRXB~relations similar to local starbursts with no significant evolution \citep[e.g.,][]{REDDY06,REDDY12,HEINIS13,TO14,BOURNE17,FUDAMOTO17}, however, stacking analyses at $z>3$ suggest significant deviations from this relation for luminous and young Lyman Break galaxies \citep[][]{LEE12,COPPIN15,FUDAMOTO17}.
	While the \textit{Hubble Space Telescope} (HST) and ground based facilities provide us with accurate measurements on the rest-frame UV properties of high-redshift galaxies, the high sensitivity of millimeter-wave interferometers such as ALMA, the \textit{Plateau de Bure Interferometer} (PdBI), and its successor the \textit{Northern Extended Millimeter Array} (NOEMA) enable us to push the measurement of the FIR properties of \textit{individual} galaxies to higher and higher redshifts \citep[][]{WALTER12,KANEKAR13,OUCHI13,GONZALEZLOPEZ14,OTA14,RIECHERS14,SCHAERER15,CAPAK15,MAIOLINO15,WATSON15,WILLOTT15,ARAVENA16,BOUWENS16,CARILLI16,DUNLOP16,KNUDSEN16,MILLER16,PAVESI16,KNUDSEN17}.
	
	\citet{CAPAK15} (hereafter \citetalias{CAPAK15}) provided the first study of the \IRXB~diagram at $z>5$ with a diverse sample of $5.1 < z < 5.7$ galaxies observed in ALMA band 7 at rest-frame $\sim150\,{\rm \mu m}$. Together with deep \textit{HST} near-IR imaging providing accurate UV colors for this sample \citep[][]{BARISIC17}, these studies suggest a significant evolution of the \IRXB~relation and its scatter at $z>5$ \citep[see also][]{FUDAMOTO17}. Although some galaxies are found to be consistent with the relation of local starbursts, more than half of the galaxies show a substantial deficit in IRX at a range of UV colors compared to the samples at $z<3$. Such galaxies are curious because their location on the \IRXB~diagram is difficult to be explained with current models for dust attenuation \citep[][]{CHARLOTFALL00} even for very low dust opacities and steep internal dust attenuation curves such as observed in the metal-poor SMC.
	
	Because of the poor constraints in the FIR, a common (partial) solution to this discrepancy is to assume increasing dust temperatures towards high redshifts \citep[][]{CAPAK15,BOUWENS16,FUDAMOTO17,NARAYANAN17}. However, also a significantly altered geometric distribution of stars and dust in these galaxies could explain their low IRX values.
	
	In this paper we try to understand the causes of such low IRX values in this high-redshift sample by exploring the infrared (IR) spectral energy distribution (SED) and the distribution of dust and stars.
	
	We begin by investigating the FIR luminosities of these galaxies which are derived from only one continuum data point at $\sim150\,{\rm \mu m}$ and are therefore very poorly constrained. Importantly, the shape of the IR SED is assumed from models that are fit at $z<4$. In particular, the \textit{luminosity weighted} temperature $T$ \citep[referred to as ``temperature'' in this paper and not to be confused with the single blackbody temperature $T_{\rm peak}$, see][]{CASEY12} that is accounting for mid-IR excess emission is crucial for defining the shape of the IR SED and therefore the FIR luminosity. Hence, it has a significant impact on the \IRXB~relation. Several studies suggest and expect a higher temperature in highly star-forming high-redshift galaxies due to a stronger UV radiation field in low-metallicity environments \citep[][]{SKLIAS14,BETHERMIN15,STRANDET16,SCHREIBER17} and as a function of various galaxy properties \citep[][]{CHAPMAN03,MAGDIS12,MAGNELLI14}. Such an evolving temperature could bring these galaxies in agreement with local \IRXB~relations \citep[][]{CAPAK15,BOUWENS16,FUDAMOTO17,NARAYANAN17}.
	A direct measurement of this temperature is, however, very difficult to obtain for main-sequence star-forming galaxies at $z>5$ (see Appendix~\ref{sec:ALMA}). Hence, our best chance to make progress is to investigate statistically correlations between temperature and other physical properties in large samples of low-redshift galaxies together with the study low-redshift analogs of such $z>5$ galaxies. We stress that this is by no means a bulletproof approach, but it allows us to gain a  picture of galaxies in the early Universe and will lead and define follow-up explorations of these galaxies with future facilities that can refine the conclusions of this work.
	
	Furthermore, the geometry of the dust distribution in high-redshift galaxies could be substantially different due to their turbulent nature -- something that is mostly not considered in recent studies. As we will show in this paper, this can lead naturally to a low IRX value and a large range of UV colors an thus can explain extreme cases in the galaxy population at high redshifts.

This work is organized as follows:
In Section~\ref{sec:LOWZ}, we present FIR measurements and their correlation with other physical properties for local galaxies from the literature along with three $z\sim0.3$ \lya~emitting galaxies which appear to be good analogs of high-redshift ($z>5$) galaxies.
	We then investigate the impact of different galaxy properties on the \IRXB~relations from a model and observational point-of-view, which will allow us to explain empirically a possible evolution of the \IRXB~relation to high redshifts (Section~\ref{sec:IRXB}).
	In Section~\ref{sec:HIGHZ}, we revisit the \citetalias{CAPAK15} galaxies, derive FIR luminosities using the priors on IR SED shape from the low-redshift samples, and present an updated \IRXB~diagram at $z\sim5.5$.  Finally, we propose a simple analytical model for the dust distribution in high-redshift galaxies that can describe the observed deficit in IRX by taking into account a non-uniform distribution of dust and stars (Section~\ref{sec:model}). 
	
	Throughout this work, we assume a flat cosmology with $\Omega_{\Lambda,0} = 0.7$, $\Omega_{m,0} = 0.3$, and $h = 0.7$. Stellar masses and SFRs are scaled to a \citet[][]{CHABRIER03} initial mass function (IMF) and magnitudes are quoted in AB \citep{OKE74}. Metallicities are quoted in the \citet[][]{PETTINI04} calibration unless specified differently.

\section{FIR properties of local and low-redshift galaxies}\label{sec:LOWZ}

One way of understanding the high-redshift universe is to assemble a sample of low-redshift objects which are ``analogs" to the high-redshift population.  These analogs typically can provide much more detailed data, and can be used to construct priors on the physical properties of high-redshift galaxies.  However, a key shortcoming of this technique is that the low-redshift sample is often an analog in only a few parameters, and so extrapolations must be made.  As such it is essential to carefully study the low-redshift sample and determine how well it matches the higher redshift one.  

Here we start with several samples of local galaxies that have a wealth of FIR data from the \textit{Herschel} observatory and other sources.  We explore  correlations between their FIR properties and other physical parameters which are measurable in our $z>5$ sample and distinguish them the most from galaxies at lower redshifts.  We then study three $z\sim0.3$ \lya~emitters that are analogs to the $z>5$ sample in rest-frame UV properties and explore their FIR properties in comparison to the \textit{Herschel} sample.  To ensure consistency we re-analyze the low-redshift samples using the same techniques we apply at higher redshift. 

\subsection{The sample of \textit{Herschel} observed local galaxies}
In the following, we consider local galaxies from the \textit{KINGFISH} sample \citep[][]{KENNICUTT11}, the \textit{GOALS} sample \citep[][]{SANDERS03,ARMUS09}, and the \textit{Dwarf Galaxy Survey} \citep[DGS,][]{MADDEN13}. Combining these samples allows us to cover a wide range in galaxy properties. Here, we briefly summarize the main properties of these samples. 	We note that we have removed all galaxies with a known Active Galactic Nucleus (AGN) from these three samples, however, we cannot rule out the presence of heavily obscured AGNs (see discussion in Section~\ref{sec:whywarm}).

\begin{itemize}

	\item The \textit{KINGFISH} sample consists of $61$ nearby galaxies within $30\,{\rm Mpc}$ with $9 < \logm < 10$ but also contains a handful of less massive dwarf galaxies at $10^7\,\Msol$ \citep{SKIBBA11}. Their sSFRs are predominantly $0.1-1.0\,{\rm Gyr^{-1}}$ and the metallicities in the range of $\oabund\sim 8.5-8.8$ \citep{KENNICUTT11}. All \textit{KINGFISH} galaxies are observed with \textit{Spitzer}/IRAC ($3.6\,{\rm \mu m}$, $4.5\,{\rm \mu m}$, $5.8\,{\rm \mu m}$, and $8.0\,{\rm \mu m}$) and \textit{Spitzer}/MIPS ($24\,{\rm \mu m}$ and $70\,{\rm \mu m}$) as well as \textit{Herschel}/PACS ($70\,{\rm \mu m}$, $110\,{\rm \mu m}$, and $160\,{\rm \mu m}$) and \textit{Herschel}/SPIRE ($250\,{\rm \mu m}$, $350\,{\rm \mu m}$, and $500\,{\rm \mu m}$) \citep[][]{TEMI09,DALE12,BENDO12}.
	
	\item The \textit{GOALS} sample consists of $\sim200$ of the most luminous infrared-selected galaxies including merging systems in the nearby Universe (LIRGs) out to $400\,{\rm Mpc}$ \citep[][]{KENNICUTT11}. In particular, about $20$ of these are ultra-luminous Infrared Galaxies (ULIRGs). The \textit{GOALS} galaxies have stellar masses mostly between $10 < \logm < 11.5$ and are therefore the most massive galaxies of these three samples \citep[][]{U12}. Furthermore, the galaxies span a wide range in sSFR from $0.1\,{\rm Gyr^{-1}}$ up to $10\,{\rm Gyr^{-1}}$ for the lowest masses ($\logm\sim9.5$). There are only a handful of metallicity measurements for these galaxies and they are in general around $\oabund=8.5$ \citep[][]{RUPKE08,RICH12}. Most of the \textit{GOALS} galaxies are observed by \textit{Spitzer}/IRAC and \textit{Spitzer}/MIPS as well as  \textit{Herschel}/PACS and \textit{Herschel}/SPIRE \citep[][]{SANDERS03,U12,DIAZSANTOS14,CHU17}.
	
	\item The \textit{DGS} sample consists of $50$ low-mass ($7 < \logm < 10$) dwarf galaxies in the nearby Universe out to $200\,{\rm Mpc}$ \citep{MADDEN13}. The galaxies populate mostly low metallicities \citep[$\oabund\sim7.8-8.7$,][]{REMYRUYER13} and are therefore similar to high-z galaxies and our low-redshift analogs, with the exception that the \textit{DGS} galaxies have a lower sSFR ($0.1-1.0\,{\rm Gyr^{-1}}$) similar to the \textit{KINGFISH} sample \citep{MADDEN13,CORMIER15}. All of the \textit{DGS} galaxies are observed with \textit{Spitzer}/IRAC and  \textit{Spitzer}/MIPS as well as \textit{Herschel}/PACS and \textit{Herschel}/SPIRE \citep[][]{BENDO12,REMYRUYER13}.
	
	\end{itemize}

	\subsubsection{Parameterization of the IR SED and fitting of temperature}\label{sec:irsedparameters}
	
	We measure the FIR properties of the local galaxies using the parameterization introduced by \citet{CASEY12} and their public IR photometry from $24\,{\rm \mu m}$ to $500\,{\rm \mu m}$. We will use this parameterization throughout this work to ensure consistent measurements for all the galaxy samples discussed in the following.
	
	The \citet{CASEY12} parameterization combines a single graybody (accounting for the reprocessed, cold to warm emission) with a mid-IR power-law (approximating the warm to hot dust components from AGN heating, hot star-forming regions, and optically thin dust). As shown in their study, this method results in a better fit to the rest-frame mid-IR part of the spectrum by marginalizing over multiple warm to hot dust components of a galaxy (see their figure 1 for an illustration) compared to a single graybody. Mathematically, the parametrization depends on the luminosity weighted temperature ($T$), the slope of the mid-IR power-law component ($\alpha$), the emissivity ($\beta_{\rm IR}$), and a normalization.

	Note that $T$ represents the average temperature marginalized over all warm temperature components combined in the IR SED and is therefore a \textit{luminosity weighted} temperature (in the following referred to as just ``temperature'' for convenience). This temperature should not be confused with the \textit{peak} temperature $T_{\rm peak}$\footnote{E.g., computed using Wien's displacement law $T_{\rm peak} = b/\lambda_{{\rm peak}}$ with $b=2898\,{\rm \mu m\,K}$.} that is quoted in some other studies and is only equivalent to $T$ only in the case of a single blackbody shaped IR SED \citep[in general it is always $T>T_{\rm peak}$, see detailed discussion in][]{CASEY12}. In this work we study the luminosity weighted temperature $T$, however, for convenience, we quote equivalent $T_{\rm peak}$ temperatures whenever possible using the relation shown in \citet{CASEY12}.
	Importantly, $T$ defines the shape of the IR SED by taking into account the mid-IR emission. It is not only affected by the temperature of the optically thick dust, but also by the opacity of the dust in the ISM of a galaxy as well as dust grain properties, which can significantly change the mid-IR slope $\alpha$ as shown by \citet[][]{SCOVILLE76} and further discussed in Section~\ref{sec:whywarm}. Therefore changes in temperature $T$ are not only limited to changes in the dust temperature, but also include changes in the opacity of the dust or its geometry. Hence, we refer with $T$ to all these effects describing the shape of the IR SED rather than solely the dust temperature.  

	For the fitting of the IR SED using the above parameters, we fix the wavelength at which the optical depth is unity to $\lambda_{0}=200\,{\rm \mu m}$ \citep{DRAINE06,CONLEY11,CASEY12,RIECHERS13}. This is somewhat arbitrary as $\lambda_{0}$ is a priori unknown and is likely to change for different samples of galaxies, although \citet{RIECHERS13} find a consistent value close to $200\,{\rm \mu m}$ for $z>4$ SMGs. Using $\lambda_{0}=100\,{\rm \mu m}$ would result in $\sim5\,{\rm K}$ cooler temperatures, but does not change our results for the total FIR luminosity, so the choice is irrelevant as long as it is treated consistently.
	For all fits, we assume a freely varying $\beta_{\rm IR}$. However, because $\beta_{\rm IR}$ and $T$ can be degenerate, we also fit the photometry with a fixed $\beta_{\rm IR}$ of $1.6$ and $2.0$ but we do not find any significant differences in the following results and conclusions.

\begin{figure}
\centering
\includegraphics[width=1.1\columnwidth, angle=0]{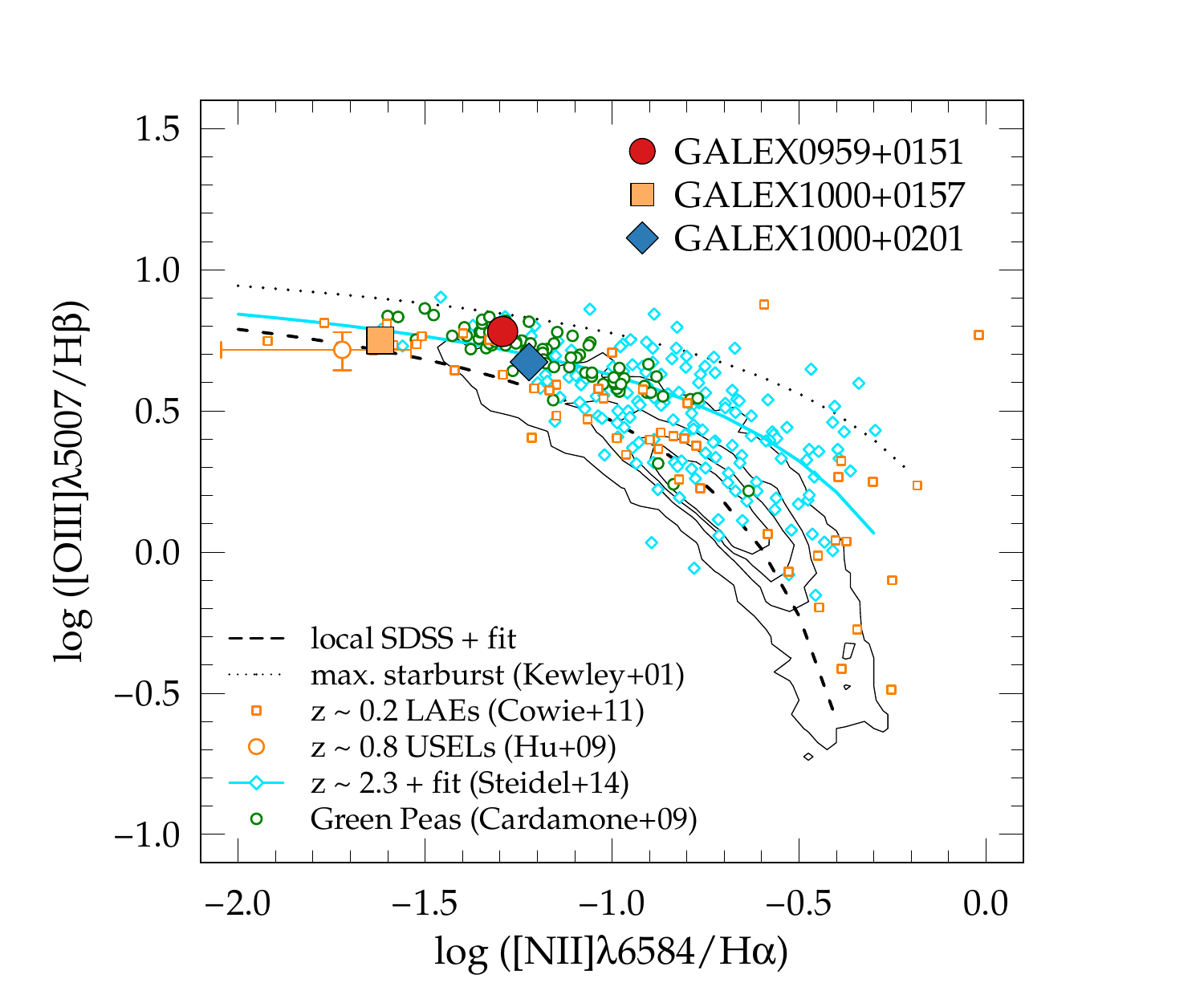}
\caption{BPT diagram \citep{BALDWIN81} with our three $z\sim0.3$ analogs (\Ga, red point; \Gb, orange square; \Gc, blue diamond). Other samples of analogs are also shown (``Green Peas'' \citep[green circles,][]{CARDAMONE09}, $z\sim0.8$ ultra-strong emission line galaxies \citep[orange circle,][]{HU09}, and $z\sim0.2$ \lya~emitters \citep[orange squares,][]{COWIE11}). The black contours show local galaxies in SDSS and the cyan diamonds show $z\sim2.3$ galaxies (including best fit) from \citet[][]{STEIDEL14}.
Our analogs show increased \oiii/\hbeta and decreased \nii/\halpha ratios as expected for galaxies in the early Universe. Furthermore, the location of the analogs does not suggest the presence of an AGN.
\label{fig:bpt}}
\end{figure}

\subsection{Low redshift analogs of high-z galaxies}
	In addition to the \textit{Herschel} samples we consider three galaxies at low redshift ($z\sim0.3$) that were selected as UV analogs to $z>5$ systems with archival FIR and ALMA observations.
	Such galaxies are extremes amongst low-redshift galaxies but show very similar properties as high-redshift galaxies in terms of sSFRs, optical emission lines, metallicity, and morphology. Low-redshift analogs, such as ``Green Peas'' \citep[][]{CARDAMONE09} or ultra strong emission line galaxies \citep[][]{HU09}, are therefore used by many studies to understand in detail the population of high-redshift galaxies, for which much less observational data exist \citep[][]{STANWAY14,BIAN16,FAISST16a,FAISST16c,GREIS16,MASTERS16,ERB16}. A very common property of these galaxies are their high \halpha equivalent-width (equivalent to sSFR). In fact, as demonstrated by \citet{FAISST16c}, the \halpha equivalent-width allows a very clean selection of low-redshift analogs with very similar properties of high-redshift galaxies.
	
	The three analogs of high-redshift galaxies (hereafter named as \Ga, \Gb, and \Gc) were originally selected from the \textit{Galaxy Evolution Explorer} \citep[\textit{GALEX},][]{MARTIN05} \lya~emitter sample described in \citet[][]{COWIE11} \citep[see also][]{COWIE10}. 

	This \lya~emitter sample contains galaxies that are spectroscopically preselected to have rest-frame $\ewlya > 15\,{\rm \AA}$ in the \textit{GALEX} far-UV or near-UV grism.  Galaxies with a clear AGN signature were removed based on a cut in \lya~FWHM and another cut in the high-ionization optical emission lines (see also Figure~\ref{fig:bpt}). All three galaxies are in the main \textit{Cosmic Evolution Survey} \citep[COSMOS,][]{SCOVILLE07}\footnote{\url{http://cosmos.astro.caltech.edu}} area and thus are imaged by ground and space based facilities in more than $30$ photometric bands ranging from UV to radio \citep[][]{LAIGLE16}. These are the only galaxies in COSMOS that are analogs of $z>5$ systems with a wealth of ancillary data in the UV, optical, and FIR. Rest-frame optical spectroscopy is available from the zCOSMOS-bright spectroscopic survey \citep[][]{LILLY07}\footnote{\url{http://archive.eso.org/cms/eso-data/data-packages/zcosmos-data-release-dr1.html}} and two of the analogs (\Ga~and \Gb) are observed in the UV with the \textit{Cosmic Origin Spectrograph} \citep[COS,][]{GREEN12} on \textit{HST} (Scarlata et al., in prep).

	The three galaxies appear to be good analogs of high redshift systems. Specifically, they show optical emission line properties and \lya~properties similar to galaxies at $z\sim5$. The \halpha equivalent-widths ($\ewha = 200-400\,{\rm \AA}$) are comparable to those measured in $z\sim5-6$ systems using \textit{Spitzer} colors \citep[][]{SHIM11,FAISST16a,RASAPPU16} and their specific SFRs based on combined UV and FIR luminosities are between $1.6\,{\rm Gyr}^{-1}$ and $8.8\,{\rm Gyr}^{-1}$. This is $2-15$ times higher than average galaxies on the star-forming main-sequence at $z\sim0.3$ at stellar masses $\logm\sim9.0-10.0$ \citep[e.g.,][]{SPEAGLE14,SCHREIBER15,LEE15}, but similar to those of $z\sim5-6$ galaxies \citep[e.g.,][]{STEINHARDT14,CAPAK15,FAISST16a}. Furthermore, the \oiii/\hbeta ratios are high ($4.5$ to $6.0$) and \nii/\halpha ratios are low ($0.025$ to $0.060$) compared to $z\sim0.3$ galaxies at similar stellar masses, but again comparable to $z\sim5-6$ systems \citep[see Figure~\ref{fig:bpt} and][]{KEWLEY13,FAISST16a,FAISST16c,MASTERS16}.
	Finally, these line ratios suggest gas-phase metallicities of $\oabund<8.3$ in any calibration, which is $0.3\,{\rm dex}$ below the average metallicity of $z\sim0.3$ galaxies at similar stellar mass but comparable to galaxies at $z>3.5$ \citep[see][]{ANDO07,MAIOLINO08,KEWLEY08,FAISST16b}.

	The detailed spectroscopic properties of the three analogs are listed in Table~\ref{tab:specprop}.

\begin{figure}
\centering
\includegraphics[width=1.0\columnwidth, angle=0]{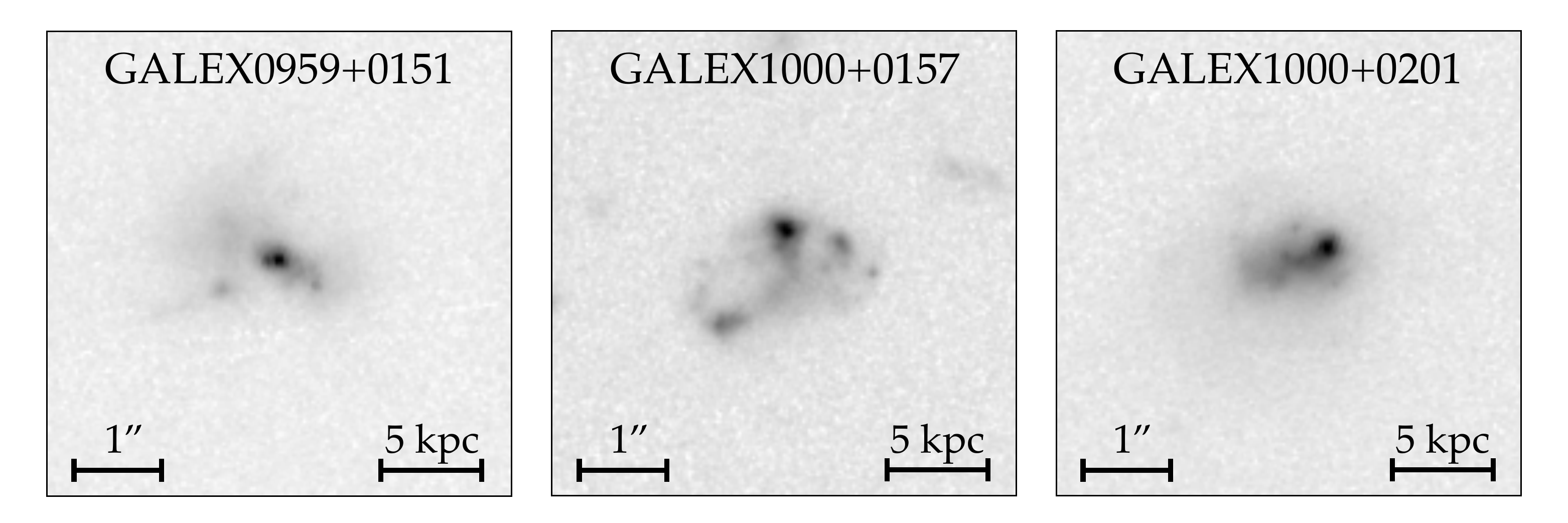}
\caption{
\textit{HST} ACS/F814W (rest-frame optical) portraits of the three $z\sim0.3$ analogs drizzled to a resolution of $0.03\arcsec/{\rm pixel}$ ($\sim 200\,{\rm pc}$). The three galaxies show a disturbed and clumpy structure with indications of recent or ongoing star formation similar to what is expected and observed in high-redshift galaxies. The compact nuclei are of the order of $0.8\,{\rm kpc}$ in diameter but the diffuse components extend out to $\sim6\,{\rm kpc}$.
\label{fig:hststamps}}
\end{figure}
	
\begin{figure*}
\centering
\includegraphics[width=2.1\columnwidth, angle=0]{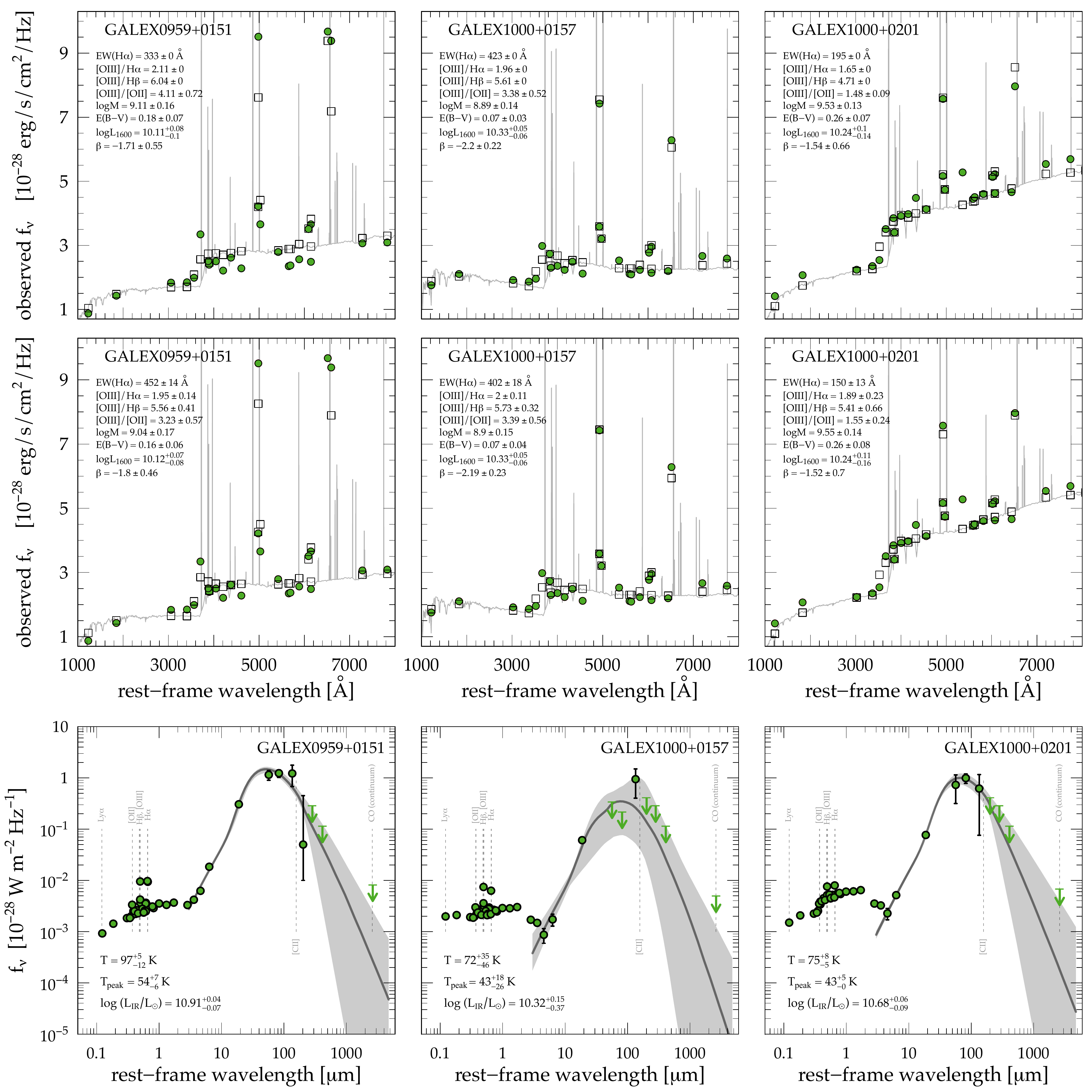}
\caption{Compilation of UV to optical and IR SEDs of the three $z\sim0.3$ analogs (green symbols). \textit{Top:} Optical SED fit with \oiii/\halpha and \oiii/\hbeta line ratios as well as $\ewha$ fixed to the spectroscopic measured values. The open squares show the photometry measured on the best-fit SED. \textit{Middle:} Same as top panel but with variable line ratios and $\ewha$. \textit{Bottom:} IR SED fit from $3-3000\,{\rm \mu m}$ to the photometry given in Table~\ref{tab:photoprop}. The green arrows marks upper limits. Optical and FIR emission lines are indicated by vertical lines. The uncertainty of the fit is indicated in gray. The best fit parameters for the UV to optical and IR fits are given in Tables~\ref{tab:optsedprop} and \ref{tab:irsedprop}, respectively.
\label{fig:sed}}
\end{figure*}

	\subsubsection{HST morphology}
	Figure~\ref{fig:hststamps} shows the optical images by \textit{HST}'s \textit{Advanced Camera for Survey} (ACS) in F814W ($I$-band) at $200\,{\rm pc}$ resolution. The bulk of the optical light is emitted from a compact nucleus of less than $0.8\,{\rm kpc}$ in diameter but diffuse components extend out to $6\,{\rm kpc}$ in all of the galaxies. Several clumps of UV light in the diffuse component with same or smaller size than the nucleus suggest regions of vigorous star formation.

	\subsubsection{UV to mid-IR data}	

	The UV to mid-IR photometry of the galaxies is extracted from the \textit{COSMOS2015} catalog \citep{LAIGLE16}\footnote{\url{ftp://ftp.iap.fr/pub/from_users/hjmcc/COSMOS2015/}}.
	Next to deep imaging in the UV and optical, the catalog includes deep \textit{Spitzer}/IRAC mid-IR data at $3.6\,{\rm \mu m}$ and $4.5\,{\rm \mu m}$ from the \textit{Spitzer Large Area Survey with Hyper-Suprime-Cam} \citep[SPLASH,][]{STEINHARDT14}\footnote{\url{http://splash.caltech.edu}} as well as \textit{Spitzer}/MIPS $24\,{\rm \mu m}$ (all three galaxies are detected). The photometry has been extracted using the positional priors from the optical bands and reliable de-blending algorithms \citep[][]{LAIGLE16}. The UV to optical SEDs are shown in the top and middle panels of Figure~\ref{fig:sed}.
	Each of the galaxies has been observed with \textit{Spitzer}/MIPS at $70\,{\rm \mu m}$ as well as \textit{Herschel}/PACS ($110\,{\rm \mu m}$ and $170\,{\rm \mu m}$) and \textit{Herschel}/SPIRE ($250\,{\rm \mu m}$, $350\,{\rm \mu m}$, and $500\,{\rm \mu m}$). 
	Due to the large point spread functions (PSF) and confusion, we choose to measure the IR photometry on a galaxy-by-galaxy basis. In brief, we perform aperture photometry with an appropriate correction to total magnitudes. The uncertainties are based on random parts of background. For galaxies with close neighbors, we apply either \textit{Spitzer}/IRAC or \textit{Spitzer}/MIPS positions as priors and use \texttt{GALFIT} \citep[version 3.0.5,][]{CHIEN11} to carefully subtract the flux of the neighboring galaxies before performing aperture photometry on the main object.
	
	The detailed IR photometry of the three analogs is listed in Table~\ref{tab:photoprop}.

	\subsubsection{Physical properties from optical SED fitting}

	We fit the optical SED of our low-redshift analogs to obtain their stellar masses and rest-frame $1600\,{\rm \AA}$ monochromatic luminosities ($\LUV$).
	We use templates based on \citet{BRUZUALCHARLOT03} and include constant, exponentially declining and increasing star formation histories (SFH) with variable metallicities from 1/20 of solar to solar. The stellar population ages range from $100\,{\rm Myrs}$ to a few Gyrs and dust is parametrized by extinction laws based on local starbursts \citep{CALZETTI00} and the SMC \citep[][]{PREVOT84,PETTINI98}. However, we found the different extinction laws do not significantly change the measured physical parameters. The redshift is fixed to the spectroscopic redshift during the fitting.
	We add strong optical emission lines coupled to \halpha with variable ratios \oiii/H$\alpha$, H$\alpha$/\hbeta\footnote{We compute \hbeta via case B recombination and the fitted \ebmv~value assuming a stellar-to-gas dust ratio of unity as suggested by recent studies \citep{REDDY15,SHIVAEI15,DEBARROS16}.}, and \oiii/\oii. Weak emission lines (such as N, S, or He) are added with a constant ratio with respect to \hbeta~for a sub-solar metallicity \citep{ANDERS03}. The \textit{GALEX} far-UV photometry is corrected for \lya~line emission derived from the spectra prior to fitting as we do not include the \lya~line in our SED models.
	The fit is performed using a Levenberg-Marquardt algorithm, as part of the \texttt{R/minpack.lm} package\footnote{\url{https://cran.r- project.org/web/packages/minpack.lm/ index.html}} and errors are obtained from a Monte Carlo sampling taking into account the error in the photometry and models.
	The UV luminosity ($\LUV$) is measured at rest-frame $1600\,{\rm \AA}$ and the UV continuum slope is fitted in the window between $1600\,{\rm \AA}$ and $2600\,{\rm \AA}$.
	We perform two fits for accessing the uncertainties in the fitted parameters. First, we fix $\ewha$, \oiii/\halpha and \oiii/\hbeta to the spectroscopic value and in a second run we leave these as free parameters. The fits are shown in Figure~\ref{fig:sed} on the top and middle panels, respectively, and we find an agreement on the order of $25\%$ or better between the photometrically and spectroscopically measured \halpha EW and line ratios.
	We find stellar masses of $\logm = 9-9.5$, moderate to low dust attenuation ($\ebmv<0.3$), and UV luminosities on the order of $\log(\LUV)\sim10.2$. The photometrically derived \oiii/\oii~ratios (ranging from $1.5$ to $3.5$) are factors $3-6$ higher compared to $z\sim0.3$ galaxies on the main-sequence but fit well the expected \oiii/\oii~ratios for $z\sim5$ galaxies \citep[e.g.,][]{FAISST16c}, which is an additional verification of their validity as analogs of high-redshift galaxies.
	We compare the UV slope $\beta$ measured from SED fitting with the direct measurement from the observed \textit{GALEX} and CFHT/u$^*$ fluxes and find a good agreement within the uncertainties.
	Since we want to study the location of these galaxies on the \IRXB~diagram, we want the measurement of $\beta$ to be independent from any assumed dust attenuation law. This is not the case for $\beta$ measurements from SED fitting. We therefore prefer to use the $\beta$ measurement obtained from the observed fluxes.
	
	The physical properties obtained from the optical SED fits are listed in Table~\ref{tab:optsedprop}.

	\subsubsection{ALMA measurements}

	The three analogs of high-redshift galaxies have been observed with ALMA in Band 3 (ID: 2012.1.00919.S, PI: Y. Kakazu) centered at an observed frequency of $92\,{\rm GHz}$, which corresponds to the rest-frame wavelength of \CO~emission ($2.6\,{\rm mm}$). The integration times vary between $1255\,{\rm s}$ and $9852\,{\rm s}$ and the number of antennae between 24 and 36 (see Table~\ref{tab:photoprop}). The line observations are performed in ``frequency division mode'' at a bandwidth of $0.9375\,{\rm GHz}$ in one spectral window and the continuum observations use the ``time division mode'' with a bandwidth of $2.0\,{\rm GHz}$ in each of the three line-free spectral windows. The synthesized beam sizes are on average $3.5\arcsec \times 2.5\arcsec$. The data are reduced using the standard \texttt{CASA} ALMA calibration pipeline.
	All three galaxies are undetected in the continuum around rest-frame $2.6\,{\rm mm}$ at $\lesssim50\,{\rm \mu Jy}$ at $3\sigma$ but we note a tentative \CO~detection at the redshift of \Ga~at a $1-2\sigma$ level with a peak emission spatially offset by $1.8\arcsec$ or $7\,{\rm kpc}$.

	\subsubsection{IR SEDs and temperature of the low-z analogs}\label{sec:firanalogs}

	We estimate the FIR properties of the analogs in the same way as for the \textit{Herschel} sample (see Section~\ref{sec:irsedparameters}). The bottom panels of Figure~\ref{fig:sed} show the best-fit IR SEDs to the three analogs (best fit parameters are listed in Table~\ref{tab:irsedprop}). We find temperatures ranging from $70\,{\rm K}$ to $100\,{\rm K}$ with uncertainties of $\sim 10\,{\rm K}$ (corresponding to $40\,{\rm K}$ to $60\,{\rm K}$ with uncertainties of $\sim 5\,{\rm K}$ in $T_{\rm peak}$). The temperature for \Gb~is only poorly constrained because only upper limits for the FIR photometry are available. 
	
\begin{figure}
\centering
\includegraphics[width=1\columnwidth, angle=0]{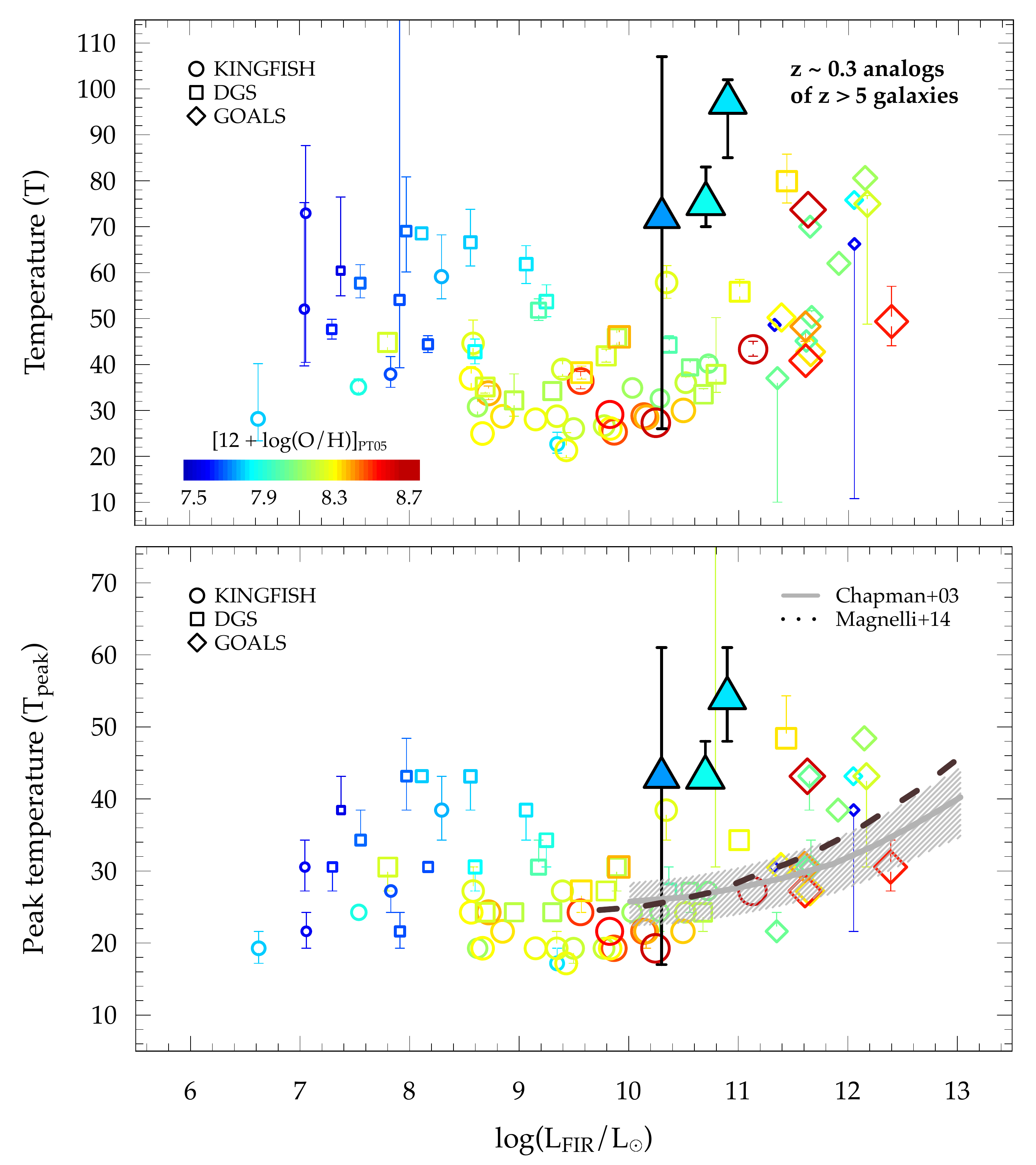}
\caption{The $\LIR$ vs. $T$ relation (top) and $\LIR$ vs. $T_{\rm peak}$ relation (bottom) colored by metallicity based on the \textit{KINGFISH} (circles), \textit{GOALS} (squares), and \textit{DGS} (diamonds) local galaxy samples. There are indications that $T$ and $T_{\rm peak}$ are enhanced in metal-poor galaxies. The correlations by \citet{CHAPMAN03} ($z\sim0$, gray hatched region) and \citet[][]{MAGNELLI14} ($0.2 < z < 0.5$, dashed black line) are indicated for $T_{\rm peak}$. Our three $z\sim0.3$ analogs are shown as big symbols color-coded by their metallicity. We use the metallicity calibration of \citet{PILYUGIN05} here for a consistent comparison. 
These local relations indicate that high-redshift galaxies (characterized by high sSFR and low metallicity) have a ``warmer'' IR SED than average galaxies at lower redshifts.
\label{fig:tdependences}}
\end{figure}

\begin{figure*}
\centering
\includegraphics[width=2\columnwidth, angle=0]{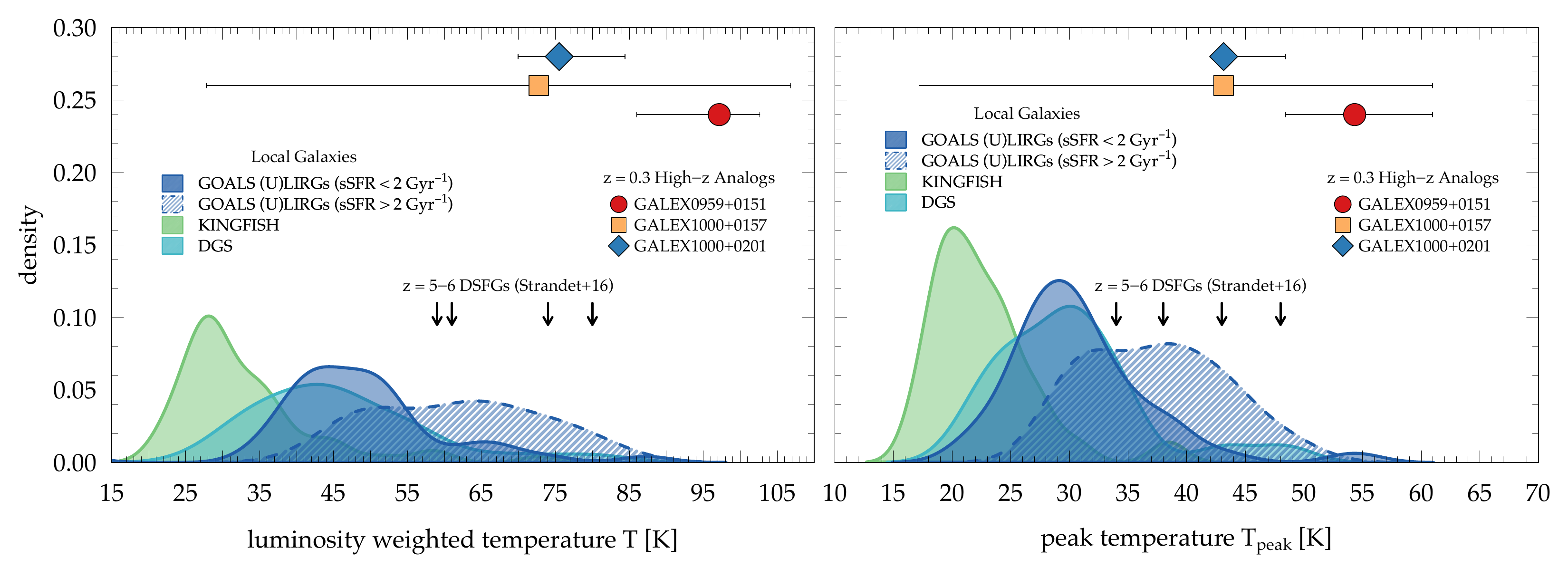}
\caption{Distribution of the luminosity weighted temperature ($T$, left) and peak temperature ($T_{\rm peak}$, right) for \textit{KINGFISH} (green), \textit{GOALS} (blue), and \textit{DGS} (cyan) nearby galaxies. The \textit{GOALS} sample is split in galaxies with high ($>2\,{\rm Gyr^{-1}}$) and low ($<2\,{\rm Gyr^{-1}}$) sSFR to emphasize the correlation between peak wavelength and sSFR. The high sSFR local galaxies show similar temperatures as our $z\sim0.3$ analogs (symbols) and the $z\sim5-6$ lensed DSFGs from \citet{STRANDET16}.
\label{fig:tempcomparison}}
\end{figure*}

\subsection{Correlations of FIR properties with metallicity and sSFR in low-redshift galaxies}\label{sec:tempdependencelocal}

Metallicity and sSFR are the two most prominent ways high-redshift galaxies differ from lower redshift ones. We therefore begin by investigating the dependency of the shape of the IR SED (parameterized by $T$) on these physical quantities.

	Figure~\ref{fig:tdependences} shows the $\LIR$ vs. $T$ and  $\LIR$ vs. $T_{\rm peak}$ relation for our local samples with open symbols color-coded by the metallicities of the galaxies. The different symbols indicate the \textit{KINGFISH} (circles), \textit{DGS} (squares), and \textit{GOALS} (diamonds) samples. The metallicities are quoted in the \citet{PILYUGIN05} calibration for a consistent comparison of the different samples.
	At intermediate to high total FIR luminosities ($\logLIR\gtrsim9$) we recover the positive correlation between $\LIR$ and $T$ (and $T_{\rm peak}$) similar to what was found in studies of luminous FIR galaxies \citep[e.g.,][]{DUNNE00,CHAPMAN03,MAGNELLI14}. However, contrarily to expected low temperatures at very low FIR luminosities, the FIR faint and metal-poor dwarf galaxies from the \textit{DGS} sample suggest increasing temperatures at $\logLIR\lesssim9$. This has not be seen in previous studies as these do not go this low in total FIR luminosity. It indicates that galaxies at low $\LIR$ and low metallicity as well as galaxies at high $\LIR$ and high metallicity are characterized by a warm IR SED based on our local samples.

	Figure~\ref{fig:tempcomparison} shows the (peak) temperature probability densities for our local samples. The \textit{KINGFISH} sample, containing the most mature and metal-rich galaxies, shows the lowest distribution in $T$ and $T_{\rm peak}$ of all the samples \citep[see also figure 8 in][]{REMYRUYER13}. The dwarf galaxies in \textit{DGS}, having significantly lower metallicities, show a $T$ distribution similar to the (U)LIRGs in the \textit{GOALS} sample with ${\rm sSFR}<2\,{\rm Gyr^{-1}}$. Notably, galaxies with the highest sSFR ($>2\,{\rm Gyr^{-1}}$ in the \textit{GOALS} sample) show an overall warmer IR SED, i.e., higher $T$. Taken at face value, this suggests that, in addition to a trend in metallicity, galaxies with higher sSFR have warmer IR SEDs. This is not entirely unexpected since sSFR is correlated with the total FIR luminosity via the SFR, which can introduce a positive correlation with $T$ as shown above. However, strictly speaking, sSFR is more similar to a FIR luminosity \textit{density} and is therefore the preferred tool over the absolute $\LIR$ to compare galaxies at different redshifts. 
	
	The three low-metallicity analogs at $z\sim0.3$ show $10-30\,{\rm K}$ ($10-20\,{\rm K}$ in $T_{\rm peak}$) higher temperatures compared to the average of galaxies at similar metallicity, such as the dwarf galaxies (Figure~\ref{fig:tdependences}). Hence, they seem to be outliers in the nearby galaxy samples. However, note that the analogs have relatively high sSFR (between $1.6\,{\rm Gyr}^{-1}$ and $8.8\,{\rm Gyr}^{-1}$) compared to these local samples which is expected to boost their temperature further. Indeed, they show similar temperatures as the \textit{GOALS} (U)LIRGs at similar sSFRs (Figure~\ref{fig:tempcomparison}).
	We also note that the total FIR luminosities of the analogs do not match up with the local samples. Specifically, for their given metallicity, the analogs show $\sim1-2$ orders of magnitude higher $\LIR$ compared to the \textit{DGS} galaxies. The same trends have been seen in other samples of analog galaxies and are also indicated in our $z\sim5.5$ sample ($\logLIR\sim10$ at metallicities of $\oabund\sim7.5-8.0$ in the \citet{PILYUGIN05} calibration similar to the analogs -- by definition). This simply means that galaxies with the same $\LIR$ but at different redshifts are not comparable, which is reasonable given their vastly different structure and stellar mass. Instead, the sSFR (a proxy for the FIR luminosity density) or equivalently \halpha surface density are better parameters to compare low and high redshift galaxies as well as to provide better analog galaxies \citep[][]{FAISST16c,MASTERS16}.
	
	Summarizing, the local samples provide evidence that galaxies at low metallicity and high sSFRs are characterized by a warmer IR SED. The UV analog galaxies (extreme cases of low-redshift galaxies but with similar properties as high-redshift galaxies) strengthen these conclusions and suggest that high-redshift galaxies (characterized by low metallicity and high sSFRs) have a warmer IR SED compared to average galaxies at lower redshifts.

		\subsection{Possible reasons for a ``warm'' IR SED}\label{sec:whywarm}
	
	In the previous section, we have outlined evidence that galaxies with low metallicity and high sSFR show a ``warmer'' IR SED (i.e., higher $T$ and  $T_{\rm peak}$) compared to the average population. Ultimately, this may suggests that the average population of high-redshift galaxies (characterized by low metallicity and high sSFRs) have increased temperatures compared to typical galaxies at lower redshifts, which will be important to characterize their FIR properties as it is shown later. In the following, we list possible reasons that can cause a warm IR SED.
	
	First of all, the presence of an AGN is one possible cause of a mid-IR excess, but known AGN have been removed from the samples discussed above and the location of our three analogs on the BPT diagram suggest no AGN component.
	A detailed analysis of  X-ray emission from a dusty star forming galaxy at $z\sim5.6$ (SPT0346-52; $T=74\,{\rm K}$, $T_{\rm peak}=43\,{\rm K}$) using Chandra also suggests that its FIR emission originates from vigorous star formation activity instead of an AGN \citep[][]{MA16}. This is strengthened by a recent analysis of a sample of four ``Green Peas'' (low-redshift analogs of $z\sim3$ galaxies) that arrives with the same conclusion \citep[][]{LOFTHOUSE17}. 
	However, the presence of an obscured AGN cannot be ruled out and in fact it can be easily missed by traditional selections as shown in the case of GN20 at $z\sim4$ \citep[][]{RIECHERS14}, or other examples at lower redshifts \citep[][]{POPE08,COPPIN10}. This obscured AGN can even dominate the mid-IR continuum but it is shown by these studies that it has a minor impact on the FIR part of the SED and therefore only a modest effect on the temperature.
	
	We have found evidence that galaxies with low metallicity as well as high sSFR have increased $T$. This suggests that there might be at least two mechanisms that cause such a warm IR SED. These might be working at the same time, especially in low-metallicity and high sSFR analogs and likely also high-redshift galaxies.
	On one hand, a high sSFR is indicative of a high density of intense UV radiation fields originating from young star-forming regions that are heating up the surrounding ISM. In the local Universe, this might be the case for compact FIR luminous galaxies such as the ULIRGS in \textit{GOALS} (Figure~\ref{fig:tempcomparison}). At high redshifts, however, this could generally be the case as these galaxies are very compact and reveal a higher population averaged sSFR. This picture is generally supported by the detection of extended \Cii~emission from the warm ISM around local LIRGs \citep[][]{DIAZSANTOS14} and in samples of high-redshift galaxies (Section~\ref{sec:UVvsFIRmorph}). 
	  On the other hand, a likely explanation for the warm IR SEDs could be a lower optical depth in the ISM due to lower dust column densities at lower metallicities. As shown by \citet{SCOVILLE76}, the mid-IR shape of the SED is not only affected by the temperature but also by the mean optical depth of the ISM and the dust mass \citep[see also][]{SCOVILLE13}. In detail, for optically thin dust, the exponential mid-IR black-body tail is modified to a power-law and the peak wavelength is shifted blue-wards, both causing an increase in $T$ as well as $T_{\rm peak}$. The same trend but weaker is expected for a decreasing dust mass. At the same time, harder UV radiation might be prevalent in low-metallicity environments with massive stars caused by less efficient ISM cooling. Interestingly, \citet{PRESTWICH13} suggest also a higher number of X-ray sources at lower metallicity that could be responsible for heating the ISM. In the local Universe, these effects are expected in metal-poor and FIR faint dwarf galaxies (Figure~\ref{fig:tdependences}). At high redshifts, this could be commonly the case as these galaxies have metallicities significantly below solar.

\section{The \IRXB~diagnostic tool}\label{sec:IRXB}

	The general idea behind the \IRXB~correlation is that dust in the ISM and circum-galactic medium around galaxies absorbs the blue light from young O and B stars and re-emits it in the FIR. Therefore, the ratio of FIR to UV luminosity ($\LIR/\LUV={\rm IRX}$) is sensitive to the total dust mass and the geometry of dust grains. On the other hand, the UV color (measured by the UV continuum spectral slope $\beta$) is sensitive to the line-of sight opacity \citep[e.g.,][]{BOUWENS12}\footnote{Note that $\beta$ also depends on the age of the stellar population, the star-formation history, and metallicity. However, for young stellar populations as it is the case in young star-forming galaxies especially at high redshifts, the intrinsic UV slope is expected to be invariant \citep[][]{SCOVILLE15} and hence we can assume $\beta$ to be a good proxy for line-of-sight opacity.}. The  \IRXB~diagram combines these two measures and can therefore be used for studying the dust and gas properties and distribution in galaxies with a minimal amount of input.
	
\begin{figure}
\centering
\includegraphics[width=1.0\columnwidth, angle=0]{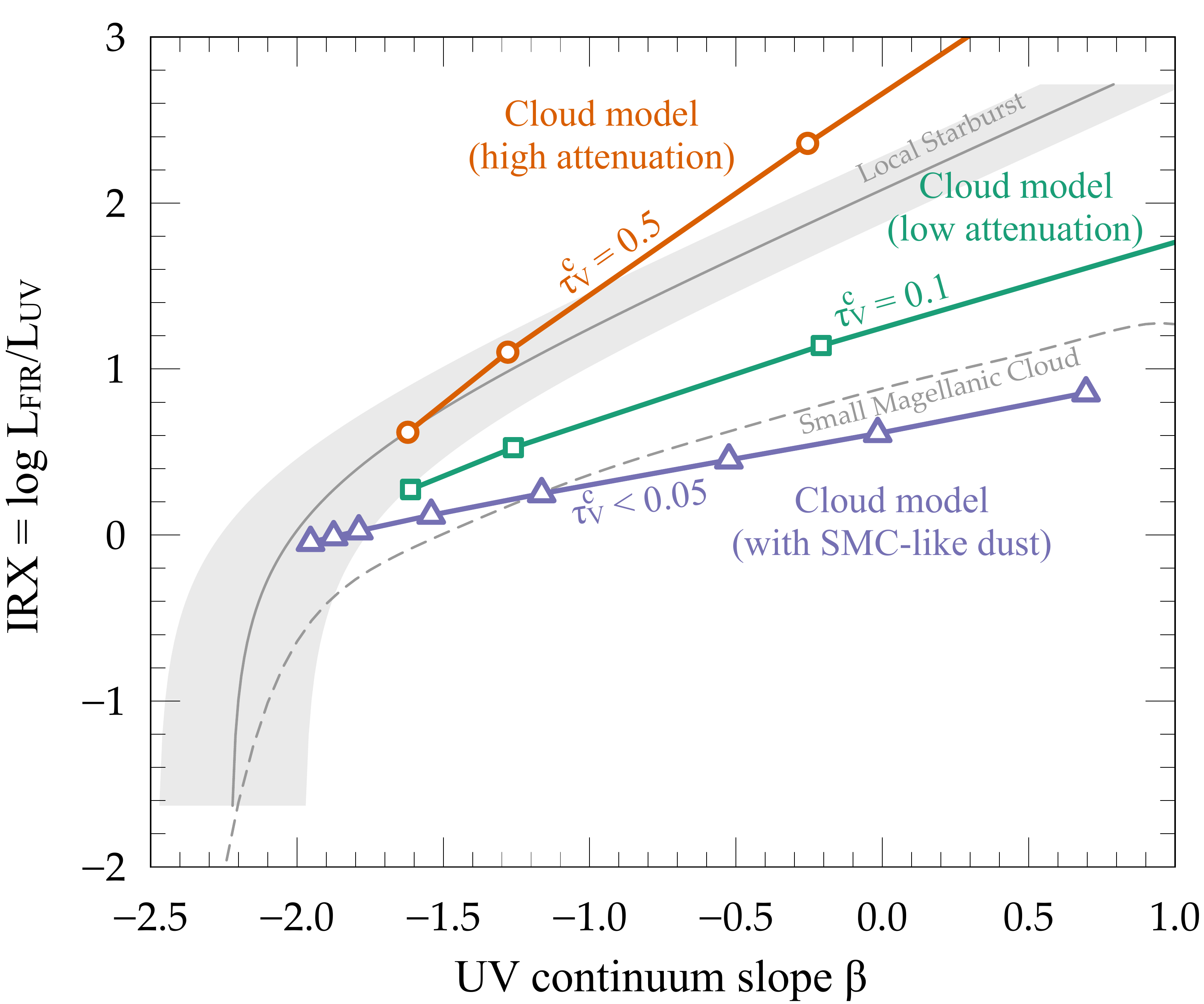}
\caption{Different models for the absorption of star light by dust in galaxies from \citet{CHARLOTFALL00}. The orange circle-line and the green square-line show models for a Poisson distribution of on average $1$ to $10$ clouds along the line-of-sight embedded in the ambient ISM with a wavelength dependent opacity $\tau(\lambda)\propto\lambda^{-1.3}$ and $\tau_{V}^{c}=0.5\;{\rm and}\;0.1$, respectively. The blue solid triangle-line shows the same model with the steeper $\tau(\lambda)$ relation of the SMC for $\tau_{V}^{c}<0.05$.
\label{fig:CF00simple}}
\end{figure}

 \subsection{Analytical models to explain \IRXB~in typical local to intermediate redshift galaxies}\label{sec:c00model}
 
 	\citet[][]{CHARLOTFALL00} \citep[see also][]{KONG04} describe a simple analytical model for the absorption of starlight in the ISM of galaxies, which can explain the location of typical galaxies on the \IRXB~diagram up to $z\sim3-4$ where detailed data exist. 
 	The model assumes dust in thermal equilibrium that is distributed in the ambient ISM in one of the following three geometrical configurations: \textit{(i)} a uniform foreground screen characterized by a single optical depth $\tau_{\lambda}^{sc}$, \textit{(ii)} a ``mixed slab'' characterized by a uniform mixture of dust and stars and optical depth $\tau_{\lambda}^{sl}$, and \textit{(iii)} a Poisson distribution of discrete clouds characterized by an optical depth per cloud ($\tau_{\lambda}^{c}$) and the average number of clouds mixed with stars along the line-of-sight to the observer ($\bar{n}$).
	Moreover, the models assume that the young stars are enshrouded in birth-clouds that dissolve after a finite lifetime of $t_{{\rm bc}}=10\,{\rm Myrs}$ \citep[see also][]{VALLINI17}. The total transmission for UV light is therefore time-varying, namely the product of the transmission function through the birth clouds \textit{and} the ambient ISM for $t<t_{\rm bc}$ and the transmission function \textit{alone} through the ISM otherwise.
 	
	These different models successfully explain the typical galaxy population on the \IRXB~diagram. Specifically, \citet{CHARLOTFALL00} find that the model featuring discrete clouds in the ISM represents best the IRX and UV color distribution of typical local starburst galaxies by assuming a simple power-law relation $\tau_{\lambda}^{c}= \tau_{V}^{c} (\lambda/5500\,{\rm \AA})^{-1.3}$ for the wavelength dependence of the optical depth of the ambient ISM. Such simple forms for $\tau_{\lambda}$ with a logarithmic slope of $-1.3$ are expected from various detailed measurements of the dust extinction curves in the Milky Way galaxy \citep[][]{SEATON79}, the Large Magellanic Cloud \citep{FITZPATRICK86}, local starburst galaxies \citep[][]{CALZETTI00}, and even normal star-forming galaxies at $z=2-4$ \citep[][]{SCOVILLE15}. For the birth-clouds, a similar optical depth is assumed with $\tau_{\lambda}^{bc}= 0.7 (\lambda/5500\,{\rm \AA})^{-0.7}$, but we note that the exact relation does not impact the following conclusions. 
	We show such a model in Figure~\ref{fig:CF00simple} for a stellar population of $300\,{\rm Myrs}$ and $\bar{n}$ running from $1$ to $10$ with $\tau_{V}^{c}=0.5$ (orange circles) and $\tau_{V}^{c}=0.1$ (green squares)\footnote{We assume a simple stellar population published with \texttt{GALAXEV} \citep[][]{BRUZUALCHARLOT03} for half-solar metallicity, a constant star-formation history, and a \citet{CHABRIER03} IMF.}.

	This simple model does successfully explain galaxies with similar dust properties as the local starbursts but it fails to reproduce galaxies with lower IRX values at a fixed UV color even for the lowest cloud optical depths (i.e., $\tau_{V}^{c}$). This part of the parameter space can be covered within this model by steepening the dust extinction curve, i.e., change the form of the wavelength dependent optical depth $\tau_{\lambda}^{c}$. The blue triangles in Figure~\ref{fig:CF00simple} show the cloud model with the more complex $\tau_{\lambda}^{c}$ of the metal-poor SMC parameterized by a 3-fold broken power-law \citep[e.g.,][]{PREVOT84} and assuming $\bar{n}=3$ and $\tau_{V}^{c}$ up to $0.05$. With this modification, the model can reproduce galaxies with lower IRX values at a given UV color for SMC-like dust. 
	At the same time, this shows how we can study the dust properties of galaxies with this very simple model of dust absorption and a minimal amount of data; the location of galaxies on the \IRXB~diagram tells us about their total amount of dust, the optical depth of their clouds in the ISM, and the wavelength dependence of their dust attenuation curve. These properties ultimately depend on other physical properties of the galaxies, for example their metallicity as shown in the case of SMC-like dust.

	\subsection{Correlation between location on the \IRXB~diagram and physical properties}\label{sec:physicsonIRXB}
	
	The \IRXB~diagram has also been studied thoroughly from an observational point of view and it is found that the galaxy properties change significantly as a function of position on this diagram \citep[e.g.,][]{HOWELL10,REDDY10,BOQUIEN12,CASEY14,ALVAREZMARQUEZ16}.
	
	For example, more mature, metal-rich, and dust-rich galaxies are expected to show high values of IRX, thus are found in the upper part of the \IRXB~diagram. Such galaxies are pictured as compact, dust-enshrouded, IR luminous star-forming systems that have likely experienced a recent starburst after a merger event \citep[e.g.,][]{LARSON16}. The blue colors of some of these galaxies can be explained by a patchy dust screen \citep[][]{CASEY14} or tidally stripped unobscured young stars or a small spatially unresolved and unobscured star-forming satellite as suggested by some IR luminous ($\logLIR>12$) local galaxies \citep[e.g.,][]{HOWELL10}.
	 An alternative scenario that does not involve starbursts or mergers is predicted by hydrodynamical simulations. These suggest that isolated normal main-sequence galaxies with $\logLIR<12$ can populate the upper left part of the \IRXB~diagram naturally due to their high gas fractions and high metal enrichment \citep[e.g.,][]{HOPKINS10,SAFARZADEH16}. The observed increase in gas fraction towards higher redshifts \citep[][]{TACCONI10,GENZEL15} would make such a configuration likely for the most metal rich galaxies in the early Universe.
	 	
\begin{figure}
\centering
\includegraphics[width=1\columnwidth, angle=0]{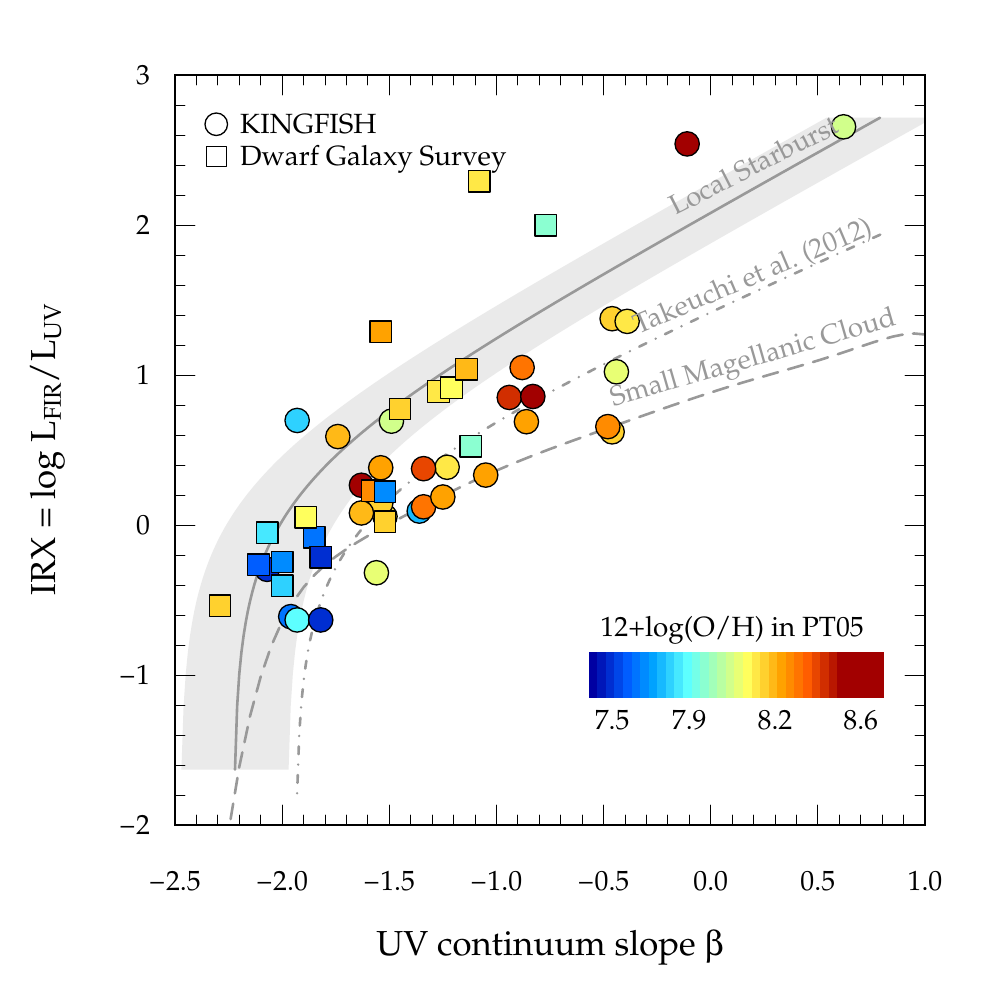}\\
\caption{
Correlation between metallicity and position on the \IRXB~diagram based on nearby galaxies from the \textit{KINGFISH} and \textit{DGS} samples. The low-metallicity dwarf galaxies are preferentially located at blue UV spectral slopes and low IRX values suggesting that galaxies at high redshift are located at similar locations.
\label{fig:irxmet}}
\end{figure}
		
	On the other hand, younger, metal-poor, and dust-poor galaxies are expected to show low values of IRX and blue UV colors and thus populate the lower parts on the \IRXB~diagram (see Figure~\ref{fig:irxmet}). They could be characterized by a steeper dust attenuation curve as it is seen in the metal-poor SMC, which causes a flatter \IRXB~relation for these galaxies \citep[see Section~\ref{sec:c00model} and discussion in][]{REDDY10}. Galaxies in the early Universe are expected to be young, dust-poor, and less metal enriched on average \citep[][]{MAIOLINO08,MANNUCCI10,FELDMANN15,FAISST16b,BIAN16,POPPING16}, thus are expected to reside on the lower parts of the \IRXB~diagram as predicted by the \citet{CHARLOTFALL00} model with a SMC-like dust attenuation curve. This has been directly observed in some FIR detected $z\sim3$ galaxies \citep[e.g.,][]{BAKER01,SIANA09,REDDY10} and there are also hints of this in our small sample of $z\sim0.3$ analogs; the analogs \Ga~and \Gc~(both sitting on the \IRXB~relation of local starbursts) have $\sim 0.2\,{\rm dex}$ higher metallicities than \Gb, which is located on the relation of the SMC.
	From the correlation with metallicity, we would also expect a correlation between position on the \IRXB~diagram and IR SED shape (Section~\ref{sec:tempdependencelocal}). As detailed in Section~\ref{sec:whywarm}, a warm IR SED can be due to heating of the ISM by a high amount of star-formation or due to a low dust optical depth in the ISM as shown by \citet[][]{SCOVILLE76}. The latter is supported by the simple dust model discussed in Section~\ref{sec:c00model}, which predicts a very low optical depth ($\tau_{V}^{c}<0.05$) for the clouds in galaxies close to the SMC relation.
	
	Extrapolating our knowledge gained from our low-redshift samples, we expect galaxies in the early Universe to show warm SEDs as well as high rates of star formation, which will make them appear close to the SMC relation.

	\section{The \IRXB~diagram at $z\sim5.5$}\label{sec:HIGHZ}
The advent of ALMA has allowed us to begin populating the \IRXB~diagram at $z>5$ \citep[e.g.,][]{OTA14,CAPAK15,MAIOLINO15,ARAVENA16,BOUWENS16}.
	\citetalias{CAPAK15} presents today's largest and most representative sample of normal main-sequence $z\sim5-6$ galaxies with detected \Cii~emission and continuum detections or limits at $158\,{\rm \mu m}$ along with robust UV measurements \citep[][]{BARISIC17}.  From the reasoning in Sections~\ref{sec:c00model} and \ref{sec:physicsonIRXB} we expect that most of the high-redshift galaxies reside in the lower part of the \IRXB~diagram, close to the relation of the SMC. However, we also expect a large variation in the metallicity and star-forming properties of high-z galaxies \citep[e.g.,][]{STEINHARDT14,FAISST16b,BARISIC17}. We therefore also expect a significant scatter in \IRXB.  
	
	The \citetalias{CAPAK15} sample consists of 9 star-forming Lyman Break galaxies at $1-4$ $L^*_{\rm UV}$ (the characteristic knee of the galaxy luminosity function) and one low luminosity quasar (\textit{HZ5}) at redshifts $5.1 < z < 5.7$. All galaxies are located in the 2 square-degree COSMOS field and therefore benefit from a multitude of ground and space-based photometric and spectroscopic data.
	
	The galaxies are spectroscopically selected via their rest-frame UV absorption features from spectra obtained with the \textit{Deep Extragalactic Imaging Multi-Object Spectrograph} \citep[DEIMOS,][]{FABER03} and represent a large range in stellar masses, SFRs, and UV luminosities at $z\sim5.5$.  A detailed analysis of their DEIMOS optical spectra and \textit{Spitzer} photometry properties suggests gas-phase metallicities around $\oabund \sim 8.5$, which is representative of the average metallicity of $\logm\sim10$ star-forming galaxies at $z\sim5$ \citep[][]{ANDO07,FAISST16b}.	
	
	All galaxies have been observed with ALMA at rest-frame $\sim150\,{\rm \mu m}$ and are detected in \Cii~emission at $158\,{\rm \mu m}$ \citep[][]{RIECHERS14,CAPAK15}. Four of them (\textit{HZ4}, \textit{HZ6}, \textit{HZ9}, and \textit{HZ10}) are detected in continuum. Furthermore, \textit{HZ6} and \textit{HZ10} have been detected in \nii~line emission and continuum at $205\,{\rm \mu m}$ \citep[][]{PAVESI16}. Both the low luminosity quasar (\textit{HZ5}) as well as \textit{HZ8} have a \Cii~detected companion (\textit{HZ5a} and \textit{HZ8W}) at their corresponding redshifts without significant detection in ground-based near-IR imaging data. \textit{HZ6} is a system of three galaxies (\textit{HZ6a}, \textit{HZ6b}, \textit{HZ6c})\footnote{These are called \textit{LBG-1a} through \textit{LBG-1c} in \citet{RIECHERS14} and \citet[][]{PAVESI16}.} with detected \Cii~emission and $158\,{\rm \mu m}$ continuum between \textit{HZ6a} and \textit{HZ6b}. \textit{HZ10} has a low luminosity companion (\textit{HZ10W}) that is, however, not spectroscopically confirmed in either optical or FIR lines.

	All UV properties (morphology, luminosity, UV color) have been derived accurately from deep near-IR imaging that has been recently obtained by \textit{HST} and is explained in detail in \citet{BARISIC17}.
	Table~\ref{tab:c15prop} provides a summary of the UV and IR properties of the \citetalias{CAPAK15} sample.

\begin{figure}
\centering
\includegraphics[width=1.05\columnwidth, angle=0]{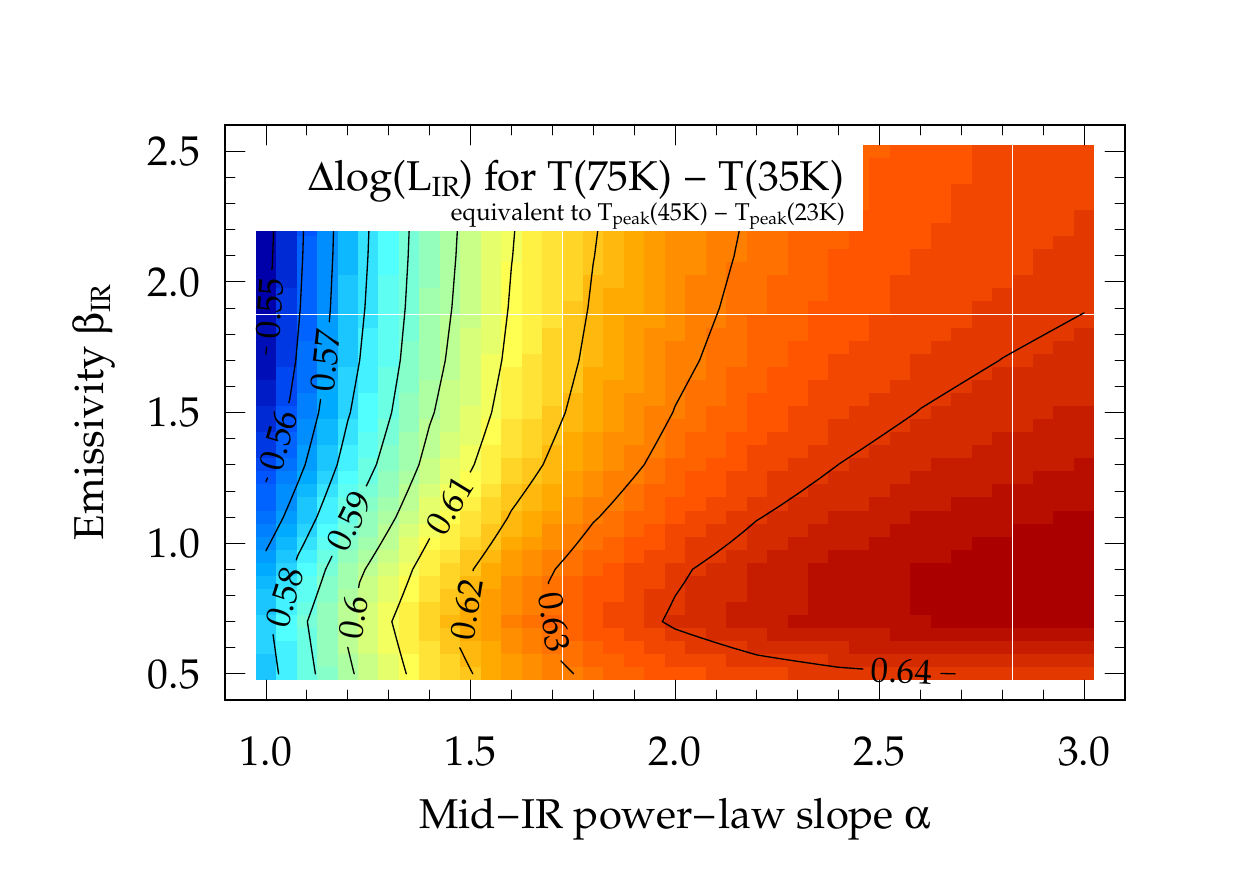}
\caption{
Impact of a ``warm'' IR SED on the total FIR luminosity measurements in the \citet{CASEY12} parameterization.
The color and contours show the \textit{increase} in $\log\LIR$ for a change in luminosity weighted temperature from $T=35\,{\rm K}$ to $T=75\,{\rm K}$ (corresponding to a change in $T_{\rm peak}$ from $23\,{\rm K}$ to $45\,{\rm K}$) as a function of the mid-IR slope $\alpha$ and the emissivity $\beta_{\rm IR}$. The total luminosity increases by $0.56-0.64\,{\rm dex}$ across the range in $\alpha$ and $\beta_{\rm IR}$ for $\Delta T=40\,{\rm K}$. Therefore $T$ is the main parameter that defines $\LIR$ and its measurement is important to obtain accurate $\LIR$ measurements in high-redshift galaxies where only a few data-points are available in the FIR.
\label{fig:dLIR}}
\end{figure}
	
	\subsection{FIR luminosity of high-z galaxies}\label{sec:firmeasurementhighz}
	
	The IR part of the SED is essential for studying the dust properties of galaxies. Although ALMA provides us with the necessary tools to observe the FIR part of the  electro-magnetic spectrum in high-redshift galaxies, it is very poorly constrained in normal (i.e., typical main-sequence) high-redshift galaxies due to their faintness, which makes such observations time consuming (see Appendix~\ref{sec:ALMA}). While good wavelength coverage in the FIR exists for a handful of bright sub-mm galaxies or FIR luminous lensed systems at $z>5$ \citep[][]{RIECHERS13,RIECHERS14,STRANDET16}, only one continuum data point at rest-frame $150\,{\rm \mu m}$ (close to the \Cii~line) is commonly measured for typical main-sequence galaxies at these redshifts.
	
	Particularly, the total FIR luminosity ($\LIR$), one of the most important parameters to constrain the location of galaxies on the \IRXB~diagram, is uncertain by a factor $>3$ because of the poorly constrained IR SED \citep[e.g.,][]{SCHAERER15}.
	The current lack of accurate studies of the IR SEDs at high redshifts makes it necessary to assume an SED shape for these galaxies. The temperature $T$ is a primary parameter for characterizing the IR SED shape in our parameterization (see Section~\ref{sec:irsedparameters}).
	
	The total FIR luminosity at $z>5$ is usually obtained by integrating over an IR SED assuming priors on temperature and other parameters from the $z<4$ universe, which is then normalized to the $158\,{\rm \mu m}$ continuum measurement of the galaxies. These parameters are typically $T=25-45\,{\rm K}$ ($T_{\rm peak} \sim 18-28\,{\rm K}$) and a range of $\alpha$ and $\beta_{\rm IR}$ corresponding to the average temperature of local (U)LIRGs (see \citetalias{CAPAK15}).  However, as shown in Section~\ref{sec:tempdependencelocal} by correlations in our local galaxy samples and analogs, there is evidence that this assumption might be incorrect for the average population of star forming galaxies at $z>5$ which have lower metallicities and higher sSFRs\footnote{Note that the constant background temperature emitted by the \textit{Cosmic Microwave Background} (CMB) is increased by a factor of seven to $\sim19\,{\rm K}$ at $z=6$. However, we expect this effect to increase the temperatures by less than $5\,{\rm K}$ on average at these redshifts \citep[][]{DACUNHA13}, which is small given the uncertainties of our measurements.}.
	
	Figure~\ref{fig:dLIR} illustrates the change in total FIR luminosity ($\Delta\log(\LIR)$) if the temperature $T$ or $T_{\rm peak}$ is underestimated by the typical difference between average $z<4$ (U)LIRGs and the low-metallicity and high sSFR analogs in our local samples.  This difference can be up to $\sim40\,{\rm K}$ ($T=35\,{\rm K}$ instead of $T=75\,{\rm K}$)\footnote{This corresponds to a difference between $T_{\rm peak} \sim 23\,{\rm K}$ and $T_{\rm peak} \sim 45\,{\rm K}$, i.e., $\Delta T_{\rm peak} \sim 22\,{\rm K}$.}, which would lead to a significant underestimation of $\LIR$ of $\sim0.6\,{\rm dex}$ as shown by the color gradient and black contours. The figure also shows that $T$ is the main parameter of uncertainty as $\Delta\log(\LIR)$ only changes by $\pm 0.04\,{\rm dex}$ over a large and conservatively estimated range of mid-IR slopes $\alpha$ and emissivities $\beta_{\rm IR}$.

	The assumption of a warmer IR SED of high-redshift galaxies is supported by our study of local galaxies and low-redshift analogs, but also by several direct measurements in the literature.
	
	\begin{itemize}

	\item \citet{ALVAREZMARQUEZ16} stack FIR SED of $z\sim3$ galaxies and find peak temperatures $T_{\rm peak} \sim 30-40\,{\rm K}$ ($T \sim 45-65\,{\rm K}$), which are at the warm end of the distribution of temperatures of local galaxies.
	
	\item \citet{SKLIAS14} measure the temperatures of \textit{Herschel} detected star-forming galaxies at $1.5 < z < 3.0$ and find peak temperatures of $35\,{\rm K} < T_{\rm peak} < 50\,{\rm K}$ ($55\,{\rm K} < T < 85\,{\rm K}$) in good agreement with our analogs. Also, \citet{SCHREIBER17} \citep[see also][]{SCHREIBER15} and \citet{KNUDSEN17} \citep[see also][]{WATSON15} find temperatures of $T_{\rm peak} \sim 40\,{\rm K}$ ($T \sim 65\,{\rm K}$) for average $z\sim4$ and one galaxy at $z\sim7.5$, respectively.
	
	\item \citet{STRANDET16} \citep[see also][]{GREVE12} present several strongly lensed galaxies at $z\sim5-6$ that are detected and selected in the sub-mm by the South Pole Telescope (SPT) survey \citep[][]{WEISS13} and observed between $100\,{\rm \mu m}$ and $3000\,{\rm \mu m}$ by SPT, ALMA, APEX, \textit{Herschel}, and \textit{Spitzer}.  We re-fit the FIR SEDs of these galaxies using the same method as our local samples and analogs and find temperatures of $60\,{\rm K} < T < 80\,{\rm K}$ ($37\,{\rm K} < T_{\rm peak} < 48\,{\rm K}$), which is in good agreement with our local analogs (see Table~\ref{tab:irsedprop} and black arrows in Figure~\ref{fig:tempcomparison}).

	\item The analysis of the starburst \textit{AzTEC-3} at $z=5.3$ \citet{RIECHERS14} also results in a temperature of $88^{+10}_{-10}\,{\rm K}$ ($T_{\rm peak}\sim51\,{\rm K}$) and an SED similar to the low-redshift analogs.
	
	\item Finally, \citet{PAVESI16} extended the FIR wavelength coverage of \textit{HZ6} and \textit{HZ10} from the \citetalias{CAPAK15} sample by measuring fluxes in the two sidebands at $158\,{\rm \mu m}$ and $205\,{\rm \mu m}$ independently with ALMA. Assuming a uniform prior for temperature ($10-100\,{\rm K}$, i.e., $5-60\,{\rm K}$ in $T_{\rm peak}$) and $\beta_{\rm IR}=1.7\pm0.5$, they derive temperatures of $60^{+35}_{-27}\,{\rm K}$ ($T_{\rm peak}\sim35\,{\rm K}$) and $36^{+25}_{-10}\,{\rm K}$ ($T_{\rm peak}\sim25\,{\rm K}$) for \textit{HZ6} and \textit{HZ10}, respectively. As the authors discuss, these temperature measurements are very uncertain as the current data does not cover wavelengths near the IR peak that are sensitive to warm dust. An additional measurement at bluer wavelengths (e.g., rest-frame $122\,{\rm \mu m}$, see Appendix~\ref{sec:ALMA}) would increase the confidence in the estimated temperature.
	
	\end{itemize}

	Following the above reasoning, we re-compute the total FIR luminosities of the 12 $z\sim5.5$ systems from \citetalias{CAPAK15} assuming a temperature prior of $60\,{\rm K} < T < 90\,{\rm K}$ ($T_{\rm peak} \sim 35-50\,{\rm K}$) for the IR SED. As in \citetalias{CAPAK15}, we create IR SEDs using the \citet{CASEY12} parameterization that we normalize to the observed $158\,{\rm \mu m}$ continuum emission of the galaxies, which is the only data-point available in the IR. The result is marginalized over a grid of mid-IR slopes and emissivities ($1.5 < \alpha < 2.5$, $1.0 < \beta_{\rm IR} < 2.0$).  The total FIR luminosity is derived by integration between $3\,{\rm \mu m}$ and $1100\,{\rm \mu m}$ and the errors are derived from the uncertainty in the photometric measurement at $158\,{\rm \mu m}$ and the marginalization over the assumed range in $\alpha$ and $\beta_{\rm IR}$.
	We list the updated $\LIR$ values in Table~\ref{tab:irsedprop}, which, as expected, are increased by $\sim0.6\,{\rm dex}$ compared to the previous results assuming a temperature prior with $25\,{\rm K} < T < 45\,{\rm K}$ ($T_{\rm peak} \sim 18-28\,{\rm K}$) as in \citetalias{CAPAK15}.

\begin{figure*}
\centering
\includegraphics[width=2\columnwidth, angle=0]{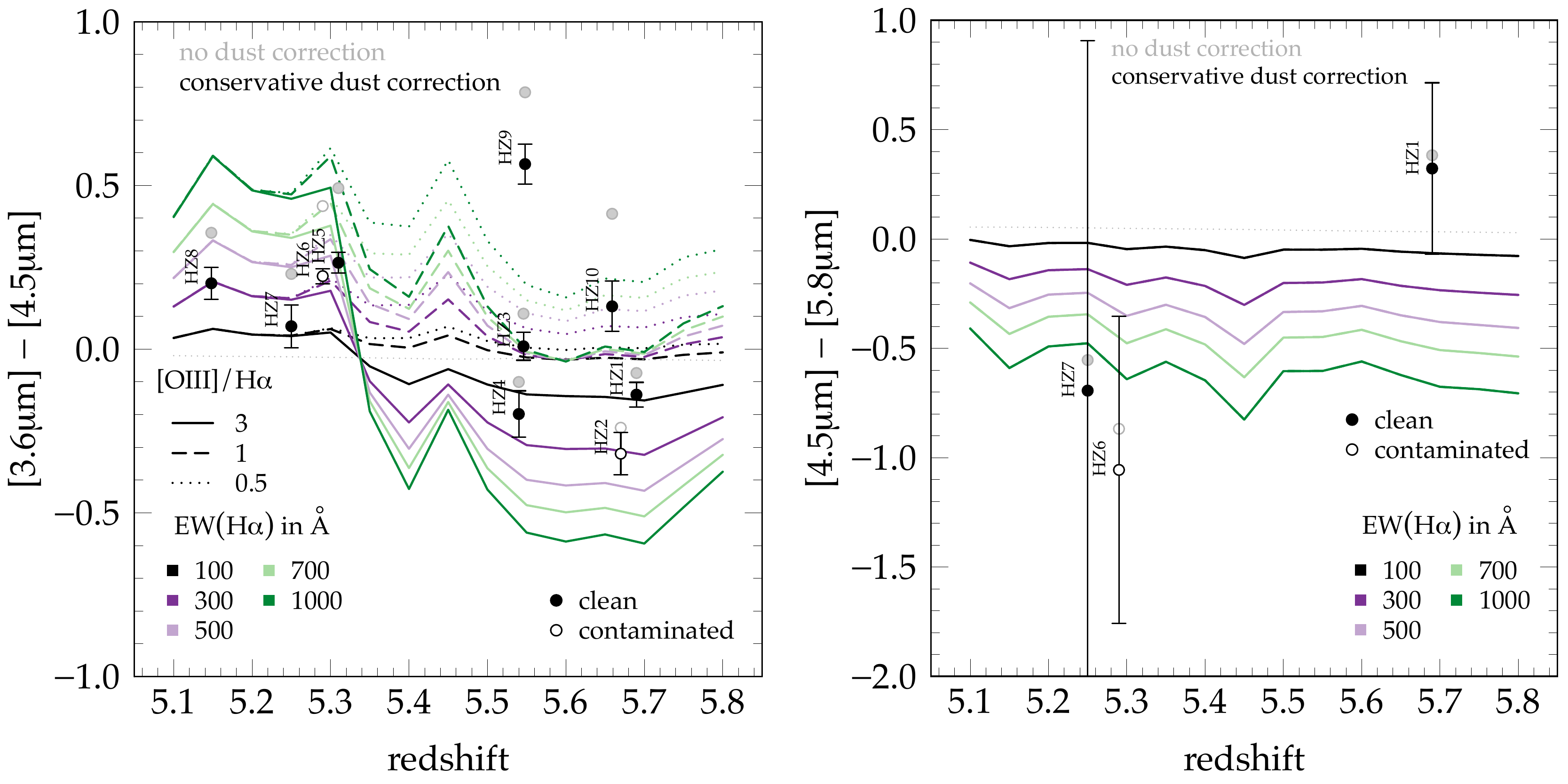}
\caption{Estimation of $\ewha$ and \oiii/\hbeta line ratios from \textit{Spitzer}/IRAC photometry for the $z\sim5.5$ galaxies. The two panels show \textit{Spitzer} $[3.6\,{\rm \mu m}]-[4.5\,{\rm \mu m}]$ (left) and $[4.5\,{\rm \mu m}]-[5.8\,{\rm \mu m}]$ (right) for a suite of models with different $\ewha$ and \oiii/\halpha ratios (see text). The black points show the data of $z\sim5.5$ galaxies corrected for dust (using the maximal $\ebmv$ derived from either IRX or $\beta$, see text), the gray points show the observed data (no dust correction). We indicate galaxies with contaminated \textit{Spitzer} photometry, e.g., due to multiple components (empty points).
\label{fig:ewhamodels}}
\end{figure*}

	\subsection{Constraints on metallicity and sSFR as derived from \textit{Spitzer} colors}\label{sec:spitzermeasure}
	
	The strong emission lines in high redshift galaxies can alter their observed photometry that in turn allows us to study them. In particular, \textit{Spitzer} observations allow us to constrain the fluxes of strong optical lines such as \oiii~and \halpha of galaxies at $z>3$ and thus can put strong constraints on their (specific) SFR, metal content, hence evolutionary stage of these galaxies \citep[e.g.,][]{SHIM11,FAISST16a,RASAPPU16}.
		
	Figure~\ref{fig:ewhamodels} shows the expected $[3.6\,{\rm \mu m}]-[4.5\,{\rm \mu m}]$ and $[4.5\,{\rm \mu m}]-[5.8\,{\rm \mu m}]$ color in the redshift range $5.1 < z < 5.8$ for different \halpha EWs and \oiii/\halpha ($\propto$~\oiii/${\rm H\beta}$) emission line ratios\footnote{We assume templates from the \citet{BRUZUALCHARLOT03} library with half-solar (stellar) metallicity and a $500\,{\rm Myr}$ old constant SFH. As shown in \citet{FAISST16a}, the choice of these parameters, as long as reasonable, has little impact on the intrinsic model color ($< 0.15\,{\rm mag}$ for $z > 5$).}. The black symbols show the \textit{dust corrected}\footnote{For dust correction we apply the maximum of the $\ebmv$ calculated directly from $\LIR/\LUV$ using the relation between IRX and $A_{1600}$ \citep{MEURER99} and from $\beta$ using the relation between $A_{1600}$ and $\beta$ \citep{CALZETTI00}.} colors of the galaxies in the \citetalias{CAPAK15} sample with their names indicated.
	Note that while all galaxies have reliable measurements at $3.6\,{\rm \mu m}$ and $4.5\,{\rm \mu m}$, only three galaxies have measured $5.8\,{\rm \mu m}$ fluxes due to the only shallow coverage at $5.8\,{\rm \mu m}$. Furthermore, the latter measurements have large uncertainties and are therefore are only meaningful for \textit{HZ1} and \textit{HZ6}.
	For galaxies at $5.1<z<5.3$, the $[3.6\,{\rm \mu m}]-[4.5\,{\rm \mu m}]$ color is a good estimator of the \halpha equivalent-width \citep[e.g.,][]{FAISST16a,RASAPPU16}, $\ewha$, while at higher redshifts it becomes degenerated with the \oiii/\halpha~ratio.	
	Both \textit{Spitzer} $[3.6\,{\rm \mu m}]-[4.5\,{\rm \mu m}]$ and $[4.5\,{\rm \mu m}]-[5.8\,{\rm \mu m}]$ colors suggest $\ewha > 500\,{\rm \AA}$ integrated over the three sub-components of \textit{HZ6}, which could indicate a recent starburst event maybe induced by the tidal interaction between these three galaxies. In comparison, these colors suggest $\ewha < 500\,{\rm \AA}$ for \textit{HZ1}, \textit{HZ7}, and \textit{HZ8}. 
	In addition, we estimate \oiii/\halpha$\lesssim1$ (or \oiii/\hbeta$<2.86$) for \textit{HZ9} and \textit{HZ10} and \oiii/\halpha$\gtrsim1$ (or \oiii/\hbeta$>2.86$) for \textit{HZ1}, \textit{HZ2}, and \textit{HZ4} from their $[3.6\,{\rm \mu m}]-[4.5\,{\rm \mu m}]$ colors.
	Assuming all these galaxies sit on the BPT ''main-sequence`` locus of high-redshift galaxies \citep[e.g.,][]{KEWLEY13,STEIDEL14,MASTERS16,SANDERS16,STROM16}, we can relate the \oiii/\halpha line ratio to \nii/\halpha that is a proxy for the gas-phase metallicity. Hence, we would expect \nii/\halpha$ > -0.6$ for \textit{HZ9} and \textit{HZ10}, which translates into a gas-phase metallicity of $\oabund \gtrsim 8.5$ according to \citet[][]{CURTI16}. The same reasoning leads to $\oabund \lesssim 8.5$ for \textit{HZ1}, \textit{HZ2}, and \textit{HZ4}.
	Note that these metallicity estimates are in good agreement with the strong UV absorption features seen in these galaxies \citep{BARISIC17}.

Overall this suggests that low-metallicity, high sSFR low-redshift galaxies are indeed good analogs for high-redshift systems and that we should be adopting higher temperatures for the FIR SEDs.  Interestingly, the exception is HZ10, which red \textit{Spitzer} colors and rest-UV spectrum (Section~\ref{sec:spitzermeasure}) indicate a more mature system with higher metallicity and lower sSFR in line with the cooler temperature estimated by \citet{PAVESI16} and being the strongest FIR continuum emitter in the \citetalias{CAPAK15} sample.

\begin{figure*}
\centering
\includegraphics[width=2.1\columnwidth, angle=0]{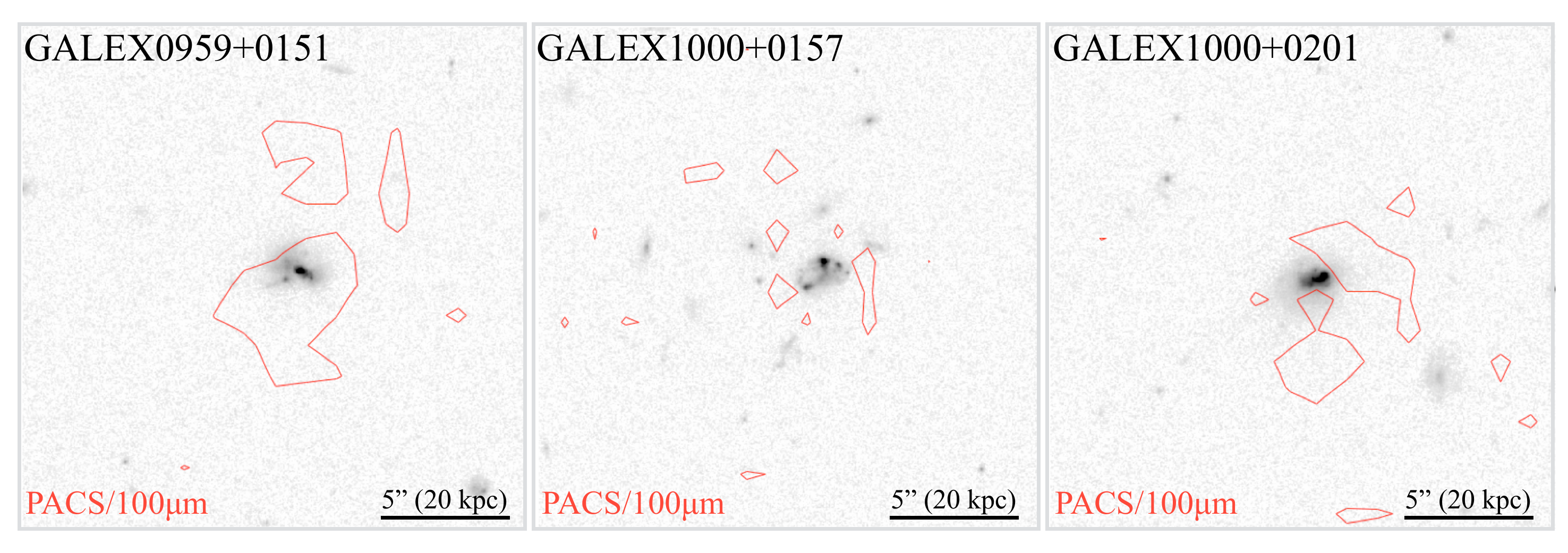}\\
\includegraphics[width=2.1\columnwidth, angle=0]{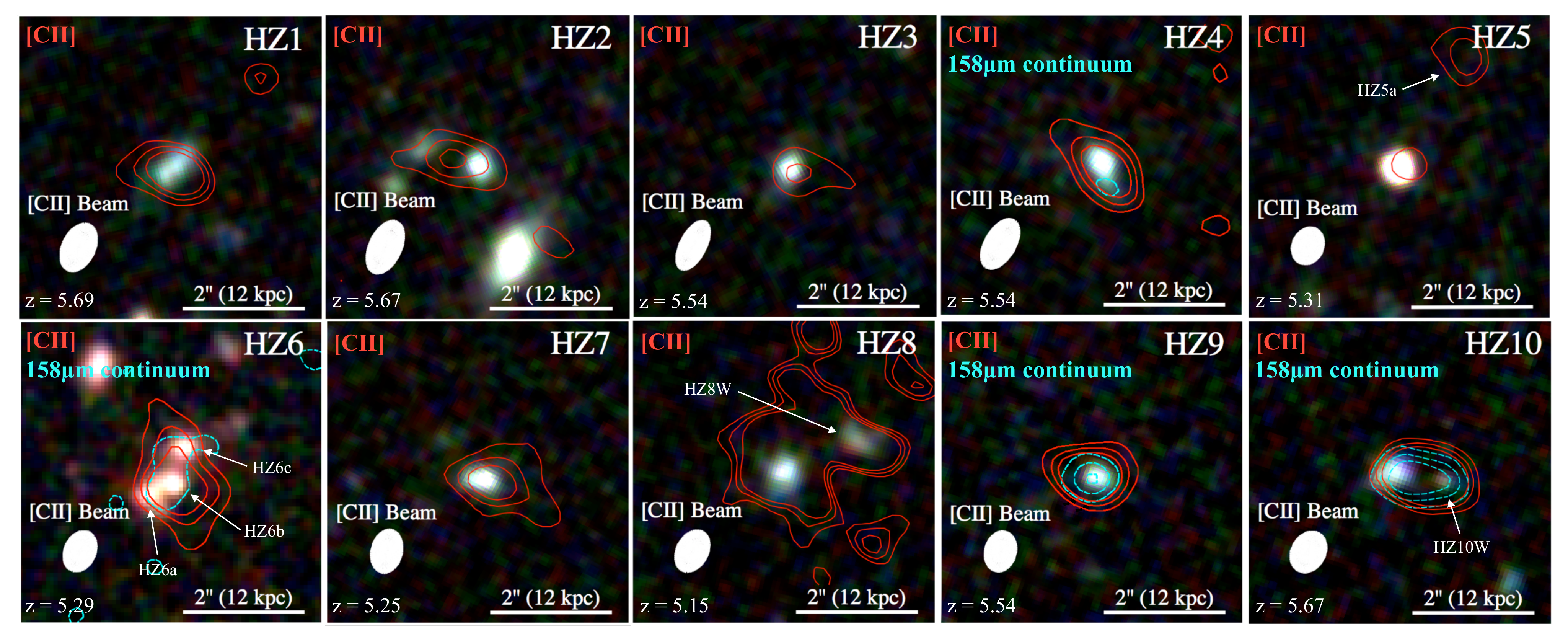}
\caption{
Spatial location of UV and FIR emission of the three analogs (upper large panels) and $z\sim5.5$ galaxies from \citetalias{CAPAK15} (lower small panels). For the analogs we show the ACS/F814W (gray scale) together with the PACS/$110\,{\rm \mu m}$ emission (red contours) for $2$, $6$, and $10\,\sigma$. For the $z\sim5.5$ galaxies we show the F105W$+$F125W$+$F160W color cutouts together with the FIR \Cii~line emission (red contours) and the continuum at $158\,{\rm \mu m}$ (cyan dashed contours) both shown for $2$, $6$, and $10\,\sigma$. The \Cii~beam FWHM is indicated by the white ellipse. Note the offset in \Cii~from the UV emission (indicating regions of unobscured star-formation) and the extended \Cii~emission, potentially indicating a turbulent and warm ISM due to the strong UV radiation of young stars.
\label{fig:offset}}
\end{figure*}
	
	\subsection{Comparison of UV, FIR, and \Cii~morphology}\label{sec:UVvsFIRmorph}
	
	The galaxies in \citetalias{CAPAK15} are all resolved in the ALMA FIR line and continuum observations at a resolution of $0.5\arcsec$, which allows, together with their high-resolution near-IR \textit{HST} images, a spatial study of rest-frame UV (a proxy for unobscured star formation) and FIR continuum (a proxy for the dust distribution and obscured star formation) in typical $z>5$ galaxies for the first time (Figure~\ref{fig:offset}). 
	
	The FIR continuum detected galaxy \textit{HZ9} shows no offset between UV and FIR emission, indicative of a dust-obscured compact central star-formation region. In contrast, the FIR emission of \textit{HZ4} and \textit{HZ10} is offset by up to by up to $5\,{\rm kpc}$ from the location of the unobscured star formation, which could indicate a substantial re-distribution of dust clouds due to UV radiation pressure, tidal interactions by a recent or on-going merger event or the production and growth of dust caused by a recent starburst \citep[c.f.,][]{SCHNEIDER16}. Such an offset of FIR continuum and unobscured star formation is also seen in the two local analogs (\Gb, \Gc) with the lowest metallicities and lowest dust content (i.e., IRX) as shown in the upper panels of Figure~\ref{fig:offset}.
	
	Emission from singly ionized carbon is commonly originating from the photo-dissociation regions (PDRs) on the surfaces of molecular clouds. The \Cii~luminosity ($\LCII$) is therefore used as a proxy for the star formation in galaxies, therefore it should coincide well the locations of star formation \citep[e.g.,][]{CARILLI13,DELOOZE14}. 
	The sample of \citetalias{CAPAK15}, however, shows offsets between \Cii~and UV emission of up to $5\,{\rm kpc}$ as well as a very extended \Cii~emission over $5-10\,{\rm kpc}$, a factor of $\sim5$ larger than the typical UV size the galaxies (Figure~\ref{fig:offset}).
	The extended \Cii~emission could indicate that a non-negligible fraction of \Cii~is originating from the warm neutral gas in the diffuse ISM that is heated by the strong UV radiation of young massive stars in high-redshift galaxies \citep[e.g.,][]{PINEDA13,DIAZSANTOS14,HERRERACAMUS15,VALLINI15}.
	Furthermore, recent hydrodynamical simulations suggest that offsets of up to $10\,{\rm kpc}$ between \Cii~emission and regions of on-going star formation are expected in young primeval galaxies at high redshift due to their strong stellar radiation pressure \citep[][]{VALLINI13,MAIOLINO15}.
	
	Overall, the FIR, \Cii, and UV morphologies would support a picture of a turbulent, low dust-column density ISM with a greater optically thin fraction which would cause a higher observed temperature.  This is consistent with our analysis of the low metallicities and high sSFRs local galaxies (Section~\ref{sec:tempdependencelocal}) and could suggest a change in the relation between SFR and $\LCII$ found in galaxies at lower redshifts \citep[e.g.,][]{ARAVENA16}.
	
		Alternatively, the offsets described above could be due to differential dust obscuration as it is commonly observed in SMGs and DSFGs \citep[][]{RIECHERS14,HODGE15,HODGE16}. For example, \Cii~is much less affected by dust obscuration and can therefore be detected at places where the UV light is completely absorbed, i.e., places of obscured star formation. On the other hand, the \Cii~must then be destroyed or expelled at locations of unobscured star formation.

\begin{figure*}
\centering
\includegraphics[width=2\columnwidth, angle=0]{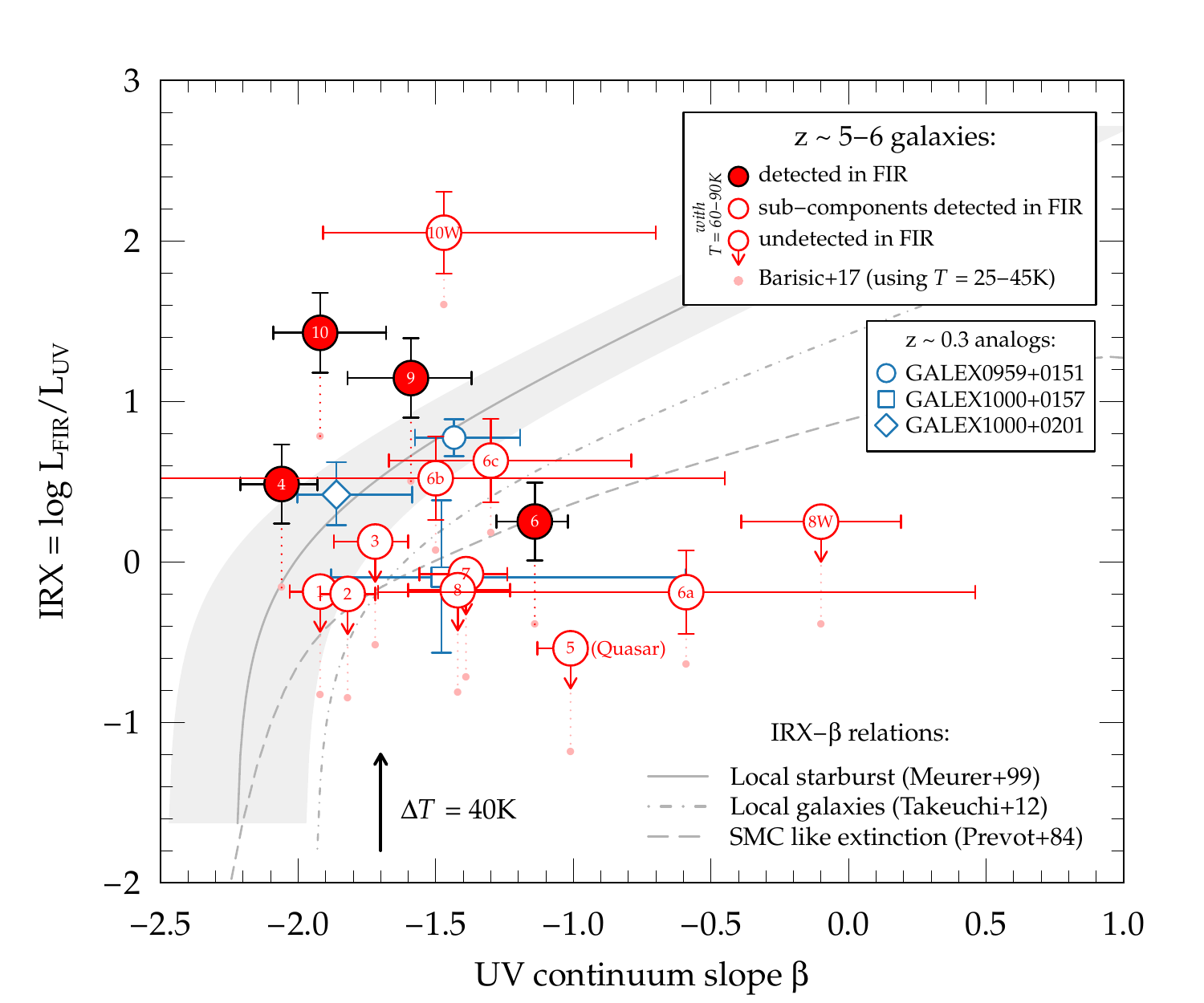}
\caption{The \IRXB~diagram at $z\sim5.5$ for a maximally warm IR SED prior with luminosity weighted temperature $T=60-90\,{K}$ ($T_{\rm peak} \sim 35-50\,{\rm K}$). We show the $z=5-6$ galaxies from \citet{BARISIC17} with FIR detection (red filled circles) and without FIR detection (empty circles, upper limits) as well as the FIR detected sub-components as described in their paper (empty circles). The position of the galaxies with a cooler IR SED prior ($T=25-45\,{\rm K}$, i.e., $T_{\rm peak} \sim 18-28\,{\rm K}$) as assumed previously \citep[e.g.,][]{CAPAK15} is indicated as faint red dots. The arrow indicates the change in IRX for $\Delta T=40\,{\rm K}$. Our three $z\sim0.3$ analogs is shown in blue and are consistent with the location of the high-redshift galaxies. Common \IRXB~relations for local starbursts (\citet[][]{MEURER99}, with scatter), \citet{TAKEUCHI12}, and galaxies with SMC-like dust \citep[][]{PREVOT84,PETTINI98} are indicated in as solid, dot-dashed, and long-dashed gray lines, respectively. Note that the upper limits of the FIR undetected $z=5-6$ galaxies are mostly consistent with the \IRXB~relation expected for SMC-like dust using the maximally warm IR SED prior. However, some galaxies still show a deficit in IRX compared to the SMC relation and cannot be explained by common analytical models of dust attenuations in galaxies (see Section~\ref{sec:model}).
\label{fig:final}}
\end{figure*}

	\subsection{Updated IRX$-\beta$ diagram of $z\sim5.5$ galaxies}\label{sec:irxb6}
	
	Figure~\ref{fig:final} shows the \IRXB~diagram at $z\sim5.5$ using the \textit{HST}-based $\beta$ measurements \citep{BARISIC17} and $\LIR$ derived for a maximally warm IR SED prior ($60\,{\rm K} < T < 90\,{\rm K}$, $T_{\rm peak} \sim 35-50\,{\rm K}$) for the galaxies in \citetalias{CAPAK15}. The FIR continuum detected high-redshift galaxies are shown as solid red symbols, while the non-detections are shown as empty red symbols with arrows. FIR detected companion galaxies are shown as empty red symbols. Previous measurements (if available) that assume a cooler IR SED prior of $T=25-45\,{\rm K}$ ($T_{\rm peak} \sim 18-28\,{\rm K}$) are shown as light red points.
	
	 If the warmer IR SED priors are correct, this would push the galaxies up by $\sim0.6\,{\rm dex}$ in IRX. This brings the galaxies at low IRX in better agreement with known dust properties measured in local starbursts \citep[][]{MEURER99} \citep[and its re-examination by][]{TAKEUCHI12} and models adopted for the metal-poor SMC \citep[][]{PETTINI98,REDDY06}.
	  We note that the applied warm IR SED prior is motivated by the extrapolation of trends found in local galaxy samples as well as three low-redshift analogs which we think are reasonable representatives of a large fraction of high-redshift galaxies. However, the true temperature distribution of typical high-redshift galaxies is still unknown and we therefore argue that our temperature prior is a very reasonable upper limit and so are the updated IRX values. Taking the early measurements by \citet{PAVESI16} as well as the trends in the local sample at face value, a temperature gradient across the \IRXB~diagram is not unexpected. For example, lower temperatures for \textit{HZ9} and \textit{HZ10}, which would reduce their IRX and bring them in better agreement with the local starburst and the \citet{TAKEUCHI12} relation, are entirely possible. More direct measurements of the IR SED at high redshifts and larger samples of low redshift analogs are necessary to study such effects.

	 We find that galaxies at $z\sim5.5$ occupy a large fraction of the \IRXB~diagram with significant scatter, which indicates a large variety in their physical properties already 1 billion years after the Big Bang. This scatter has to be understood in order to use the \IRXB\ relation to predict the dust properties of high-redshift galaxies from UV colors only as it is commonly done. The current sample of galaxies is too small to make such statistical predictions as well as point out the precise trends on the \IRXB\ diagram as a function of galaxy properties. In the following, we can therefore only derive a patchy understanding of how galaxies behave in relation to $\beta$ and IRX.

	 We find galaxies with blue UV colors ($\beta< -1.5$) and large IRX values on or above the curve for local starbursts (\textit{HZ9}, \textit{HZ10}). As discussed in Section~\ref{sec:spitzermeasure}, these are likely amongst the most metal-rich and mature  main-sequence galaxies at these redshifts. The blue UV color of these galaxies could be explained by a recent perturbation of the dust clouds by tidal interactions or a patchy optically thick dust screen (Section~\ref{sec:physicsonIRXB}). The former could be the case for \textit{HZ10} showing a large offset between UV and FIR emission, while the latter might be the case for \textit{HZ9} for which UV and FIR emission coincide (Section~\ref{sec:UVvsFIRmorph}). We note that \textit{HZ4}, which is also elevated somewhat above the location of local galaxies but not as much as the others, is expected to be less metal enriched. Also, its morphology (Figure~\ref{fig:offset}) shows that its FIR emission is offset in a tail pointing to north-east. The large fraction of unobscured UV emission along the line-of-sight causes its blue UV color ($\beta\lesssim -2$) but its substantial dust mass causes a high IRX value. 
	 
	 Next to galaxies showing increased IRX values, we find galaxies with low in IRX values (even \textit{upper limits}) and red UV colors (e.g., \textit{HZ6a}, \textit{HZ7}, \textit{HZ8}, or \textit{HZ8W}). This is even the case for a maximally warm IR SED prior that is assumed here and would be even more significant if no temperature evolution is assumed. Such galaxies are expected to have a significant line-of-sight dust attenuation but only little dust FIR emission, i.e., a low total dust mass. The recent \textit{HST} observations provide very accurate $\beta$ measurements of these galaxies as shown by the simulations in \citet[][]{BARISIC17}. Therefore we conclude that these galaxies have a deficit in IRX and are not biased in $\beta$.  Also, the UV spectra and \textit{Spitzer} colors strongly suggest that these galaxies are young and not dominated by old stellar populations that could cause a red UV color \citep[e.g.,][]{HOWELL10,NARAYANAN17}.
	 These galaxies are curious because such a configuration of IRX and $\beta$ is difficult to explain with current models of dust attenuation in galaxies. As shown in Section~\ref{sec:c00model}, the simple analytical model of \citet{CHARLOTFALL00} cannot cover that part of the parameter space even assuming low dust opacities and a steep SMC-like dust attenuation curve.
	From the \textit{Spitzer} colors (Section~\ref{sec:spitzermeasure}) as well as UV spectra \citep[][]{BARISIC17}, we would expect these galaxies to be primeval and metal-poor. The large offsets between \Cii~emission and star forming regions and the extent of \Cii~emission could be indicative of vigorous star formation, strong stellar radiation pressure, and substantial heating of the diffuse ISM. We suggest that the enhanced turbulence in these galaxies paired with merger events significantly alters their spatial dust distribution.
	 In the next section, we propose an updated model for dust absorption in the ISM of these galaxies that can explain the location of these peculiar galaxies on the \IRXB~diagram by taking into account geometrical effects of the dust and star distribution.

\begin{figure}
\centering
\includegraphics[width=1.0\columnwidth, angle=0]{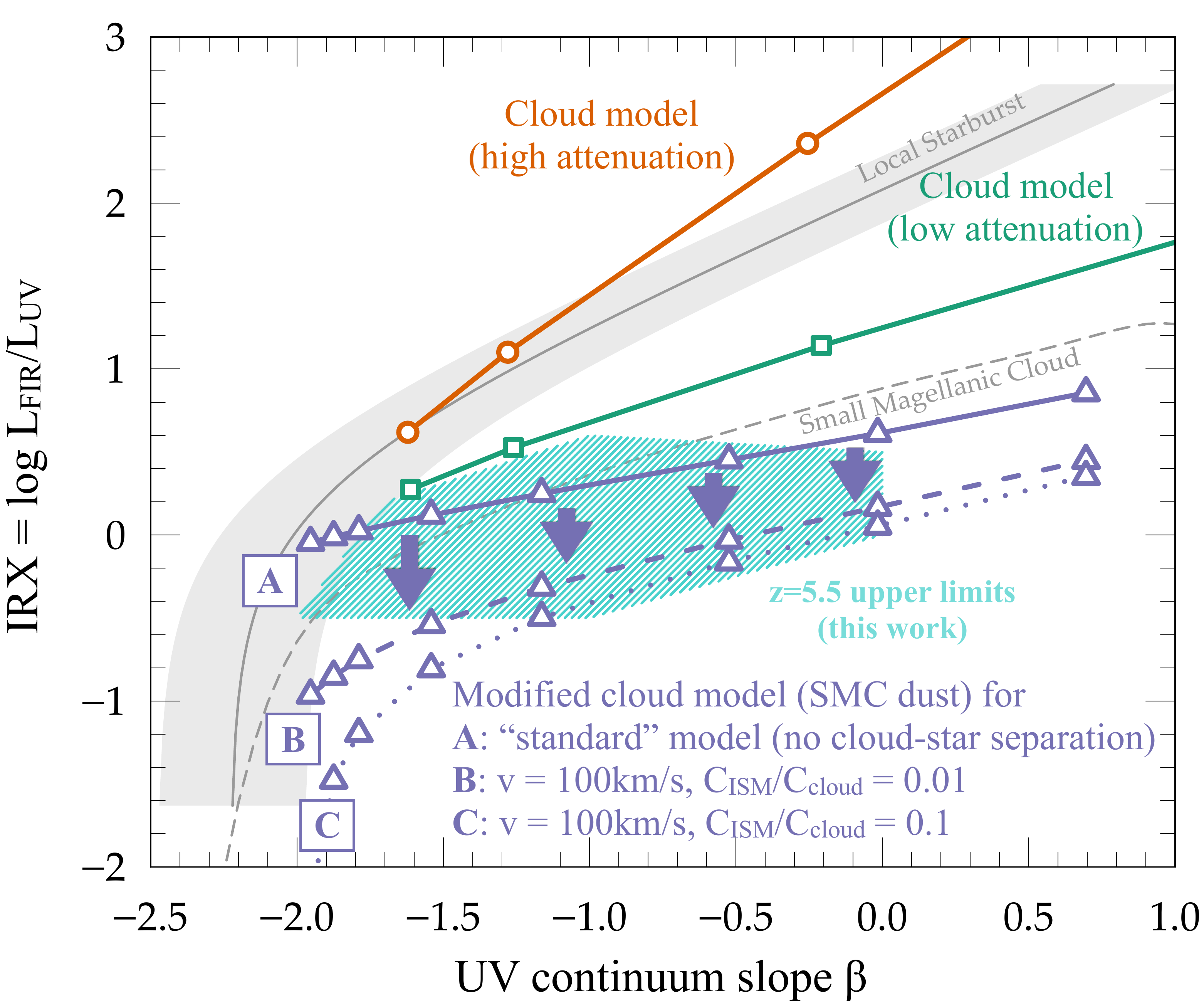}
\caption{Same as Figure~\ref{fig:CF00simple}, but here including our proposed updated model that can explain the low IRX values of the FIR undetected $z>5$ galaxies (cyan hatched region). The revised model takes into account the separation of clouds and young stars due to the increased turbulent motion in high-redshift galaxies. Shown are two different cloud column density contrasts of 0.01 (long-dashed blue triangle-line) and 0.1 (dotted blue triangle-line) and a turbulent motion of $v~\approx~\sigma_{\rm ISM}~\approx~100\,{\rm km/s}$ of the surrounding warm ISM is assumed.
\label{fig:model}}
\end{figure}

	 \subsection{Dust model for FIR faint high-z galaxies: Separating clouds from stars}\label{sec:model}
	 
	As discussed in Section~\ref{sec:c00model}, a simple analytical model for dust absorption in galaxies with well mixed dust and stars in thermal equilibrium \citep{CHARLOTFALL00} is able to explain the majority of galaxies on the \IRXB~diagram. However, it is not able to explain the FIR faint galaxies at very low IRX at a given UV color found at $z>5$ even for low dust opacities and a steep SMC-like dust attenuation curve. In order to explain the location of these galaxies, the dust clouds have to be in a configuration that allows them to attenuate the UV light enough to cause the red $\beta$ but at the same time the FIR emission has to be decreased by $0.5-1.0\,{\rm dex}$ with respect to SMC-like galaxies.
	
	 We propose that this can be achieved by a spatial separation of the clouds and the star-forming regions that naturally occurs as the birth clouds are disrupted by the radiation pressure of the stars and turbulences in the ISM. Instead of letting the birth-clouds disappear after $10\,{\rm Myrs}$ (the typical life time of O and B stars), we modify the \citet{CHARLOTFALL00} prescription such that the birth-clouds are separated from the star-forming regions by $\delta s(t) = t\times\delta v$ with $t<10\,{\rm Myrs}$.  The total FIR luminosity is decreased roughly by a factor of $\propto 1/\delta s^2$ (where $\delta s$ is the distance between star-forming region and cloud) due to the smaller angle along the line-of-sight covered by the cloud, while the UV color from the obscured regions is not significantly changed. 

	We assume that initially, the star clusters have the same velocity as the birth-cloud. After that, the cloud is subject to hydrodynamic forces and pressures, e.g., caused by the radiation of young stars or merging events, which causes the clouds to run into the nearby low-density warm ISM. The velocity change of the clouds can be estimated by a momentum conservation argument and depends on the column density ratio of the clouds and the surrounding ISM as well as the relative velocity $v$ of the external medium
	
	\begin{equation}
	\delta v \approx \left( \frac{C_{\rm ISM}}{C_{\rm cloud}} \right) \times v.
	\end{equation}
	
	The turbulent velocity dispersion of the ISM ($\sigma_{\rm ISM}$) provides a reasonable estimate for $v$. From the \Cii~emission line widths of the $z\sim5.5$ galaxies measured by \citetalias{CAPAK15} we find $\sim100\,{\rm km/s}$ and we take this value to approximate $\sigma_{\rm ISM}$ in the following. Note that a significant part of the \Cii~line width may be due to rotation as it is suggested for \textit{HZ9} and \textit{HZ10} \citep[][]{JONES17}. Therefore, this approximation might lead to an upper limit for $\sigma_{\rm ISM}$.
	Our simple modified model is shown in Figure~\ref{fig:model} by the dashed and dotted blue lines for $C_{\rm ISM}/C_{\rm cloud}$ of $0.01$ and $0.1$, respectively, assuming an SMC-like wavelength dependence of the optical depth of the ambient ISM.

	Our model can quantitatively explain the low IRX values of the $z\sim5.5$ galaxies and also can naturally lead to the large observed range in $\beta$ at a fixed IRX value. Depending on the geometry and the viewing angle of the observer, the dust clouds might be spatially offset from the line-of-sight and the UV light can escape the galaxy without being attenuated, which would cause blue UV colors but very similar IRX values. 
	
	Qualitatively, this model is also consistent with other observed characteristics of high-redshift galaxies such as the increased turbulent motion, increased radiation pressure, merger events, and strong gas inflows. In all cases, this leads to dust that is not well mixed with the stars, with isolated ISM clouds spatially offset from the regions of star formation. At the same time, low opacity dust screens due to tidal disruption or from the diffuse ISM can be created in front of the stars, producing the observed red UV colors. This setup could create lines-of-sight with attenuated UV light of the star-forming regions as well as ones that are clear, at a fixed IRX.
	In fact we see hints of such behavior in local galaxies.  NGC\,5253, NGC\,1568, and Zwicky 403 are examples of local galaxies that have similar morphological properties, although the $z>5$ population is likely more extreme \citep[c.f.,][]{GILDEPAZ07}. All three galaxies show confined isolated clouds of gas and dust along with clear lines of sight to the star forming regions with a complex morphology.

	 \subsection{Other possible geometric configurations}\label{sec:othergeom}
	 
	There are other geometric configurations that can cause the observed low IRX values and red UV spectral slopes. Radiation pressure and tidal forces can stir up the ISM in these young primeval galaxies and could result in a large fraction of dust and gas that is expelled into their circum-galactic regions. 
	The low luminosity quasar \textit{HZ5} suggests evidence for an extended halo of dust and gas out to several kilo-parsecs around its \Cii~confirmed (foreground) companion \textit{HZ5a}. The spectrum of the quasar shows absorption features at the \Cii~redshift of \textit{HZ5a} as well as a narrow \lya~emission line possibly caused by absorption in the foreground galaxy \citep[][]{CAPAK11}.
	Such circum-galactic dust can attenuate the UV light of a companion galaxy in the background and cause a red $\beta$. At the same time, its IRX value is determined by the foreground galaxy's dust.
	The galaxy \textit{HZ8W} (a \Cii~confirmed companion of \textit{HZ8}) could be an example of such a geometrical configuration. It is characterized by a red spectral slope of $-0.1$ and an upper limit in IRX of only $0.2$. The UV light of this galaxy could be substantially attenuated by a foreground screen of dust and gas originating from the larger and more massive galaxy \textit{HZ8} in several kpc projected distance. This is also indicated by its very extended \Cii~emission. The \textit{HZ6} system could show a similar configuration.

\begin{figure}
\centering
\includegraphics[width=1.0\columnwidth, angle=0]{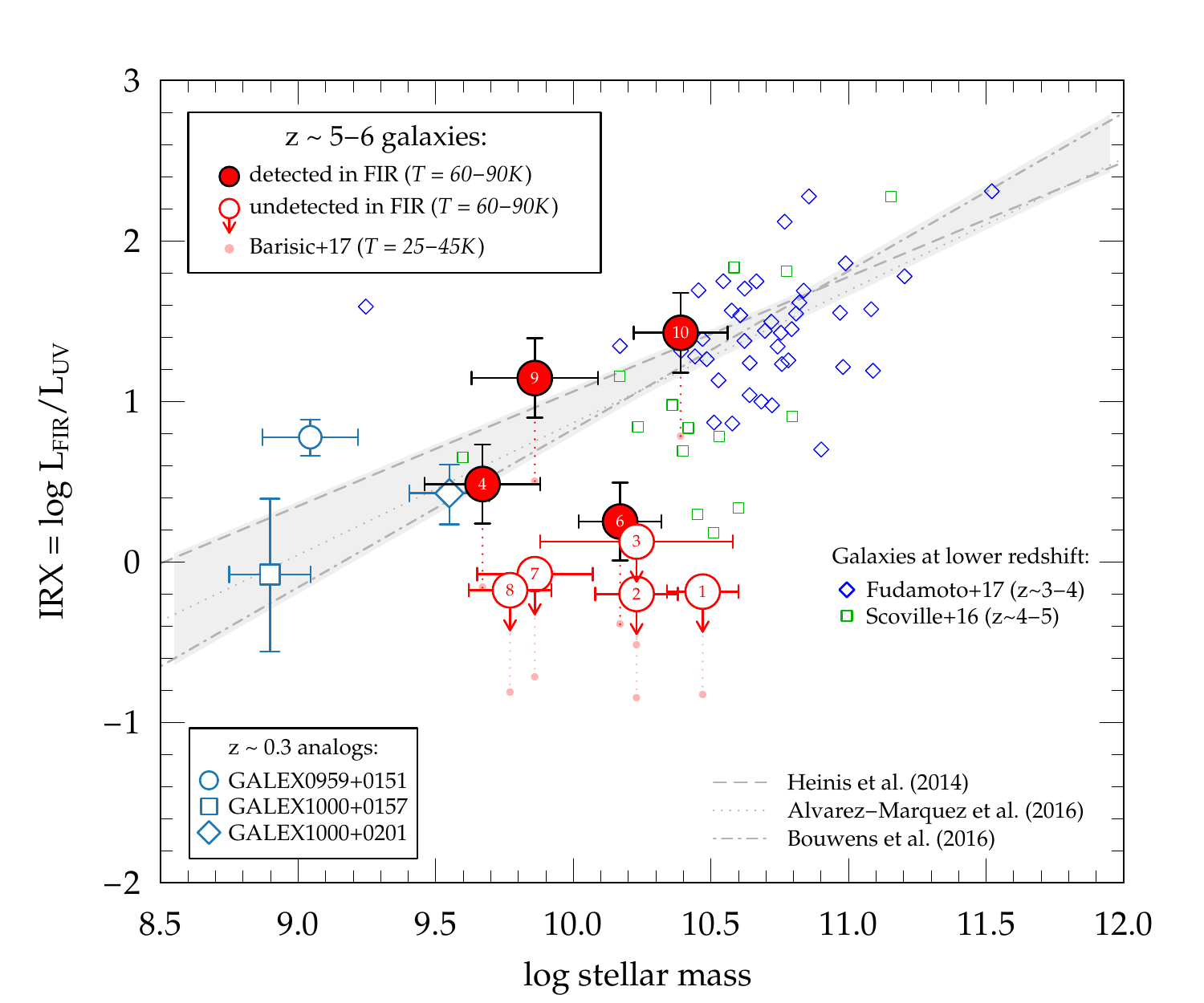}\\
\caption{
IRX vs. stellar mass diagram for galaxies at $z=5-6$ assuming a maximally warm IR SED (large red symbols) compared to galaxies at $z=3-4$ \citep[blue diamonds,][]{FUDAMOTO17}, $z=4-5$ \citep[green squares,][]{SCOVILLE16}, and our three analogs (large blue symbols). Relations derived at $z\sim2-4$ are shown in gray (dashed: \citet{HEINIS14}; dotted: \citet{ALVAREZMARQUEZ16}; dot-dashed: \citet{BOUWENS16}). 
\label{fig:IRXM}}
\end{figure}

\subsection{Correlation with stellar mass}

	Several studies indicate a relation between IRX and stellar mass for galaxies at $z<4$ \citep[e.g.,][]{PANNELLA09,HEINIS14,PANNELLA15,ALVAREZMARQUEZ16}.
	The growing samples at $z>4$ show that this relation might break down at higher redshifts \citep[e.g.,][]{FUDAMOTO17}. Specifically, it is found that these galaxies show a lower IRX at a fixed stellar mass than expected from the relations at lower redshifts. Part of the reason for this discrepancy is the unknown IR SED of the galaxies \citep[][]{BOUWENS16,FUDAMOTO17}.
	Figure~\ref{fig:IRXM} shows the IRX vs. stellar mass diagram for the $z=5-6$ galaxies assuming a warmer IR SED prior of $T=60-90\,{\rm K}$ ($T_{\rm peak} \sim 35-50\,{\rm K}$) together with data and relations at lower redshift taken from the literature. Even with a warmer IR SED, the FIR undetected galaxies still fall below the expected IRX vs. stellar mass relation found at $z<4$.
	Assuming the stellar masses are robust, this suggests that the stellar mass becomes uncorrelated with IRX (i.e., total dust) at $z>5$ similar to the relation between IRX and $\beta$. This could be due to mergers, which would increase the stellar mass without necessarily changing the IRX, or a modified geometric distribution of dust as discussed here. Alternatively it could be that at these redshifts the production of dust is lagging behind the mass growth of the galaxies which happens on very short timescales.
	
	As mentioned in Section~\ref{sec:irxb6}, such defined relations between IRX and $\beta$ as well as stellar mass are important to predict the dust properties of large samples of high-redshift galaxies from quantities measured in the rest-frame UV and optical. Unfortunately, we currently lack large enough galaxy samples to define such relations accurately. Large ALMA programs targeting the FIR of high-redshift galaxies will be crucial in the future to progress.

\begin{figure*}
\centering
\includegraphics[width=2.0\columnwidth, angle=0]{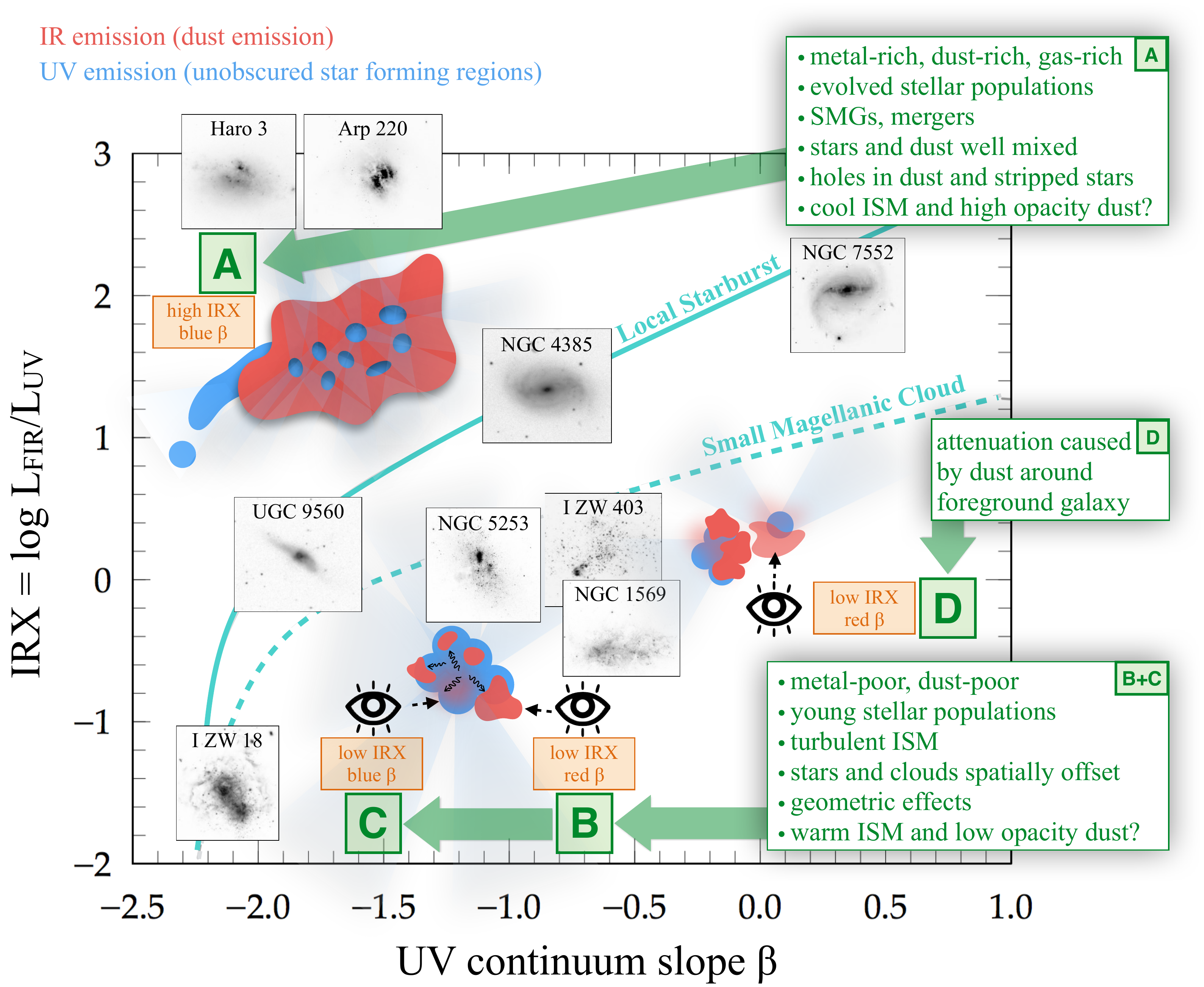}\\
\caption{
Schematic of different populations of galaxies at $z\sim5.5$ on the \IRXB~diagram and their likely properties suggested by \Cii~emission, FIR continuum, UV continuum, \textit{Spitzer} colors, and UV spectroscopy (Sections~\ref{sec:spitzermeasure}, \ref{sec:UVvsFIRmorph}, and \ref{sec:irxb6}).
The insets show multi-color images of local galaxies placed at their approximate position on the \IRXB~diagram \citep[e.g.,][]{GILDEPAZ07}. IR emission (dust) is color-coded in red, UV emission (unobscured star-forming regions) in blue.
\textbf{Configuration A:} Galaxies with high IRX values and blue UV colors are likely dust-rich, metal-rich, and consist of evolved stellar populations. The stars and gas is well mixed in a spatially compact configuration. The blue UV colors could be due to holes in the dust cover, tidally stripped young stars or faint blue satellite galaxies.
\textbf{Configurations B and C:} Galaxies with low IRX values and a range of UV colors have dust-poor and metal-poor young stellar populations. The dust clouds and star-forming regions are spatially separated in the turbulent ISM of these galaxies due to stellar radiation pressure, mergers, and gas inflows. Our proposed model (Section~\ref{sec:model}) can explain the low IRX values of these galaxies. Such a configuration causes obscured sight-lines (B) as well as clear sight-lines (C), hence the UV color depends significantly on geometry and viewing angle. This way, red UV colors at low IRX values can be reached.
\textbf{Configuration D:} In some cases, low IRX and red UV colors could be created by attenuation of the UV light of a background galaxy by the thin layer of dust that is expelled into the circum-galactic medium around a foreground galaxy (Section~\ref{sec:othergeom}). 
\label{fig:schematic}}
\end{figure*}
	
\section{Summary and Conclusion}\label{sec:end}

	Observations of the \IRXB~relation across cosmic time is important for understanding the evolution of ISM and dust properties of galaxies in the early Universe. Recent studies of the \IRXB~relation of main-sequence galaxies at $z>3$ have reported a significant evolution of this relation. Because the UV colors are accurately measured by HST, this evolution is attributed to a deficit in IRX at fixed $\beta$ compared to galaxy samples at lower redshifts with similar stellar masses. These low IRX values and large ranges in UV color are curious, since they cannot be explained by standard models of dust attenuation in galaxies.
	The total FIR luminosity of these high-redshift samples is very poorly constrained often with only one data point at these wavelengths. A common solution to this discrepancy is therefore an evolving temperature, which is the most important parameter to define the IR SED. On the other hand, a different geometric distribution of dust and stars can be responsible for the evolution of the \IRXB~relation to high redshifts.
	In this work, we have investigated both ideas and find that also with a ``maximally'' warm IR SED the distribution of dust and stars likely needs to be altered to explain the location of some of the $z>5$ galaxies on the \IRXB~plane.
	
	We first infer the IR SED shape of high-redshift galaxies by using correlations of the \textit{luminosity weighted} temperature (in this work referred as ``temperature'') with metallicity and sSFR in samples of nearby galaxies, which we extend with the extreme cases of three $z\sim0.3$ analogs of $z>5$ galaxies. This provides us with good evidence that the IR SEDs of high-redshift galaxies show a warm emission component as suggested by temperatures of up to $T=60-90\,{\rm K}$ ($T_{\rm peak} \sim 35-50\,{\rm K}$), which is significantly warmer than typical $z < 3$ galaxies ($T \sim 25-45\,{\rm K}$, $T_{\rm peak} \sim 18-28\,{\rm K}$). We believe the warmer IR SEDs are due to a combination of lower ISM optical depth and denser and stronger UV radiation fields associated with more intense star formation and low metallicity environment.  A comparison of the UV, \Cii, and FIR morphologies along with metallicities in high and low redshift systems may support this picture.
		
	Assuming such maximally warm IR SEDs would increase the FIR luminosity (and therefore IRX) by up to $0.6\,{\rm dex}$ (factor of 4) with respect to previous measurements that assume the on average cooler IR SEDs of local galaxies. This reduces the deficit in IRX and brings most of the galaxies at $z>5$ in agreement with the relation measured at lower and local redshifts, however, with significant scatter. Even with the assumption of such warm IR SEDs is correct, a population of galaxies with red UV colors and low IRX remains. We suggest that the geometric distribution of dust and stars is altered in high-redshift galaxies.
	
	Common dust attenuation models assuming a well mixed ISM \citep[][]{CHARLOTFALL00} are not able to explain the galaxies with low IRX values and large range in UV colors. The large spatial offsets between star-forming regions and \Cii~emission as well as the \textit{Spitzer} colors suggest that these galaxies are young, metal-poor, and likely have enhanced cloud motions due to stellar radiation pressure. Based on this, we propose an updated dust attenuation model in which the dust clouds are separated from the young stars by their radiation pressure. We show that such a separation can be easily achieved within the lifetime of massive O and B stars in young star-forming galaxies at high redshifts. Such a model can successfully reproduce the location of the FIR faint galaxies with low IRX. Furthermore, it can naturally reproduce the large scatter in UV colors due to viewing-angle effects.
	
	We note that the evidence of a warmer IR SED in high-redshift galaxies is indirectly derived from correlations in local galaxies and three exemplary analogs of $z>5$ galaxies and the fact that high-redshift galaxies are metal-poor and have high sSFRs. However, we stress that a direct measurement of, e.g., the temperature $T$ is absolutely crucial to verify these trends. Unfortunately, such measurements turn out to be time consuming and only possible for some of the brightest high-redshift galaxies we know to-date (see Appendix~\ref{sec:ALMA}). The first estimates of the temperature of two normal $z\sim5.5$ galaxies suggest large variations, highlighting the importance of such direct measurements \citep[][]{PAVESI16}. The two galaxies occupy two different regions on the \IRXB~diagram and likely are different in terms of metallicity and sSFR despite having comparable UV emission. This hints towards a large diversity in the galaxy population at $z>5$, only 1 billion years after the Big Bang that is waiting to be explored with further observations.

	\acknowledgements
The authors would like to thank Nick Scoville, Caitlin Casey, Ranga-Ram Chary, Rychard Bouwens, Kirsten Larson, Shoubaneh Hemmati, and Lee Armus for valuable discussions which improved this manuscript. Furthermore, the authors thank the anonymous referee for the valuable feedback.
	D.R. and R.P. acknowledge support from the National Science Foundation under grant number AST-1614213 to Cornell University. R.P. acknowledges support through award SOSPA3-008 from the NRAO. 
	We thank the ALMA staff for facilitating the observations and aiding in the calibration and reduction process. This paper makes use of the following ALMA data: ADS/JAO.ALMA\#2011.0.00064.S, ADS/JAO.ALMA\#2012.1.00523.S, ADS/JAO.ALMA\#2012.1.00919.S, ADS/JAO.ALMA\#2015.1.00928.S.
	ALMA is a partnership of ESO (representing its member states), NSF (USA) and NINS (Japan), together with NRC (Canada) and NSC and ASIAA (Taiwan) and KASI (Republic of Korea), in cooperation with the Republic of Chile. The Joint ALMA Observatory is operated by ESO, AUI/NRAO and NAOJ.
	This work is based in part on observations made with the Spitzer Space Telescope and the W.M. Keck Observatory, along with archival data from the NASA/ESA Hubble Space Telescope, the Subaru Telescope, the Canada-France-Hawaii-Telescope and the ESO Vista telescope obtained from the NASA/IPAC Infrared Science Archive.
	This research has made use of the NASA/IPAC Extragalactic Database (NED) which is operated by the Jet Propulsion Laboratory, California Institute of Technology, under contract with the National Aeronautics and Space Administration.

\begin{deluxetable*}{lccc  ccccccc}
\tabletypesize{\scriptsize}
\tablecaption{UV and optical spectroscopic properties of the 3 analog galaxies\label{tab:specprop}}
\tablewidth{0pt}
\tablehead{
\colhead{Name} & \colhead{R.A.} & \colhead{Decl.} & \colhead{z} & \colhead{\oiii/\halpha} & \colhead{\oiii/\hbeta} & \colhead{\nii/\halpha}& \colhead{$\ewlya$} & \colhead{$\ewha$} & \colhead{$\oabund$$^a$} & \colhead{SFR$_{{\rm H}\alpha}$}\\
\colhead{} & \colhead{(J2000.0)} & \colhead{(J2000.0)} & \colhead{(optical)} & \colhead{} & \colhead{} & \colhead{} & \colhead{(\AA)} & \colhead{(\AA)} & \colhead{} & \colhead{($\Msol$yr$^{-1}$)} \\[-0.2cm]
}
\startdata
\Gaa & 09:59:40.34 & 01:51:21.30 & $0.2506$ & $2.11\pm 0.03$ & $6.04\pm 0.03$ & $0.051\pm 0.001$ & $21\pm 3$ & $333\pm 4$ & $8.16\pm 0.01$ & $45$\\
\Gbb & 10:00:27.85 & 01:57:03.60 & $0.2647$ & $1.96\pm 0.04$ & $5.61\pm 0.04$ & $0.024\pm 0.002$ & $43\pm 2$ & $423\pm 2$ & $7.98\pm 0.02$ & $17$\\
\Gcc & 10:00:35.76 & 02:01:13.50 & $0.2653$ & $1.65\pm 0.03$ & $4.71\pm 0.03$ & $0.060\pm 0.001$ & $19\pm 2$ & $195\pm 2$ & $8.20\pm 0.01$ & $11$\\
\enddata
\tablenotetext{a}{This uses the \citet{PETTINI04} calibration based on \nii. The \citet{MAIOLINO08} calibration yields $7.90\pm 0.01$, $7.90\pm 0.01$, and $8.17\pm 0.01$, respectively, but we find that this estimate is not reliable because it is based on the location where the upper and lower branch separate.}
\end{deluxetable*}

\begin{deluxetable*}{lccc  ccccc}
\tabletypesize{\scriptsize}
\tablecaption{Photometric IR properties of the 3 low-redshift analogs\label{tab:photoprop}}
\tablewidth{0pt}
\tablehead{
\colhead{Name} & \multicolumn{1}{c}{\textit{Spitzer}/MIPS} & \multicolumn{2}{c}{\textit{Herschel}/PACS} & \multicolumn{3}{c}{\textit{Herschel}/SPIRE} & \multicolumn{2}{c}{ALMA$^a$}\\[-0.2cm]
\colhead{} & \multicolumn{1}{c}{} & \multicolumn{2}{c}{------------------------} & \multicolumn{3}{c}{------------------------------------} & \multicolumn{2}{c}{-----------------------------------}\\[-0.2cm]
\colhead{} & \colhead{$70\,{\rm \mu m}$} & \colhead{$110\,{\rm \mu m}$} & \colhead{$160\,{\rm \mu m}$} & \colhead{$250\,{\rm \mu m}$} & \colhead{$350\,{\rm \mu m}$} & \colhead{$500\,{\rm \mu m}$} & \colhead{$S_{\rm cont}$} & \colhead{$T_{\rm exp}$ (\#Antenna)}\\
\colhead{} & \colhead{(mJy)} & \colhead{(mJy)} & \colhead{(mJy)} & \colhead{(mJy)} & \colhead{(mJy)} & \colhead{(mJy)} & \colhead{(mJy)} & \colhead{(seconds)}\\[-0.4cm]
}
\startdata
\Gaa & $11.5\pm 2.5$ & $12.4\pm 2.1$ & $12.2\pm 5.4$ & $0.5\pm 4.0$ & $<2.8^{\dagger}$ & $<1.1^{\dagger}$ & $< 0.0793^{\ddagger}$ & 1255.87 ($27$)\\
\Gbb & $<3.3^{\dagger}$ & $< 2.1^{\dagger}$ & $9.4\pm 5.4$ & $<4.0^{\dagger}$ & $<2.8^{\dagger}$ & $<1.1^{\dagger}$ & $<0.0484^{\ddagger}$ & 9216.43 ($36$)\\
\Gcc & $7.3\pm 4.2$ & $9.9\pm2.1$ & $6.2\pm 5.4$ & $<4.0^{\dagger}$ & $<2.8^{\dagger}$ & $<1.1^{\dagger}$ & $<0.0658^{\ddagger}$ & 9852.10 ($24$)\\
\enddata
\tablenotetext{a}{ALMA observations at the position of CO (observed frequency $\sim 90\,{\rm GHz}$). No CO emission is detected in any of the three galaxies with high significance ($L^{\prime}_{\rm CO} < 5.5\times 10^{8}\,{\rm K\,km\,s^{-1}\,pc^{2}}$ at $3\sigma$).}
\tablenotetext{{\dagger}}{$1\sigma$ upper limits on fluxes.}
\tablenotetext{^{\ddagger}}{$3\sigma$ upper limits on fluxes.}
\end{deluxetable*}

\begin{deluxetable*}{lccc  ccccccc}
\tabletypesize{\scriptsize}
\tablecaption{Fitted UV and optical properties of the 3 low-redshift analogs\label{tab:optsedprop}}
\tablewidth{0pt}
\tablehead{
\colhead{Name} & \colhead{$\ewha$} & \colhead{\oiii/\halpha} & \colhead{\oiii/\hbeta} & \colhead{\oiii/\oii} & \colhead{$\logm$} & \colhead{\ebmv} & \colhead{$\log(\LUV$/$\Lsol)$} & \colhead{$\beta$~$^a$} & \colhead{SFR$_{{\rm UV+FIR}}$$^b$}\\
\colhead{} & \colhead{(\AA)} & \colhead{} & \colhead{} & \colhead{} & \colhead{} & \colhead{(mag)} & \colhead{} & \colhead{(UV cont. slope)} & \colhead{($\Msol$yr$^{-1}$)}\\[-0.4cm]
}
\startdata
\Gaa & $452\pm 14$ & $1.95\pm 0.14$ & $5.56\pm 0.41$ & $3.23\pm 0.57$ & $9.04\pm 0.17$ & $0.16\pm 0.06$ & $10.12^{+0.07}_{-0.08}$ & $-1.80\pm 0.46$ & 8.8\\
 & (333)$^{\dagger}$ & (2.11)$^{\dagger}$ & (6.04)$^{\dagger}$ & $4.11\pm 0.72$ & $9.11\pm 0.16$ & $0.18\pm 0.07$ & $10.11^{+0.08}_{-0.10}$ & $-1.71\pm 0.55$ & \\
  &  &  &  &  &  & \multicolumn{2}{c}{$\beta$ from GALEX: }  & $-1.43^{+0.24}_{-0.14}$ &  \\[0.2cm]
\Gbb & $402\pm 18$ & $2.00\pm 0.11$ & $5.73\pm 0.32$ & $3.39\pm 0.56$ & $8.9\pm 0.15$ & $0.07\pm 0.04$ & $10.33^{+0.05}_{-0.06}$ & $-2.19\pm 0.23$ & $4.9$ \\
 & (423)$^{\dagger}$ & (1.96)$^{\dagger}$ & (5.61)$^{\dagger}$ & $3.38\pm 0.52$ & $8.89\pm 0.14$ & $0.07\pm 0.03$ & $10.33^{+0.05}_{-0.06}$ & $-2.20\pm 0.22$ & \\
   &  &  &  &  &  & \multicolumn{2}{c}{$\beta$ from GALEX: }  & $-1.48^{+0.87}_{-0.40}$ &  \\ [0.2cm]
\Gcc & $150\pm 13$ & $1.89\pm 0.23$ & $5.41\pm 0.66$ & $1.55\pm 0.24$ & $9.55\pm 0.14$ & $0.26\pm 0.08$ & $10.24^{+0.11}_{-0.16}$ & $-1.52\pm 0.70$ & $6.5$ \\
 & (195)$^{\dagger}$ & (1.65)$^{\dagger}$ & (4.71)$^{\dagger}$ & $1.48\pm 0.09$ & $9.53\pm 0.13$ & $0.26\pm 0.07$ & $10.24^{+0.10}_{-0.14}$ & $-1.54\pm 0.66$ & \\
   &  &  &  &  &  & \multicolumn{2}{c}{$\beta$ from GALEX: }  & $-1.86^{+0.28}_{-0.14}$ &  \\[0.2cm]
\enddata
\tablenotetext{{a}}{Note that $\beta$ derived from SED fitting is strongly correlated to the assumed dust attenuation curve (SMC or local starburst in this case). For the final \IRXB~diagram, we therefore use the $\beta$ independently computed from the \textit{GALEX} photometry, corrected by the \lya~emission contaminating the FUV filter.}
\tablenotetext{{b}}{UV measured at $1600\,{\rm \AA}$ and FIR from $\LIR$, see Table~\ref{tab:irsedprop}.}
\tablenotetext{{\dagger}}{These parameters are fixed to the spectroscopic value during the SED fitting.}
\end{deluxetable*}

\begin{deluxetable*}{lcc cc  cc}
\tabletypesize{\scriptsize}
\tablecaption{FIR properties of the 3 low-z analogs and high-z galaxies from the literature.\label{tab:irsedprop}}
\tablewidth{0pt}
\tablehead{
\colhead{Name} & \colhead{z} & \colhead{$\alpha$} & \colhead{$\beta_{\rm IR}$} & \colhead{$T~^{a}$} & \colhead{$T_{\rm peak}$~$^{b}$} & \colhead{$\log(\LIR/\Lsol)$~$^c$}\\
\colhead{} &\colhead{} & \colhead{(MIR power-law slope)} & \colhead{(emissivity)} & \colhead{(K)} & \colhead{(K)} & \colhead{}\\[-0.4cm]
}
\startdata
	\multicolumn{7}{c}{\textbf{Low-redshift analogs (This work)}}\\
\Gaa & $0.2506$ & $2.43^{+0.07}_{-0.15}$ & $1.16^{+3.46}_{-1.04}$ & $97^{+5}_{-12}$ & $54^{+7}_{-6}$ & $10.91^{+0.04}_{-0.07}$\\
\Gbb$^\dagger$ & $0.2647$ & $2.42^{+0.47}_{-0.61}$ & $1.80^{+3.00}_{-1.47}$ & $72^{+35}_{-46}$ & $43^{+18}_{-26}$ & $10.32^{+0.15}_{-0.37}$\\
\Gcc & $0.2653$ & $2.47^{+0.09}_{-0.11}$ & $1.18^{+3.16}_{-1.28}$ & $75^{+8}_{-5}$ & $43^{+5}_{-0}$ & $10.68^{+0.06}_{-0.09}$\\
Stack & $-$ & $2.12^{+1.25}_{-0.63}$ & $2.88^{+12.87}_{-2.36}$ & $84^{+9}_{-17}$ & $48^{+6}_{-5}$ & $10.68^{+0.03}_{-0.04}$\\[0.2cm]
\multicolumn{7}{c}{\textbf{High-z galaxies from \citetalias{CAPAK15} (This work, using warm IR SED prior)}}\\
\multicolumn{7}{c}{(parameters derived for $1.5 < \alpha < 2.5$, $1.0 < \beta_{\rm IR} < 2.0$, $60\,{\rm K} < T < 90\,{\rm K}$ [$35\,{\rm K} < T_{\rm peak} < 50\,{\rm K}$])}\\
HZ1 & $5.6885$ & $-$ & $-$ & $-$ & $-$ & $<10.96^{+0.22}_{-0.25}$\\
HZ2 & $5.6697$ & $-$ & $-$ & $-$ & $-$ & $<10.95^{+0.22}_{-0.25}$\\
HZ3 & $5.5416$ & $-$ & $-$ & $-$ & $-$ & $<11.18^{+0.22}_{-0.25}$\\
HZ4 & $5.5440$ & $-$ & $-$ & $-$ & $-$ & $11.78^{+0.22}_{-0.25}$\\
HZ5$^\ast$ & $5.3089$ & $-$ & $-$ & $-$ & $-$ & $<10.95^{+0.22}_{-0.25}$\\
$\cdot\;\;$HZ5a & & $-$ & $-$ & $-$ & $-$ & $<10.95^{+0.22}_{-0.25}$\\
HZ6 (all)$^\bullet$ & $5.2928$ & $-$ & $-$ & $-$ & $-$ & $11.78^{+0.22}_{-0.25}$\\
$\cdot\;\;$HZ6a &   & $-$ & $-$ & $-$ & $-$ & $10.92^{+0.22}_{-0.25}$\\
$\cdot\;\;$HZ6b &   & $-$ & $-$ & $-$ & $-$ & $11.52^{+0.22}_{-0.25}$\\
$\cdot\;\;$HZ6c &   & $-$ & $-$ & $-$ & $-$ & $11.44^{+0.22}_{-0.25}$\\
HZ7 & $5.2532$ & $-$ & $-$ & $-$ & $-$ & $<10.99^{+0.22}_{-0.25}$\\
HZ8 & $5.1533$ & $-$ & $-$ & $-$ & $-$ & $<10.90^{+0.22}_{-0.25}$\\
$\cdot\;\;$HZ8W &  & $-$ & $-$ & $-$ & $-$ & $<10.90^{+0.22}_{-0.25}$\\
HZ9 & $5.5410$ & $-$ & $-$ & $-$ & $-$ & $12.18^{+0.22}_{-0.25}$\\
HZ10& $5.6566$ & $-$ & $-$ & $-$ & $-$ & $12.58^{+0.22}_{-0.25}$\\
$\cdot\;\;$HZ10W$^\wr$  &  & $-$ & $-$ & $-$ & $-$ & ($12.28^{+0.22}_{-0.25}$)\\[0.2cm]
\multicolumn{7}{c}{\textbf{Temperature constraints on high-z galaxies from \citet{PAVESI16}}}\\
\multicolumn{7}{c}{(parameters derived for $1.2 < \beta_{\rm IR} < 2.2$, $10\,{\rm K} < T < 100\,{\rm K}$ [$5\,{\rm K} < T_{\rm peak} < 60\,{\rm K}$], $\alpha$ is unconstraint)}\\
HZ6 (all) & $5.2928$ & $-$ & $-$ & $60^{+35}_{-27}$ & $35^{+20}_{-16}$ & $12.11^{+0.48}_{-0.48}$\\
HZ10 & $5.6566$ & $-$ & $-$ & $36^{+25}_{-10}$ & $25^{+18}_{-7}$ & $11.55^{+0.55}_{-0.66}$\\[0.2cm]
	\multicolumn{7}{c}{\textbf{Lensed high-z \textit{Herschel} detected $1.5 < z < 3.0$ galaxies \citep{SKLIAS14}}}\\
	\multicolumn{7}{c}{(modified blackbody with fixed $\beta_{\rm IR}=1.5$, corrected for magnification)}\\
A68/C0 & $1.59$ & $-$ & $-$ & $-$ & $34.5^{+1}_{-1}$ & $11.06$ \\
A68/h7 & $2.15$ & $-$ & $-$ & $-$ & $43.4^{+1}_{-1}$ & $12.26$ \\
A68/HLS115 & $1.59$ & $-$ & $-$ & $-$ & $37.5^{+1}_{-1}$ & $11.53$ \\
A68/nn4 & $3.19$ & $-$ & $-$ & $-$ & $54.9^{+1}_{-1}$ & $12.83$ \\
MACS0451 north & $2.01$ & $-$ & $-$ & $-$ & $49.2^{+1}_{-1}$ & $10.87$ \\
MACS0451 full arc & $\arcsec$ & $-$ & $-$ & $-$ & $50-80$ & $11.27$ \\
[0.2cm]
	\multicolumn{7}{c}{\textbf{Lensed high-z DSFGs \citep{STRANDET16}}}\\
	\multicolumn{7}{c}{(refitted by our method with photometry from \citet{STRANDET16}}\\
SPT2319-55$^\diamondsuit$ & $5.2929$ & $1.73^{+53.19}_{-1.17}$ & $2.46^{+0.17}_{-0.16}$ & $59^{+5}_{-7}$ & $34^{+4}_{-1}$ & $<13.79^{+0.40}_{-0.17}$$^\ddagger$\\
SPT2353-50$^\diamondsuit$ & $5.5760$ & $0.91^{+3.28}_{-0.59}$ & $2.60^{+0.18}_{-0.18}$ & $61^{+6}_{-6}$ & $38^{+1}_{-4}$ & $<14.02^{+0.36}_{-0.28}$$^\ddagger$\\
SPT0346-52$^\diamondsuit$ & $5.6559$ & $2.07^{+5.26}_{-0.72}$ & $2.52^{+2.68}_{-0.66}$ & $74^{+3}_{-3}$ & $43^{+1}_{-5}$ & $13.52^{+0.16}_{-0.23}$$^\star$\\
SPT2351-57$^\diamondsuit$ & $5.8110$ & $3.12^{+17.55}_{-2.50}$ & $2.41^{+0.17}_{-0.16}$ & $80^{+5}_{-7}$ & $48^{+1}_{-5}$ & $<13.92^{+0.41}_{-0.08}$$^\ddagger$\\[0.2cm]
\multicolumn{7}{c}{\textbf{Intense high-z starburst \textit{AzTEC-3} \citet{RIECHERS14}}}\\
AzTEC-3 & $5.2988$ & $6.17^{+2.73}_{-2.60}$ & $2.16^{+0.27}_{-0.27}$ & $88^{+10}_{-10}$ & $51^{+6}_{-6}$ & $13.34^{+0.08}_{-0.10}$\\[0.2cm]
\multicolumn{7}{c}{\textbf{High-z galaxies from \citet{WILLOTT15}}}\\
\multicolumn{7}{c}{(parameters derived for $1.5 < \alpha < 2.5$, $1.0 < \beta_{\rm IR} < 2.0$, $60\,{\rm K} < T < 90\,{\rm K}$ [$35\,{\rm K} < T_{\rm peak} < 50\,{\rm K}$])}\\
CLM1 & $6.1657$ & $-$ & $-$ & $-$ & $-$ & $11.18^{+0.22}_{-0.25}$\\
WMH5 & $6.0695$ & $-$ & $-$ & $-$ & $-$ & $11.87^{+0.22}_{-0.25}$\\[0.2cm]
\multicolumn{7}{c}{\textbf{Candidate \Cii-detected high-z galaxies from \citet{ARAVENA16} (blind search)}}\\
\multicolumn{7}{c}{(parameters derived for $1.5 < \alpha < 2.5$, $1.0 < \beta_{\rm IR} < 2.0$, $60\,{\rm K} < T < 90\,{\rm K}$ [$35\,{\rm K} < T_{\rm peak} < 50\,{\rm K}$])}\\
IDX25 & $6.357$ & $-$ & $-$ & $-$ & $-$ & $<11.16^{+0.22}_{-0.25}$\\
IDX34 & $7.491$ & $-$ & $-$ & $-$ & $-$ & $<11.66^{+0.22}_{-0.25}$\\
ID02 & $7.914$ & $-$ & $-$ & $-$ & $-$ & $<11.43^{+0.22}_{-0.25}$\\
ID04 & $6.867$ & $-$ & $-$ & $-$ & $-$ & $11.53^{+0.22}_{-0.25}$\\
ID09 & $6.024$ & $-$ & $-$ & $-$ & $-$ & $<11.15^{+0.22}_{-0.25}$\\
ID14 & $6.751$ & $-$ & $-$ & $-$ & $-$ & $<11.22^{+0.22}_{-0.25}$\\
ID27 & $7.575$ & $-$ & $-$ & $-$ & $-$ & $11.20^{+0.22}_{-0.25}$\\
ID30 & $6.854$ & $-$ & $-$ & $-$ & $-$ & $<11.42^{+0.22}_{-0.25}$\\
ID31 & $7.494$ & $-$ & $-$ & $-$ & $-$ & $11.14^{+0.22}_{-0.25}$\\
ID38 & $6.593$ & $-$ & $-$ & $-$ & $-$ & $<11.20^{+0.22}_{-0.25}$\\
ID41 & $6.346$ & $-$ & $-$ & $-$ & $-$ & $<11.20^{+0.22}_{-0.25}$\\
ID44 & $7.360$ & $-$ & $-$ & $-$ & $-$ & $<11.25^{+0.22}_{-0.25}$\\
ID49 & $6.051$ & $-$ & $-$ & $-$ & $-$ & $<11.13^{+0.22}_{-0.25}$\\
ID52 & $6.018$ & $-$ & $-$ & $-$ & $-$ & $<11.37^{+0.22}_{-0.25}$\\[0.2cm]
\enddata
\tablenotetext{a}{Luminosity weighted temperature as in \citet{CASEY12}.}
\tablenotetext{b}{Temperature measured from the wavelength of peak flux emission via Wien's displacement law.}
\tablenotetext{c}{The total FIR luminosity is integrated between rest-frame $3-1100\,{\rm \mu m}$. If necessary, the values are converted to this integration interval.}
\tablenotetext{^\dagger}{This source is IR faint and therefore these estimates are mostly based on limits causing the large uncertainties.}
\tablenotetext{^\diamondsuit}{No constraints blue-ward of rest-frame $40\,{\rm \mu m}$ are available for these sources causing the estimate for $\alpha$ to be very uncertain.}
\tablenotetext{^\ddagger}{The magnification for these sources is not known. Therefore the $\LIR$ has not been de-magnified in these cases.}
\tablenotetext{^\star}{$\LIR$ has been corrected for magnification using $\mu=5.4$ \citep[][]{STRANDET16}.}
\tablenotetext{^\ast}{This object is a low-luminosity quasar.}
\tablenotetext{^\bullet}{This object is called \textit{LBG-1} in \citet{RIECHERS14} and \citet{PAVESI16}.}
\tablenotetext{\wr}{Assuming half of the FIR luminosity of \textit{HZ10}.}
\end{deluxetable*}

\begin{deluxetable*}{l ccccc}
\tabletypesize{\scriptsize}
\tablecaption{Summary of the UV and FIR properties of the \citetalias{CAPAK15} galaxy sample from \citet{CAPAK15} and \citet{BARISIC17}.\label{tab:c15prop}}
\tablewidth{0pt}
\tablehead{
\colhead{Name} & \colhead{z} & \colhead{$\beta$} & \colhead{$\log(\LUV/\Lsol)$} & \colhead{$\log(\LCII/\Lsol)$} & \colhead{$158\,{\rm \mu m}$ continuum flux}\\
\colhead{} & \colhead{} & \colhead{(UV spectral slope)} & \colhead{} & \colhead{} & \colhead{($\mu$Jy)}\\[-0.4cm]
}
\startdata
HZ1 & 5.690 & $-1.92_{-0.11}^{+0.14}$ & 11.21 $\pm$ 0.01 & $8.40\pm0.32$ & $<30$ \\
HZ2 & 5.670 & $-1.82_{-0.10}^{+0.10}$ & 11.15 $\pm$ 0.01 & $8.56\pm0.41$ & $<29$ \\
HZ3 & 5.546 & $-1.72_{-0.15}^{+0.12}$ & 11.08 $\pm$ 0.01 & $8.67\pm0.28$ & $<51$ \\
HZ4 & 5.540 & $-2.06_{-0.15}^{+0.13}$ & 11.28 $\pm$ 0.01 & $8.98\pm0.22$ & $202\pm47$ \\
HZ5 & 5.310 & $-1.01_{-0.12}^{+0.06}$ & 11.45 $\pm$ 0.004 & $<7.20$ & $<32$ \\
$\cdot\;\;$HZ5a &  & \nodata & $<10.37$ & $8.15\pm0.27$ & $<32$ \\
HZ6 (all) & 5.290 & $-1.14_{-0.14}^{+0.12}$ & 11.47 $\pm$ 0.01 & $9.23\pm0.04$ & $220\pm36$ \\
$\cdot\;\;$HZ6a &  & $-0.59_{-1.12}^{+1.05}$ & 11.11 $\pm$ 0.07 & $8.32\pm0.06$ & $30\pm20$ \\
$\cdot\;\;$HZ6b &  & $-1.50_{-1.22}^{+1.05}$ & 11.00 $\pm$ 0.07 & $8.81\pm0.02$ & $120\pm20$ \\
$\cdot\;\;$HZ6c &  & $-1.30_{-0.37}^{+0.51}$ & 10.81 $\pm$ 0.07 & $8.65\pm0.03$ & $100\pm20$ \\
HZ7 & 5.250 & $-1.39_{-0.17}^{+0.15}$ & 11.05 $\pm$ 0.02 & $8.74\pm0.24$ & $<36$ \\
HZ8 & 5.148 & $-1.42_{-0.18}^{+0.19}$ & 11.04 $\pm$ 0.02 &  $8.41\pm0.18$ & $<30$\\
$\cdot\;\;$HZ8W &  & $-0.10_{-0.29}^{+0.29}$ & 10.57 $\pm$ 0.04 & $8.31\pm0.23$ & $<30$ \\
HZ9 & 5.548 & $-1.59_{-0.23}^{+0.22}$ & 10.95 $\pm$ 0.02 &  $9.21\pm0.09$ & $516\pm42$\\
HZ10 & 5.659 & $-1.92_{-0.17}^{+0.24}$ & 11.14 $\pm$ 0.02 & $9.13\pm0.13$ & $1261\pm44$ \\
$\cdot\;\;$HZ10W$^\dagger$ &  & $-1.47_{-0.44}^{+0.77}$ & 10.23 $\pm$ 0.05 & ($8.83\pm0.13$) & ($630\pm44$) \\
\enddata
\tablenotetext{\dagger}{Assuming half the IR flux and \Cii~line luminosity of HZ10.}
\end{deluxetable*}

\appendix

\section{Quantifying the difficulty of measuring IR SEDs at $z>5$}\label{sec:ALMA}

	In Section~\ref{sec:firmeasurementhighz} we show that the knowledge of the shape of the IR SED is crucial to measure basic IR properties like the total luminosity $\LIR$. Furthermore, it enables us to study in detail the ISM properties of galaxies such as their temperature and optical depth, which tell us about the current star-formation conditions in these galaxies.
	As pointed out earlier, IR SEDs of $z>5$ galaxies are very poorly constrained, most of the time by just one data-point at $158\,{\rm \mu m}$ if at all. Measurement of DSFGs at $z=5-6$ reveal increased $T$ and thus warm SED shapes compared to galaxies a $z<4$ \citep[][]{RIECHERS13,RIECHERS14,STRANDET16,MA16}. The first estimates of $T$ for two normal $z\sim5.5$ galaxies reveal two very different temperatures, indicative of large variations in the SED shape of high-redshift galaxies \citep{PAVESI16}. However, these measurements are uncertain as they only use wavelength data red-ward of \Cii.
	One way to investigate the diversity of IR SEDs of high-redshift galaxies statistically is to use larger samples of low-redshift analogs of these galaxies. This, however, might be dangerous as we do not know the degree at which analogs are actually representing the high-redshift galaxy population as a whole. 
	Direct measurements of galaxies at high redshifts therefore seems to be the preferred strategy. Due to the faintness of these galaxies, this is a difficult endeavor and hence an optimal and effective observing strategy is needed.
	
	In the following, we outline a possible observing strategy in order to better constrain the IR SEDs (specifically constrained by the luminosity weighted temperature $T$) of $z\sim6$ galaxies. Since observations in the IR are expensive, we focus in particular on minimizing the total observing time and maximizing the science output.

\begin{figure*}
\centering
\includegraphics[width=1\columnwidth, angle=0]{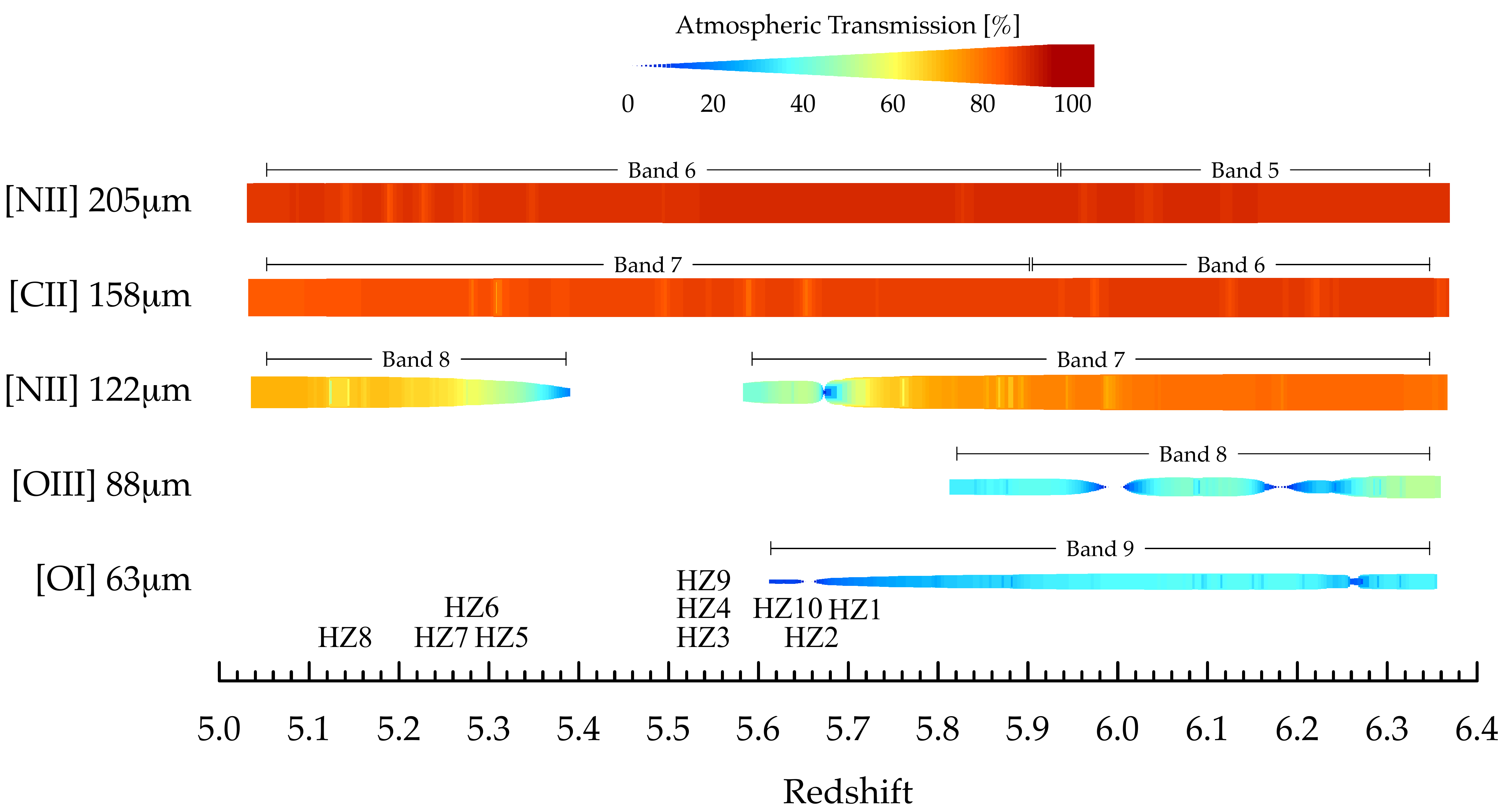}
\caption{Observability of FIR emission lines (\oi, \oiii, \nii, and \Cii) with ALMA as a function of redshift. The atmospheric transmission (assuming a precipitable water vapor column density of $1\,{\rm mm}$) is indicated in colors and the width of the bars. We also indicate the ALMA bands in which the lines can be observed. The redshift range $5.82 < z < 5.96$ is the lowest for which all FIR lines can be observed at a reasonable atmospheric transmission. Spectroscopic samples become scarce at higher redshifts. The redshifts of the \citetalias{CAPAK15} galaxies are also indicated for reference.
\label{fig:almatransmission}}
\end{figure*}

\subsection{Theoretical calculations}

	To measure the temperature of high-redshift galaxies via continuum measurements, we choose wavelengths close to bright FIR emission lines (\oi~at $63{\rm \mu m}$, \oiii~at $88{\rm \mu m}$, \nii~at $122{\rm \mu m}$ and $205{\rm \mu m}$, and \Cii~at $158{\rm \mu m}$). With this strategy, any ALMA observations will produce line diagnostics in addition to FIR continuum measurements to further investigate the ISM of these galaxies \citep[e.g.,][]{PAVESI16}.
	Figure~\ref{fig:almatransmission} shows the observability of these FIR emission lines with ALMA as a function of redshift. The color and width of the bands for each FIR emission line corresponds to the atmospheric transmission in per-cent assuming a precipitable water vapor of $1\,{\rm mm}$. An efficient strategy is to look for the lowest redshift (to have a reasonable sample of spectroscopically confirmed galaxies) at which most of the FIR lines are observable with ALMA. It turns out that the redshift range $5.82 < z < 5.96$ fulfills this criteria with Band 5 being operational in ALMA cycle 5. Note that $z\sim6.1$ and $z\sim6.3$ are other possible windows at which all five emission lines are observable. If only interested in the highest transmission wavelengths ($122\,{\rm \mu m}$, $158\,{\rm \mu m}$, and $205\,{\rm \mu m}$), all redshifts above $z\sim5.9$ are a good choice.
	
\begin{figure*}
\centering
\includegraphics[width=1\columnwidth, angle=0]{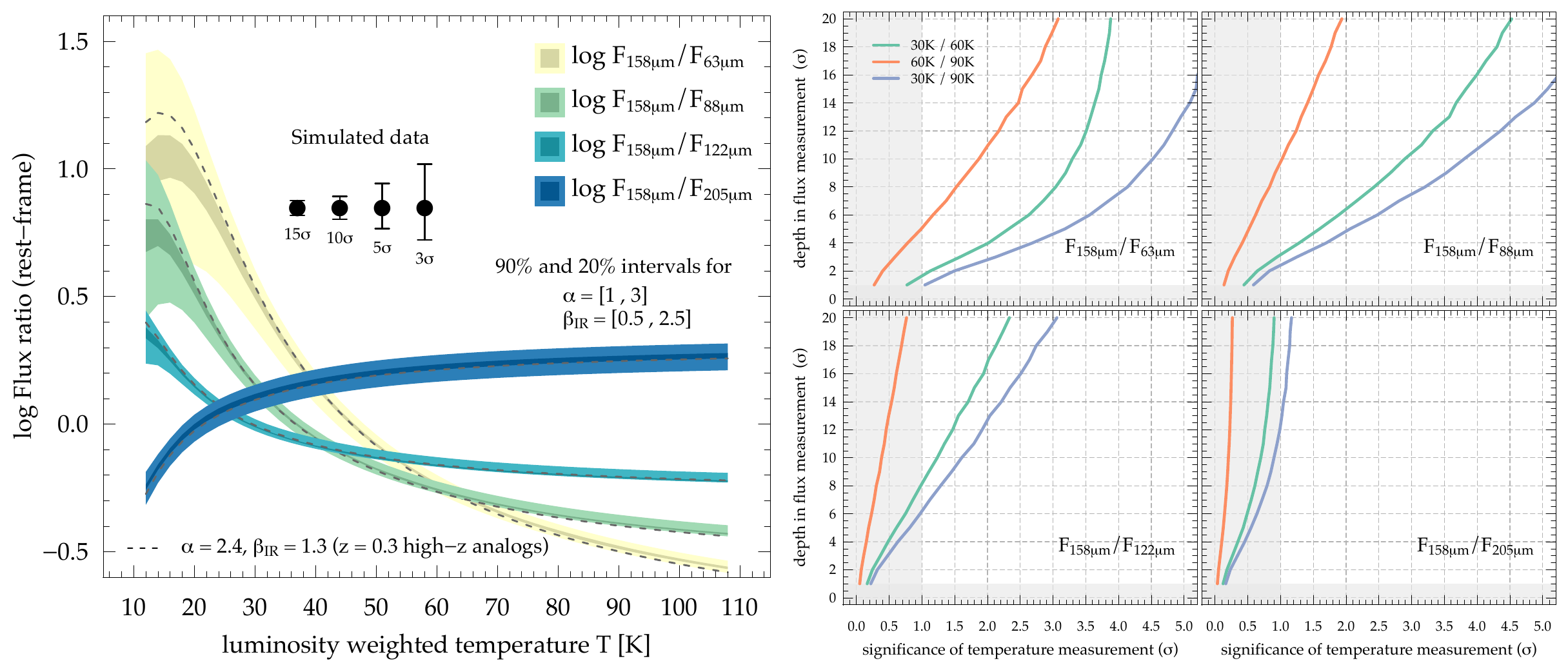}
\caption{\textit{Left panel:} Various flux ratios (with respect to $158\,{\rm \mu m}$) computed for a graybody$+$MIR power-law \citep[][]{CASEY12} as a function of luminosity weighted temperature. The width of the curves correspond to the \textit{model uncertainties} assuming a range in MIR power-law $\alpha$ and emissivity ($\beta_{\rm IR}$) as indicated. The dotted line represents the mean parameters of our $z\sim0.3$ analogs of high-z galaxies. The black symbols show the uncertainties in the flux ratios for detections of different $\sigma$ significance.
\textit{Right panels:} Necessary depth of flux observations ($\sigma$) in $63\,{\rm \mu m}$, $88\,{\rm \mu m}$, $122\,{\rm \mu m}$, and $205\,{\rm \mu m}$ (four sub-panels) to reach a temperature measurement of a given $\sigma-$confidence to distinguish three different pairs of temperatures (color coded). Temperatures above $30\,{\rm K}$ are not possible to distinguish at $>1\sigma$ with the $F_{158\,{\rm \mu m}}/F_{205\,{\rm \mu m}}$ flux ratio even with a $20\sigma$ observation at $205\,{\rm \mu m}$.
\label{fig:fluxratios}}
\end{figure*}
	
	The left-side panel in Figure~\ref{fig:fluxratios} shows FIR continuum flux ratios around these FIR emission lines as a function of the luminosity weighted temperature  derived from a graybody $+$ mid-IR power-law parametrization by \citet{CASEY12}. We choose to use the ratios with respect to rest-frame $158\,{\rm \mu m}$, which has a high atmospheric transmission ($93\%$ at $z\sim6$) and is well constrained by existing data sets in terms of detections and limits. In the following, we refer to $158\,{\rm \mu m}$ as the ``primary wavelength'' and to the other wavelengths ($63\,{\rm \mu m}$, $88\,{\rm \mu m}$, $122\,{\rm \mu m}$, and $205\,{\rm \mu m}$) as the ``secondary wavelengths''. The width of the bands visualizes the \textit{model uncertainty} ($90\%$ confidence interval) of the continuum flux ratio for a large range in emissivities ($0.5 < \beta_{\rm IR} < 2.5$) and mid-IR power-law slopes ($1 < \alpha < 3$). We note that the relation between continuum flux ratio and temperature does not significantly depend on $\alpha$ and $\beta_{\rm IR}$ and is primarily dependent on the assumed temperature $T$ (see also Section~\ref{sec:firmeasurementhighz}). The black points with error bars show simulated data points with fluxes observed at different depths for comparison. The continuum flux ratio $F_{158\,{\rm \mu m}}/F_{63\,{\rm \mu m}}$ would be the ideal choice to quantify the temperature because of its steep dependence on temperature. However, the atmospheric transparency for $z\sim5.9$ at the observed wavelength of $63\,{\rm \mu m}$~is lowest with $38\%$ compared to $40\%$, $82\%$, and $94\%$ for $88\,{\rm \mu m}$, $122\,{\rm \mu m}$, and $205\,{\rm \mu m}$, respectively. 
	The right-side panel in Figure~\ref{fig:fluxratios} shows this more qualitatively. Each sub-panel corresponds to a continuum flux ratio and shows the needed depth of the flux measurement (in $\sigma$) to reach a certain $\sigma$ of significance to distinguish three different pairs of temperature. The calculation of the latter combines the model uncertainty at a given temperature with the \textit{measurement uncertainty} (corresponding to the $y-$axis). The $F_{158\,{\rm \mu m}}/F_{63\,{\rm \mu m}}$ ratio gives the best turn-out for a given depth of flux measurement. All considered temperatures can be distinguished at $>2\sigma$ ($>3\sigma$) significance for a $10\sigma$ ($20\sigma$) detection at $63\,{\rm \mu m}$. For $F_{158\,{\rm \mu m}}/F_{88\,{\rm \mu m}}$ temperatures of $30\,{\rm K}$ and $60\,{\rm K}$ can be distinguished at $2-2.5\sigma$ significance for a $6\sigma$ detection at $88\,{\rm \mu m}$, while for $F_{158\,{\rm \mu m}}/F_{122\,{\rm \mu m}}$ a $>15\sigma$ detection in $122\,{\rm \mu m}$ is necessary to reach the same significance. Finally, the $F_{158\,{\rm \mu m}}/F_{205\,{\rm \mu m}}$ vs. temperature relation is too shallow to separate temperatures at $T>30\,{\rm K}$ with a significance of more than $1\sigma$ for any observational depth. The latter is the reason why the temperature estimates by \citet{PAVESI16} have such a large quoted uncertainty.

\begin{figure*}
\centering
\includegraphics[width=1\columnwidth, angle=0]{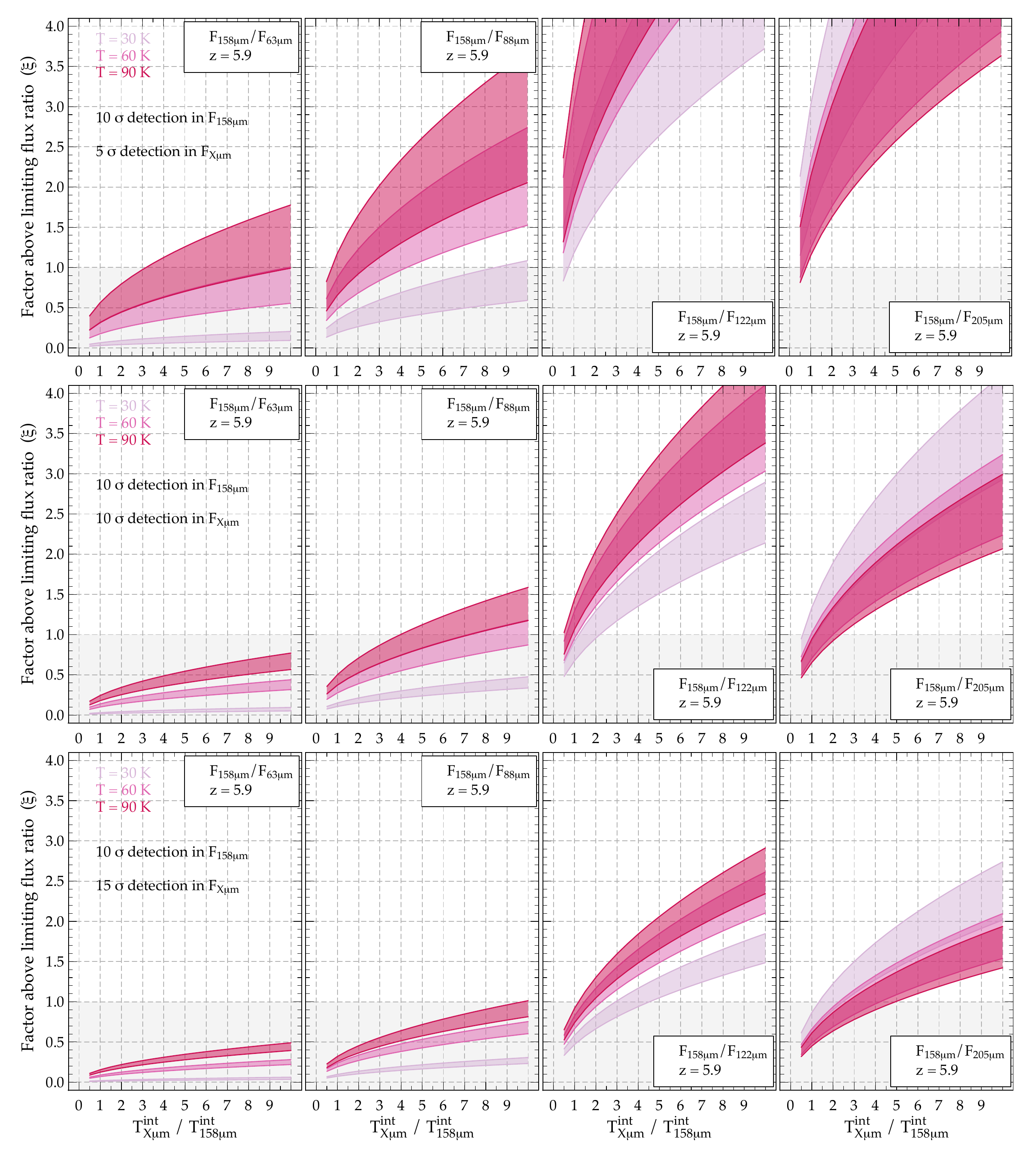}
\caption{Within each of the three sets of panels, the sub-panels show the significance to distinguish temperatures of $30\,{\rm K}$, $60\,{\rm K}$, and $90\,{\rm K}$ with flux ratios $F_{158\,{\rm \mu m}}/F_{X\,{\rm \mu m}}$, where $X$ denotes the ``secondary bands'' ($63\,{\mu m}$, $88\,{\mu m}$, $122\,{\mu m}$, and $205\,{\mu m}$ from left to right, respectively), as a function of integration time in the secondary bands with respect to the integration time at $158\,{\rm \mu m}$ ($10\,\sigma$ detection). The width of the curves include \textit{model uncertainties} (see Figure~\ref{fig:fluxratios}) as well as \textit{measurement uncertainties}, which are varied between the three sets of panels ($5\,\sigma$, $10\,\sigma$, and $15\,\sigma$ detection in the secondary bands from top to bottom). Temperatures can only be cleanly separated if the curves are not overlapping. Furthermore, the $y-$axis shows the factor above the flux ratio, $\xi$, that can be observed in a given configuration. Thus if $\xi<1$, the depth of an observation is not sufficient to measure a flux ratio at a given temperature as shown in Figure~\ref{fig:fluxratios}.
It is difficult to distinguish temperatures between $T=60\,{\rm K}$ and $T=90\,{\rm K}$, because the temperature$-$flux ratio relations are too shallow for $122\,{\rm \mu m}$ and $205\,{\rm \mu m}$ even for very high $\sigma$ detections, and, on the other hand, a too long integration time at $63\,{\rm \mu m}$ and $88\,{\rm \mu m}$ is needed to measures flux ratios at all and at the same time providing a $>5\sigma$ detection to distinguish the temperatures. In contrast, temperatures between $T=30\,{\rm K}$ and $T=60\,{\rm K}$ are easy to distinguish with $F_{158\,{\rm \mu m}}/F_{122\,{\rm \mu m}}$ and $F_{158\,{\rm \mu m}}/F_{205\,{\rm \mu m}}$ flux ratios and $>10\sigma$ detection at $122\,{\rm \mu m}$ and $205\,{\rm \mu m}$.
\label{fig:almaresults}}
\end{figure*}

\subsection{A fiducial survey at $z=5.9$}\label{sec:fiducialsurvey}

	The final success to measure a temperature depends on \textit{(i)} the ability to reach the flux ratio for a given temperature (the \textit{limiting flux ratio} $\xi$) and \textit{(ii)} the ability to split different temperatures (as discussed above).
	Figure~\ref{fig:almaresults} demonstrates these points by combining the previous results with relative integration time calculations for a fiducial redshift of $z=5.9$. The panels show the reached flux ratio above the limiting flux ratio as a function of integration in the secondary wavelengths with respect to the integration time necessary for a $10\sigma$ detection in $158\,{\rm \mu m}$. Note that $\xi<1$ only result in an upper limit in temperature for all flux ratios except for $F_{158\,{\rm \mu m}}/F_{205\,{\rm \mu m}}$ where it results in a lower temperature limit (see left panel of Figure~\ref{fig:fluxratios}). The horizontal panels show the four different flux ratios while the sets of panels (arranged vertically) show different depths for the observations at the secondary wavelengths ($5\sigma$, $10\sigma$, and $15\sigma$). The width of the curves (showing temperatures of $30\,{\rm K}$, $60\,{\rm K}$, and $90\,{\rm K}$)\footnote{Corresponding to $20\,{\rm K}$, $35\,{\rm K}$, and $50\,{\rm K}$ in $T_{\rm peak}$.} include the model and measurement uncertainties.
	In summary, we note the following.
	\begin{itemize}
	\item Temperatures cannot be distinguished above $T=60\,{\rm K}$ with more than $1\sigma$ significance using the $F_{158\,{\rm \mu m}}/F_{122\,{\rm \mu m}}$ and $F_{158\,{\rm \mu m}}/F_{205\,{\rm \mu m}}$ ratios at flux measurement depths corresponding to less than $15\sigma$ signifiance.
	\item Temperatures can be confined at $T < 30\,{\rm K}$ or $T >30\,{\rm K}$ at $>1.5\sigma$ for a $10\sigma$ detection in $122\,{\rm \mu m}$ or $205\,{\rm \mu m}$ within $1-2$ times the integration time in $158\,{\rm \mu m}$ (with $10\sigma$).
	\item The measurement of temperatures with the $F_{158\,{\rm \mu m}}/F_{63\,{\rm \mu m}}$ and $F_{158\,{\rm \mu m}}/F_{88\,{\rm \mu m}}$ ratios is not feasible within a reasonable amount of time (less than $10$ times the integration time in $158\,{\rm \mu m}$).
	\item Therefore, the optimal observing strategy to study the temperature of $z\sim6$ galaxies is to target rest-frame $158\,{\rm \mu m}$ and $122\,{\rm \mu m}$ (or $205\,{\rm \mu m}$ if $122\,{\rm \mu m}$ is affected by low transmission, but with a significantly decreased success rate), thereby requiring a depth of at least $10\sigma$.
	\end{itemize}
	
\begin{figure*}
\centering
\includegraphics[width=1\columnwidth, angle=0]{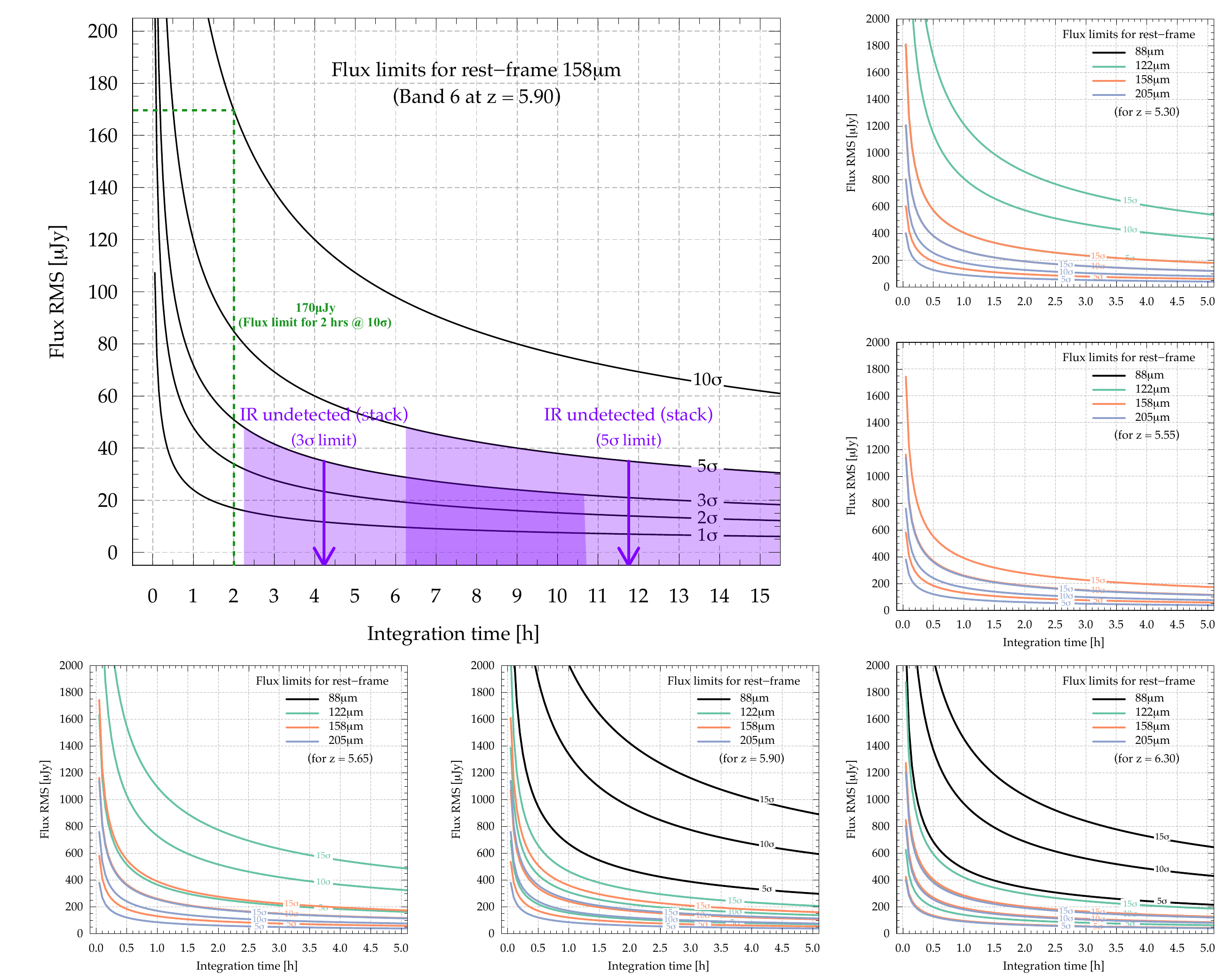}
\caption{Integration time calculations (assuming dual polarization with a bandwidth of $7.5\,{\rm GHz}$, $0.913\,{\rm mm}$ water vapor column density, and $40$ antennae).
\textit{Small panels:} The $5\sigma$, $10\sigma$, and $15\sigma$ flux RMS per beam at $88\,{\rm \mu m}$, $122\,{\rm \mu m}$, $158\,{\rm \mu m}$, and $205\,{\rm \mu m}$ (color coded) as a function of integration time. Each panel shows a different redshift (note that some wavelengths are not visible at certain redshift, see Figure~\ref{fig:almatransmission}). 
\textit{Large panel:} In-detail integration time calculations for $158\,{\rm \mu m}$ at $z=5.9$ for $1\sigma$, $2\sigma$, $3\sigma$, $5\sigma$, and $10\sigma$ detections. The arrows show a lower limit to the required integration times for a $3\sigma$ and $5\sigma$ $158\,{\rm \mu m}-$continuum detection of the IR undetected sources.
\label{fig:almaexptimes}}
\end{figure*}

	After setting up the frame-work, we now proceed to establish integration times and flux limits for which temperatures can be measured efficiently.
	Figure~\ref{fig:almaexptimes} shows the estimated flux root mean square (RMS) limits at a given $\sigma$ significance as a function of integration time for different wavelengths and redshifts. We assume a dual polarization set-up with a bandwidth of $7.5\,{\rm GHz}$, precipitable water vapor column density of $0.913\,{\rm mm}$, $40$ antennae, and a resolution of $1\arcsec$ in the following. The calculations are done for objects at $\delta(2000) = +$02:00:00 observed with ALMA and the integration times do not include overheads. 
	The large panel shows a more detailed graph for $158\,{\rm \mu m}$ at $z=5.9$. An integration time of $2\,{\rm hours}$ would result in a $10\sigma$ $158\,{\rm \mu m}$ continuum flux sensitivity of $170\,{\rm \mu Jy}$ per beam, which is similar the flux observed in \textit{HZ4} (see Table~\ref{tab:expphot}). To distinguish temperatures of $30{\rm K}$ and $60\,{\rm K}$ with $F_{158\,{\rm \mu m}}/F_{122\,{\rm \mu m}}$ for such a source at $z=5.9$ with a significance of $1\sigma$ (for which a $10\sigma$ detection at $122\,{\rm \mu m}$ is necessary), we have to double the integration time at $158\,{\rm \mu m}$, thus $4\,{\rm hours}$ at $122\,{\rm \mu m}$ (see Figure~\ref{fig:almaresults}). Alternatively, a $2\times 4 = 8\,{\rm hour}$ integration (corresponding to $15\sigma$ at $122\,{\rm \mu m}$) will result in a significance to distinguish $T=30{\rm K}$ and $T=60\,{\rm K}$ of $2\sigma$. The significance increases by roughly $30-50\%$ for temperatures of $T=30{\rm K}$ and $T=90\,{\rm K}$.

\subsection{Temperature measurement of existing sources detected in \Cii}
	
	\citetalias{CAPAK15} provides four of the IR brightest $\sim L^*$ galaxies at $z\sim5.5$ that can be followed up with our method to verify our inferred average temperature of such galaxies. Table~\ref{tab:expphot} shows their expected (and measured in case of $158\,{\rm \mu m}$) fluxes. We also include the other $7$ galaxies that are not detected in the FIR continuum. For the prediction of the fluxes we assumed a MIR power-law slope of $1.5 < \alpha < 2.5$, an emissivity of $1 < \beta_{\rm IR} < 2$ and a temperature $60\,{\rm K} < T < 90\,{\rm K}$ ($35\,{\rm K} < T_{\rm peak} < 50\,{\rm K}$). For $25\,{\rm K} < T < 45\,{\rm K}$ ($18\,{\rm K} < T_{\rm peak} < 28\,{\rm K}$), as assumed in \citetalias{CAPAK15}, fluxes at wavelengths blue-ward of \Cii~would be lower while fluxes red-ward of \Cii~would be larger because of the shift of the SED to lower temperatures. In the following, we assume the same ALMA configuration as in Section~\ref{sec:fiducialsurvey}.
	
	\textbf{HZ10.} The temperature of \textit{HZ10} at $z\sim5.65$ has already been estimated to be $T\sim30\,{\rm K}$ ($T_{\rm peak} \sim 20\,{\rm K}$) \citep[][]{PAVESI16}. However, the uncertainty of this estimate is large since only data at $>158\,{\rm \mu m}$ are available. 
	A better temperature estimate can be achieved by a $17\sigma$ measurement in  $122\,{\rm \mu m}$. This would uniquely differentiate a temperature of $T=30\,{\rm K}$ and $T=60\,{\rm K}$ ($T=90\,{\rm K}$) with $2\sigma$ ($3\sigma$) significance (Figure~\ref{fig:fluxratios}). With an estimated flux of $\sim2000\,{\rm \mu Jy}$ at $122\,{\rm \mu m}$, a $17\sigma$ measurement can be achieved in $\sim20-30\,{\rm minutes}$ of observing time in Band 7 (see Figures~\ref{fig:almaexptimes}~and~\ref{fig:almatransmission}). Assuming a temperature prior of $25\,{\rm K} < T < 45\,{\rm K}$ instead, we would expect a flux sensitivity of $\sim1400\,{\rm \mu Jy}$ and a required integration time of $\sim40-50\,{\rm minutes}$.
	
	\textbf{HZ9.} For the second brightest galaxy in FIR at $z\sim5.55$, rest-frame $122\,{\rm \mu m}$ cannot be observed as it is falling between Bands 7 and 8. Similarly, $63\,{\rm \mu m}$ and $88\,{\rm \mu m}$ are out of range. Instead, a $20\sigma$ continuum measurement at $205\,{\rm \mu m}$ can uniquely differentiate a temperature of $T=30\,{\rm K}$ and $T=60\,{\rm K}/90\,{\rm K}$ with $1\sigma$ significance. With an estimated flux of $\sim300\,{\rm \mu Jy}$ at $205\,{\rm \mu m}$, a $20\sigma$ measurement can be achieved in $\sim1.5\,{\rm hours}$ of observation in Band 6. Assuming a temperature prior of $25\,{\rm K} < T < 45\,{\rm K}$ instead, we would expect a flux of $\sim370\,{\rm \mu Jy}$, which would decrease the integration time to $\sim1\,{\rm hour}$.
	
	\textbf{HZ4.} \textit{HZ4} is at the same redshift as \textit{HZ9} and can therefore not be observed at $122\,{\rm \mu m}$.  Also, $63\,{\rm \mu m}$ and $88\,{\rm \mu m}$ are out of the wavelength range. The $20\sigma$ measurement at $205\,{\rm \mu m}$ that is needed to uniquely differentiate a temperature of $T=30\,{\rm K}$ and $T=60\,{\rm K}/90\,{\rm K}$ with $1\sigma$ significance, can only be achieved at integration times of more than $5\,{\rm hours}$ for the estimated $205\,{\rm \mu m}$ continuum flux of $\sim110\,{\rm \mu Jy}$. The same is true for the source \textit{WMH5} detected by \citet{WILLOTT15}.
	
	\textbf{HZ6.} \textit{HZ6} at $z\sim5.3$ has been observed at $205\,{\rm \mu m}$ by \citet{PAVESI16} and a temperature of $T=60^{+35}_{-27}\,{\rm K}$ ($T_{\rm peak} \sim 35\,{\rm K}$) has been derived with large uncertainty. Note that these measurements are integrated over all the three components. \textit{HZ6} can be re-observed at $122\,{\rm \mu m}$ (expected flux is $340\,{\rm \mu Jy}$) reaching a $6\,\sigma$ continuum detection in $2.5\,{\rm hours}$. This results in a splitting of $T=30\,{\rm K}$ and $T=90\,{\rm K}$ at $1\,\sigma$ level. The $20\sigma$ measurement at $205\,{\rm \mu m}$ that is needed to uniquely differentiate a temperature of $T=30\,{\rm K}$ and $T=60\,{\rm K}/90\,{\rm K}$ with $1\sigma$ significance, can only be achieved at integration times of more than $5\,{\rm hours}$ for the estimated $205\,{\rm \mu m}$ continuum flux of $\sim126\,{\rm \mu Jy}$ \citep[c.f. $110\pm4\,{\rm \mu Jy}$ as measured by][]{PAVESI16}.
	
	\textbf{FIR undetected galaxies.} The $158\,{\rm \mu m}$ continuum flux is not known for these galaxies and therefore only a lower limit in integration time can be inferred. Taking at face value the continuum flux limit of the stacked observations ($\sim35\,{\rm \mu Jy}$), an integration time of at least $12\,{\rm hours}$ is needed for a continuum detection of $5\sigma$ at $158\,{\rm \mu m}$ as shown in the large panel in Figure~\ref{fig:almaexptimes}. Note that this figure was made for $z=5.9$, but due to the flat transmission function around rest-frame $158\,{\rm \mu m}$ (Figure~\ref{fig:almatransmission}), the integration time estimates also hold for $z\sim5.5$. For a $5\sigma$ detection at $158\,{\rm \mu m}$ (instead of $10\sigma$) in Figure~\ref{fig:almaresults}, the resulting integration times would increase by a factor $4$. Thus, for a $5\sigma$ detection at $205\,{\rm \mu m}$, Figure~\ref{fig:almaresults} suggests $4-8$ times the integration time in $158\,{\rm \mu m}$, so more than $50\,{\rm hours}$.
	Thus, estimating the temperature for these FIR undetected sources is not feasible with today's capabilities. This is also true for the $158\,{\rm \mu m}$ continuum detected candidates found in ALMA blind searches \citep{ARAVENA16} as well as for CLM1 from \citet{WILLOTT15}.
	
\begin{deluxetable*}{l c c ccccccccc}
\tabletypesize{\scriptsize}
\tablecaption{Expected photometry for high-z galaxies predicted from their $158\,{\rm \mu m}$ flux and $1.5 < \alpha < 2.5$, $1 < \beta_{\rm IR} < 2$, and $60\,{\rm K} < T < 90\,{\rm K}$ ($35\,{\rm K} < T_{\rm peak} < 50\,{\rm K}$).\label{tab:expphot}}
\tablewidth{0pt}
\tablehead{
\colhead{Name} & \colhead{z} & \colhead{$\log(\LIR/\Lsol)$} & \colhead{$63\,{\rm\mu m}$} & \colhead{$88\,{\rm\mu m}$} & \colhead{$110\,{\rm\mu m}$} & \colhead{$122\,{\rm\mu m}$} & \colhead{$158\,{\rm\mu m}$$^{a}$} & \colhead{$190\,{\rm\mu m}$} & \colhead{$205\,{\rm\mu m}$} & \colhead{$220\,{\rm\mu m}$} & \colhead{Reference$^{b}$} \\
\colhead{} & \colhead{} & \colhead{} & \colhead{(${\rm \mu Jy}$)} & \colhead{(${\rm \mu Jy}$)} & \colhead{(${\rm \mu Jy}$)} & \colhead{(${\rm \mu Jy}$)} & \colhead{(${\rm \mu Jy}$)} & \colhead{(${\rm \mu Jy}$)} & \colhead{(${\rm \mu Jy}$)} & \colhead{(${\rm \mu Jy}$)} & \colhead{}
}
\startdata
HZ10 & 5.6566 & $12.58^{+0.22}_{-0.25}$ & $3100^{+583}_{-649}$ & $2773^{+268}_{-319}$ & $2234^{+120}_{-136}$ & $1947^{+74}_{-81}$ & ($1261^{+ 0}_{- 0}$) & $858^{+31}_{-28}$ & $722^{+38}_{-37}$ & $608^{+45}_{-39}$ & 1 \\ 
HZ9 & 5.5410 & $12.18^{+0.22}_{-0.25}$ & $1269^{+239}_{-266}$ & $1135^{+110}_{-130}$ & $914^{+49}_{-56}$ & $797^{+30}_{-33}$ & ($516^{+ 0}_{- 0}$) & $351^{+13}_{-11}$ & $296^{+15}_{-15}$ & $249^{+19}_{-16}$ & 1 \\ 
HZ6 (all) & 5.2928 & $11.78^{+0.22}_{-0.25}$ & $541^{+102}_{-113}$ & $484^{+47}_{-56}$ & $390^{+21}_{-24}$ & $340^{+13}_{-14}$ & ($220^{+ 0}_{- 0}$) & $150^{+ 5}_{- 5}$ & $126^{+ 7}_{- 6}$ & $106^{+ 8}_{- 7}$ & 2 \\ 
WMH5 & 6.0695 & $11.87^{+0.22}_{-0.25}$ & $536^{+101}_{-112}$ & $479^{+46}_{-55}$ & $386^{+21}_{-24}$ & $337^{+13}_{-14}$ & ($218^{+ 0}_{- 0}$) & $148^{+ 5}_{- 5}$ & $125^{+ 7}_{- 6}$ & $105^{+ 8}_{- 7}$ & 3 \\ 
HZ4 & 5.5440 & $11.78^{+0.22}_{-0.25}$ & $497^{+93}_{-104}$ & $444^{+43}_{-51}$ & $358^{+19}_{-22}$ & $312^{+12}_{-13}$ & ($202^{+ 0}_{- 0}$) & $137^{+ 5}_{- 4}$ & $116^{+ 6}_{- 6}$ & $97^{+ 7}_{- 6}$ & 1 \\ 
HZ6b & 5.2928 & $11.52^{+0.22}_{-0.25}$ & $295^{+56}_{-62}$ & $264^{+25}_{-30}$ & $213^{+11}_{-13}$ & $185^{+ 7}_{- 8}$ & ($120^{+ 0}_{- 0}$) & $82^{+ 3}_{- 3}$ & $69^{+ 4}_{- 3}$ & $58^{+ 4}_{- 4}$ & 2 \\ 
HZ6c & 5.2928 & $11.44^{+0.22}_{-0.25}$ & $246^{+46}_{-51}$ & $220^{+21}_{-25}$ & $177^{+ 9}_{-11}$ & $154^{+ 6}_{- 6}$ & ($100^{+ 0}_{- 0}$) & $68^{+ 2}_{- 2}$ & $57^{+ 3}_{- 3}$ & $48^{+ 4}_{- 3}$ & 2 \\ 
IDX34 & 7.4910 & $<11.66^{+0.22}_{-0.25}$ & $<246^{+46}_{-51}$ & $<220^{+21}_{-25}$ & $<177^{+ 9}_{-11}$ & $<154^{+ 6}_{- 6}$ & ($<100^{+ 0}_{- 0}$) & $<68^{+ 2}_{- 2}$ & $<57^{+ 3}_{- 3}$ & $<48^{+ 4}_{- 3}$ & 4 \\ 
ID04 & 6.8670 & $11.52^{+0.22}_{-0.25}$ & $202^{+38}_{-42}$ & $180^{+17}_{-21}$ & $145^{+ 8}_{- 9}$ & $127^{+ 5}_{- 5}$ & ($82^{+ 0}_{- 0}$) & $56^{+ 2}_{- 2}$ & $47^{+ 2}_{- 2}$ & $40^{+ 3}_{- 3}$ & 4 \\ 
ID52 & 6.0180 & $<11.37^{+0.22}_{-0.25}$ & $<172^{+32}_{-36}$ & $<154^{+15}_{-18}$ & $<124^{+ 7}_{- 8}$ & $<108^{+ 4}_{- 4}$ & ($<70^{+ 0}_{- 0}$) & $<48^{+ 2}_{- 2}$ & $<40^{+ 2}_{- 2}$ & $<34^{+ 3}_{- 2}$ & 4 \\ 
ID30 & 6.8540 & $<11.42^{+0.22}_{-0.25}$ & $<160^{+30}_{-33}$ & $<143^{+14}_{-16}$ & $<115^{+ 6}_{- 7}$ & $<100^{+ 4}_{- 4}$ & ($<65^{+ 0}_{- 0}$) & $<44^{+ 2}_{- 1}$ & $<37^{+ 2}_{- 2}$ & $<31^{+ 2}_{- 2}$ & 4 \\ 
ID02 & 7.9140 & $<11.43^{+0.22}_{-0.25}$ & $<135^{+25}_{-28}$ & $<121^{+12}_{-14}$ & $<97^{+ 5}_{- 6}$ & $<85^{+ 3}_{- 4}$ & ($<55^{+ 0}_{- 0}$) & $<37^{+ 1}_{- 1}$ & $<32^{+ 2}_{- 2}$ & $<27^{+ 2}_{- 2}$ & 4 \\ 
HZ3 & 5.5416 & $<11.18^{+0.22}_{-0.25}$ & $<125^{+24}_{-26}$ & $<112^{+11}_{-13}$ & $<90^{+ 5}_{- 6}$ & $<79^{+ 3}_{- 3}$ & ($<51^{+ 0}_{- 0}$) & $<35^{+ 1}_{- 1}$ & $<29^{+ 2}_{- 1}$ & $<25^{+ 2}_{- 2}$ & 1 \\ 
CLM1 & 6.1657 & $11.18^{+0.22}_{-0.25}$ & $108^{+20}_{-23}$ & $97^{+ 9}_{-11}$ & $78^{+ 4}_{- 5}$ & $68^{+ 3}_{- 3}$ & ($44^{+ 0}_{- 0}$) & $30^{+ 1}_{- 1}$ & $25^{+ 1}_{- 1}$ & $21^{+ 2}_{- 1}$ & 3 \\ 
ID41 & 6.3460 & $<11.20^{+0.22}_{-0.25}$ & $<108^{+20}_{-23}$ & $<97^{+ 9}_{-11}$ & $<78^{+ 4}_{- 5}$ & $<68^{+ 3}_{- 3}$ & ($<44^{+ 0}_{- 0}$) & $<30^{+ 1}_{- 1}$ & $<25^{+ 1}_{- 1}$ & $<21^{+ 2}_{- 1}$ & 4 \\ 
ID09 & 6.0240 & $<11.15^{+0.22}_{-0.25}$ & $<103^{+19}_{-22}$ & $<92^{+ 9}_{-11}$ & $<74^{+ 4}_{- 5}$ & $<65^{+ 2}_{- 3}$ & ($<42^{+ 0}_{- 0}$) & $<29^{+ 1}_{- 1}$ & $<24^{+ 1}_{- 1}$ & $<20^{+ 2}_{- 1}$ & 4 \\ 
ID14 & 6.7510 & $<11.22^{+0.22}_{-0.25}$ & $<103^{+19}_{-22}$ & $<92^{+ 9}_{-11}$ & $<74^{+ 4}_{- 5}$ & $<65^{+ 2}_{- 3}$ & ($<42^{+ 0}_{- 0}$) & $<29^{+ 1}_{- 1}$ & $<24^{+ 1}_{- 1}$ & $<20^{+ 2}_{- 1}$ & 4 \\ 
ID38 & 6.5930 & $<11.20^{+0.22}_{-0.25}$ & $<103^{+19}_{-22}$ & $<92^{+ 9}_{-11}$ & $<74^{+ 4}_{- 5}$ & $<65^{+ 2}_{- 3}$ & ($<42^{+ 0}_{- 0}$) & $<29^{+ 1}_{- 1}$ & $<24^{+ 1}_{- 1}$ & $<20^{+ 2}_{- 1}$ & 4 \\ 
IDX25 & 6.3570 & $<11.16^{+0.22}_{-0.25}$ & $<98^{+19}_{-21}$ & $<88^{+ 8}_{-10}$ & $<71^{+ 4}_{- 4}$ & $<62^{+ 2}_{- 3}$ & ($<40^{+ 0}_{- 0}$) & $<27^{+ 1}_{- 1}$ & $<23^{+ 1}_{- 1}$ & $<19^{+ 1}_{- 1}$ & 4 \\ 
ID44 & 7.3600 & $<11.25^{+0.22}_{-0.25}$ & $<98^{+19}_{-21}$ & $<88^{+ 8}_{-10}$ & $<71^{+ 4}_{- 4}$ & $<62^{+ 2}_{- 3}$ & ($<40^{+ 0}_{- 0}$) & $<27^{+ 1}_{- 1}$ & $<23^{+ 1}_{- 1}$ & $<19^{+ 1}_{- 1}$ & 4 \\ 
ID49 & 6.0510 & $<11.13^{+0.22}_{-0.25}$ & $<98^{+19}_{-21}$ & $<88^{+ 8}_{-10}$ & $<71^{+ 4}_{- 4}$ & $<62^{+ 2}_{- 3}$ & ($<40^{+ 0}_{- 0}$) & $<27^{+ 1}_{- 1}$ & $<23^{+ 1}_{- 1}$ & $<19^{+ 1}_{- 1}$ & 4 \\ 
HZ7 & 5.2532 & $<10.99^{+0.22}_{-0.25}$ & $<89^{+17}_{-19}$ & $<79^{+ 8}_{- 9}$ & $<64^{+ 3}_{- 4}$ & $<56^{+ 2}_{- 2}$ & ($<36^{+ 0}_{- 0}$) & $<24^{+ 1}_{- 1}$ & $<21^{+ 1}_{- 1}$ & $<17^{+ 1}_{- 1}$ & 1 \\ 
ID27 & 7.5750 & $11.20^{+0.22}_{-0.25}$ & $84^{+16}_{-17}$ & $75^{+ 7}_{- 9}$ & $60^{+ 3}_{- 4}$ & $53^{+ 2}_{- 2}$ & ($34^{+ 0}_{- 0}$) & $23^{+ 1}_{- 1}$ & $19^{+ 1}_{- 1}$ & $16^{+ 1}_{- 1}$ & 4 \\ 
HZ5 & 5.3089 & $<10.95^{+0.22}_{-0.25}$ & $<79^{+15}_{-16}$ & $<70^{+ 7}_{- 8}$ & $<57^{+ 3}_{- 3}$ & $<49^{+ 2}_{- 2}$ & ($<32^{+ 0}_{- 0}$) & $<22^{+ 1}_{- 1}$ & $<18^{+ 1}_{- 1}$ & $<15^{+ 1}_{- 1}$ & 1 \\ 
HZ5a & 5.3089 & $<10.95^{+0.22}_{-0.25}$ & $<79^{+15}_{-16}$ & $<70^{+ 7}_{- 8}$ & $<57^{+ 3}_{- 3}$ & $<49^{+ 2}_{- 2}$ & ($<32^{+ 0}_{- 0}$) & $<22^{+ 1}_{- 1}$ & $<18^{+ 1}_{- 1}$ & $<15^{+ 1}_{- 1}$ & 1 \\ 
HZ1 & 5.6886 & $<10.96^{+0.22}_{-0.25}$ & $<74^{+14}_{-15}$ & $<66^{+ 6}_{- 8}$ & $<53^{+ 3}_{- 3}$ & $<46^{+ 2}_{- 2}$ & ($<30^{+ 0}_{- 0}$) & $<20^{+ 1}_{- 1}$ & $<17^{+ 1}_{- 1}$ & $<14^{+ 1}_{- 1}$ & 1 \\ 
HZ6a & 5.2928 & $10.92^{+0.22}_{-0.25}$ & $74^{+14}_{-15}$ & $66^{+ 6}_{- 8}$ & $53^{+ 3}_{- 3}$ & $46^{+ 2}_{- 2}$ & ($30^{+ 0}_{- 0}$) & $20^{+ 1}_{- 1}$ & $17^{+ 1}_{- 1}$ & $14^{+ 1}_{- 1}$ & 2 \\ 
HZ8 & 5.1533 & $<10.90^{+0.22}_{-0.25}$ & $<74^{+14}_{-15}$ & $<66^{+ 6}_{- 8}$ & $<53^{+ 3}_{- 3}$ & $<46^{+ 2}_{- 2}$ & ($<30^{+ 0}_{- 0}$) & $<20^{+ 1}_{- 1}$ & $<17^{+ 1}_{- 1}$ & $<14^{+ 1}_{- 1}$ & 1 \\ 
HZ8W & 5.1532 & $<10.90^{+0.22}_{-0.25}$ & $<74^{+14}_{-15}$ & $<66^{+ 6}_{- 8}$ & $<53^{+ 3}_{- 3}$ & $<46^{+ 2}_{- 2}$ & ($<30^{+ 0}_{- 0}$) & $<20^{+ 1}_{- 1}$ & $<17^{+ 1}_{- 1}$ & $<14^{+ 1}_{- 1}$ & 1 \\ 
ID31 & 7.4940 & $11.14^{+0.22}_{-0.25}$ & $74^{+14}_{-15}$ & $66^{+ 6}_{- 8}$ & $53^{+ 3}_{- 3}$ & $46^{+ 2}_{- 2}$ & ($30^{+ 0}_{- 0}$) & $20^{+ 1}_{- 1}$ & $17^{+ 1}_{- 1}$ & $14^{+ 1}_{- 1}$ & 4 \\ 
HZ2 & 5.6697 & $<10.95^{+0.22}_{-0.25}$ & $<71^{+13}_{-15}$ & $<64^{+ 6}_{- 7}$ & $<51^{+ 3}_{- 3}$ & $<45^{+ 2}_{- 2}$ & ($<29^{+ 0}_{- 0}$) & $<20^{+ 1}_{- 1}$ & $<17^{+ 1}_{- 1}$ & $<14^{+ 1}_{- 1}$ & 1 \\ 
\enddata
\tablenotetext{{a}}{The $158\,{\rm \mu m}$ fluxes are measured used for predicting the fluxes at the other wavelengths.}
\tablenotetext{{b}}{List of references: 1--\citet{CAPAK15}; 2--\citet{BARISIC17}; 3--\citet{WILLOTT15}; 4--\citet{ARAVENA16}}
\tablenotetext{\dagger}{Assuming a temperature of $T\sim30\,{\rm K}$ ($T_{\rm peak} \sim 20\,{\rm K}$) for \textit{HZ10} \citep[][]{PAVESI16}, this would result in the following fluxes for $63\,{\rm \mu m}$, $88\,{\rm \mu m}$, $110\,{\rm \mu m}$, $122\,{\rm \mu m}$, $190\,{\rm \mu m}$, $205\,{\rm \mu m}$, $220\,{\rm \mu m}$: $618\,{\rm \mu Jy}$, $1185\,{\rm \mu Jy}$, $1411\,{\rm \mu Jy}$, $1434\,{\rm \mu Jy}$, $1012\,{\rm \mu Jy}$, $902\,{\rm \mu Jy}$, $800\,{\rm \mu Jy}$.}
\end{deluxetable*}




\bibliographystyle{aasjournal}
\bibliography{bibli.bib}

\begin{thebibliography}{}
\expandafter\ifx\csname natexlab\endcsname\relax\def\natexlab#1{#1}\fi

\bibitem[{{{\'A}lvarez-M{\'a}rquez} {et~al.}(2016){{\'A}lvarez-M{\'a}rquez},
  {Burgarella}, {Heinis}, {Buat}, {Lo Faro}, {B{\'e}thermin},
  {L{\'o}pez-Fort{\'{\i}}n}, {Cooray}, {Farrah}, {Hurley}, {Ibar}, {Ilbert},
  {Koekemoer}, {Lemaux}, {P{\'e}rez-Fournon}, {Rodighiero}, {Salvato}, {Scott},
  {Taniguchi}, {Vieira}, \& {Wang}}]{ALVAREZMARQUEZ16}
{{\'A}lvarez-M{\'a}rquez}, J., {Burgarella}, D., {Heinis}, S., {et~al.} 2016,
  \aap, 587, A122

\bibitem[{{Anders} \& {Fritze-v.~Alvensleben}(2003)}]{ANDERS03}
{Anders}, P., \& {Fritze-v.~Alvensleben}, U. 2003, \aap, 401, 1063

\bibitem[{{Ando} {et~al.}(2007){Ando}, {Ohta}, {Iwata}, {Akiyama}, {Aoki}, \&
  {Tamura}}]{ANDO07}
{Ando}, M., {Ohta}, K., {Iwata}, I., {et~al.} 2007, \pasj, 59, 717

\bibitem[{{Aravena} {et~al.}(2016){Aravena}, {Decarli}, {Walter}, {Bouwens},
  {Oesch}, {Carilli}, {Bauer}, {Da Cunha}, {Daddi}, {G{\'o}nzalez-L{\'o}pez},
  {Ivison}, {Riechers}, {Smail}, {Swinbank}, {Weiss}, {Anguita}, {Bacon},
  {Bell}, {Bertoldi}, {Cortes}, {Cox}, {Hodge}, {Ibar}, {Inami}, {Infante},
  {Karim}, {Magnelli}, {Ota}, {Popping}, {van der Werf}, {Wagg}, \&
  {Fudamoto}}]{ARAVENA16}
{Aravena}, M., {Decarli}, R., {Walter}, F., {et~al.} 2016, ArXiv e-prints,
  arXiv:1607.06772

\bibitem[{{Armus} {et~al.}(2009){Armus}, {Mazzarella}, {Evans}, {Surace},
  {Sanders}, {Iwasawa}, {Frayer}, {Howell}, {Chan}, {Petric}, {Vavilkin},
  {Kim}, {Haan}, {Inami}, {Murphy}, {Appleton}, {Barnes}, {Bothun}, {Bridge},
  {Charmandaris}, {Jensen}, {Kewley}, {Lord}, {Madore}, {Marshall},
  {Melbourne}, {Rich}, {Satyapal}, {Schulz}, {Spoon}, {Sturm}, {U}, {Veilleux},
  \& {Xu}}]{ARMUS09}
{Armus}, L., {Mazzarella}, J.~M., {Evans}, A.~S., {et~al.} 2009, \pasp, 121,
  559

\bibitem[{{Baker} {et~al.}(2001){Baker}, {Lutz}, {Genzel}, {Tacconi}, \&
  {Lehnert}}]{BAKER01}
{Baker}, A.~J., {Lutz}, D., {Genzel}, R., {Tacconi}, L.~J., \& {Lehnert}, M.~D.
  2001, \aap, 372, L37

\bibitem[{{Baldwin} {et~al.}(1981){Baldwin}, {Phillips}, \&
  {Terlevich}}]{BALDWIN81}
{Baldwin}, J.~A., {Phillips}, M.~M., \& {Terlevich}, R. 1981, \pasp, 93, 5

\bibitem[{{Bari{\v s}i{\'c}} {et~al.}(2017){Bari{\v s}i{\'c}}, {Faisst},
  {Capak}, {Pavesi}, {Riechers}, {Scoville}, {Cooke}, {Kartaltepe}, {Casey}, \&
  {Smol{\v c}i{\'c}}}]{BARISIC17}
{Bari{\v s}i{\'c}}, I., {Faisst}, A.~L., {Capak}, P.~L., {et~al.} 2017, ArXiv
  e-prints, arXiv:1707.02980

\bibitem[{{Battisti} {et~al.}(2017){Battisti}, {Calzetti}, \&
  {Chary}}]{BATTISTI17}
{Battisti}, A.~J., {Calzetti}, D., \& {Chary}, R.-R. 2017, ArXiv e-prints,
  arXiv:1704.07426

\bibitem[{{Bendo} {et~al.}(2012){Bendo}, {Galliano}, \& {Madden}}]{BENDO12}
{Bendo}, G.~J., {Galliano}, F., \& {Madden}, S.~C. 2012, \mnras, 423, 197

\bibitem[{{B{\'e}thermin} {et~al.}(2015){B{\'e}thermin}, {Daddi}, {Magdis},
  {Lagos}, {Sargent}, {Albrecht}, {Aussel}, {Bertoldi}, {Buat}, {Galametz},
  {Heinis}, {Ilbert}, {Karim}, {Koekemoer}, {Lacey}, {Le Floc'h}, {Navarrete},
  {Pannella}, {Schreiber}, {Smol{\v c}i{\'c}}, {Symeonidis}, \&
  {Viero}}]{BETHERMIN15}
{B{\'e}thermin}, M., {Daddi}, E., {Magdis}, G., {et~al.} 2015, \aap, 573, A113

\bibitem[{{Bian} {et~al.}(2016){Bian}, {Kewley}, {Dopita}, \& {Blanc}}]{BIAN16}
{Bian}, F., {Kewley}, L., {Dopita}, M., \& {Blanc}, G. 2016, ArXiv e-prints,
  arXiv:1611.08595

\bibitem[{{Boquien} {et~al.}(2009){Boquien}, {Calzetti}, {Kennicutt}, {Dale},
  {Engelbracht}, {Gordon}, {Hong}, {Lee}, \& {Portouw}}]{BOQUIEN09}
{Boquien}, M., {Calzetti}, D., {Kennicutt}, R., {et~al.} 2009, \apj, 706, 553

\bibitem[{{Boquien} {et~al.}(2012){Boquien}, {Buat}, {Boselli}, {Baes},
  {Bendo}, {Ciesla}, {Cooray}, {Cortese}, {Eales}, {Gavazzi}, {Gomez},
  {Lebouteiller}, {Pappalardo}, {Pohlen}, {Smith}, \& {Spinoglio}}]{BOQUIEN12}
{Boquien}, M., {Buat}, V., {Boselli}, A., {et~al.} 2012, \aap, 539, A145

\bibitem[{{Bourne} {et~al.}(2017){Bourne}, {Dunlop}, {Merlin}, {Parsa},
  {Schreiber}, {Castellano}, {Conselice}, {Coppin}, {Farrah}, {Fontana},
  {Geach}, {Halpern}, {Knudsen}, {Micha{\l}owski}, {Mortlock}, {Santini},
  {Scott}, {Shu}, {Simpson}, {Simpson}, {Smith}, \& {van der Werf}}]{BOURNE17}
{Bourne}, N., {Dunlop}, J.~S., {Merlin}, E., {et~al.} 2017, \mnras, 467, 1360

\bibitem[{{Bouwens} {et~al.}(2009){Bouwens}, {Illingworth}, {Franx}, {Chary},
  {Meurer}, {Conselice}, {Ford}, {Giavalisco}, \& {van Dokkum}}]{BOUWENS09}
{Bouwens}, R.~J., {Illingworth}, G.~D., {Franx}, M., {et~al.} 2009, \apj, 705,
  936

\bibitem[{{Bouwens} {et~al.}(2012){Bouwens}, {Illingworth}, {Oesch}, {Franx},
  {Labb{\'e}}, {Trenti}, {van Dokkum}, {Carollo}, {Gonz{\'a}lez}, {Smit}, \&
  {Magee}}]{BOUWENS12}
{Bouwens}, R.~J., {Illingworth}, G.~D., {Oesch}, P.~A., {et~al.} 2012, \apj,
  754, 83

\bibitem[{{Bouwens} {et~al.}(2016){Bouwens}, {Aravena}, {Decarli}, {Walter},
  {da Cunha}, {Labb{\'e}}, {Bauer}, {Bertoldi}, {Carilli}, {Chapman}, {Daddi},
  {Hodge}, {Ivison}, {Karim}, {Le Fevre}, {Magnelli}, {Ota}, {Riechers},
  {Smail}, {van der Werf}, {Weiss}, {Cox}, {Elbaz}, {Gonzalez-Lopez},
  {Infante}, {Oesch}, {Wagg}, \& {Wilkins}}]{BOUWENS16}
{Bouwens}, R.~J., {Aravena}, M., {Decarli}, R., {et~al.} 2016, \apj, 833, 72

\bibitem[{{Bruzual} \& {Charlot}(2003)}]{BRUZUALCHARLOT03}
{Bruzual}, G., \& {Charlot}, S. 2003, \mnras, 344, 1000

\bibitem[{{Buat} {et~al.}(2005){Buat}, {Iglesias-P{\'a}ramo}, {Seibert},
  {Burgarella}, {Charlot}, {Martin}, {Xu}, {Heckman}, {Boissier}, {Boselli},
  {Barlow}, {Bianchi}, {Byun}, {Donas}, {Forster}, {Friedman}, {Jelinski},
  {Lee}, {Madore}, {Malina}, {Milliard}, {Morissey}, {Neff}, {Rich},
  {Schiminovitch}, {Siegmund}, {Small}, {Szalay}, {Welsh}, \& {Wyder}}]{BUAT05}
{Buat}, V., {Iglesias-P{\'a}ramo}, J., {Seibert}, M., {et~al.} 2005, \apjl,
  619, L51

\bibitem[{{Buat} {et~al.}(2010){Buat}, {Giovannoli}, {Burgarella}, {Altieri},
  {Amblard}, {Arumugam}, {Aussel}, {Babbedge}, {Blain}, {Bock}, {Boselli},
  {Castro-Rodr{\'{\i}}guez}, {Cava}, {Chanial}, {Clements}, {Conley},
  {Conversi}, {Cooray}, {Dowell}, {Dwek}, {Eales}, {Elbaz}, {Fox},
  {Franceschini}, {Gear}, {Glenn}, {Griffin}, {Halpern}, {Hatziminaoglou},
  {Heinis}, {Ibar}, {Isaak}, {Ivison}, {Lagache}, {Levenson}, {Lonsdale}, {Lu},
  {Madden}, {Maffei}, {Magdis}, {Mainetti}, {Marchetti}, {Morrison}, {Nguyen},
  {O'Halloran}, {Oliver}, {Omont}, {Owen}, {Page}, {Pannella}, {Panuzzo},
  {Papageorgiou}, {Pearson}, {P{\'e}rez-Fournon}, {Pohlen}, {Rigopoulou},
  {Rizzo}, {Roseboom}, {Rowan-Robinson}, {S{\'a}nchez Portal}, {Schulz},
  {Seymour}, {Shupe}, {Smith}, {Stevens}, {Strazzullo}, {Symeonidis},
  {Trichas}, {Tugwell}, {Vaccari}, {Valiante}, {Valtchanov}, {Vigroux}, {Wang},
  {Ward}, {Wright}, {Xu}, \& {Zemcov}}]{BUAT10}
{Buat}, V., {Giovannoli}, E., {Burgarella}, D., {et~al.} 2010, \mnras, 409, L1

\bibitem[{{Buat} {et~al.}(2012){Buat}, {Noll}, {Burgarella}, {Giovannoli},
  {Charmandaris}, {Pannella}, {Hwang}, {Elbaz}, {Dickinson}, {Magdis}, {Reddy},
  \& {Murphy}}]{BUAT12}
{Buat}, V., {Noll}, S., {Burgarella}, D., {et~al.} 2012, \aap, 545, A141

\bibitem[{{Burgarella} {et~al.}(2005){Burgarella}, {Buat}, \&
  {Iglesias-P{\'a}ramo}}]{BURGARELLA05}
{Burgarella}, D., {Buat}, V., \& {Iglesias-P{\'a}ramo}, J. 2005, \mnras, 360,
  1413

\bibitem[{{Calzetti} {et~al.}(2000){Calzetti}, {Armus}, {Bohlin}, {Kinney},
  {Koornneef}, \& {Storchi-Bergmann}}]{CALZETTI00}
{Calzetti}, D., {Armus}, L., {Bohlin}, R.~C., {et~al.} 2000, \apj, 533, 682

\bibitem[{{Capak} {et~al.}(2011){Capak}, {Riechers}, {Scoville}, {Carilli},
  {Cox}, {Neri}, {Robertson}, {Salvato}, {Schinnerer}, {Yan}, {Wilson}, {Yun},
  {Civano}, {Elvis}, {Karim}, {Mobasher}, \& {Staguhn}}]{CAPAK11}
{Capak}, P.~L., {Riechers}, D., {Scoville}, N.~Z., {et~al.} 2011, \nat, 470,
  233

\bibitem[{{Capak} {et~al.}(2015){Capak}, {Carilli}, {Jones}, {Casey},
  {Riechers}, {Sheth}, {Carollo}, {Ilbert}, {Karim}, {Lefevre}, {Lilly},
  {Scoville}, {Smolcic}, \& {Yan}}]{CAPAK15}
{Capak}, P.~L., {Carilli}, C., {Jones}, G., {et~al.} 2015, \nat, 522, 455

\bibitem[{{Cardamone} {et~al.}(2009){Cardamone}, {Schawinski}, {Sarzi},
  {Bamford}, {Bennert}, {Urry}, {Lintott}, {Keel}, {Parejko}, {Nichol},
  {Thomas}, {Andreescu}, {Murray}, {Raddick}, {Slosar}, {Szalay}, \&
  {Vandenberg}}]{CARDAMONE09}
{Cardamone}, C., {Schawinski}, K., {Sarzi}, M., {et~al.} 2009, \mnras, 399,
  1191

\bibitem[{{Carilli} \& {Walter}(2013)}]{CARILLI13}
{Carilli}, C.~L., \& {Walter}, F. 2013, \araa, 51, 105

\bibitem[{{Carilli} {et~al.}(2016){Carilli}, {Chluba}, {Decarli}, {Walter},
  {Aravena}, {Wagg}, {Popping}, {Cortes}, {Hodge}, {Weiss}, {Bertoldi}, \&
  {Riechers}}]{CARILLI16}
{Carilli}, C.~L., {Chluba}, J., {Decarli}, R., {et~al.} 2016, \apj, 833, 73

\bibitem[{{Casey}(2012)}]{CASEY12}
{Casey}, C.~M. 2012, \mnras, 425, 3094

\bibitem[{{Casey} {et~al.}(2014){Casey}, {Scoville}, {Sanders}, {Lee},
  {Cooray}, {Finkelstein}, {Capak}, {Conley}, {De Zotti}, {Farrah}, {Fu}, {Le
  Floc'h}, {Ilbert}, {Ivison}, \& {Takeuchi}}]{CASEY14}
{Casey}, C.~M., {Scoville}, N.~Z., {Sanders}, D.~B., {et~al.} 2014, \apj, 796,
  95

\bibitem[{{Chabrier}(2003)}]{CHABRIER03}
{Chabrier}, G. 2003, \pasp, 115, 763

\bibitem[{{Chapman} {et~al.}(2003){Chapman}, {Helou}, {Lewis}, \&
  {Dale}}]{CHAPMAN03}
{Chapman}, S.~C., {Helou}, G., {Lewis}, G.~F., \& {Dale}, D.~A. 2003, \apj,
  588, 186

\bibitem[{{Charlot} \& {Fall}(2000)}]{CHARLOTFALL00}
{Charlot}, S., \& {Fall}, S.~M. 2000, \apj, 539, 718

\bibitem[{{Chu} {et~al.}(2017){Chu}, {Sanders}, {Larson}, {Mazzarella},
  {Howell}, {D{\'{\i}}az-Santos}, {Xu}, {Paladini}, {Schulz}, {Shupe},
  {Appleton}, {Armus}, {Billot}, {Chan}, {Evans}, {Fadda}, {Frayer}, {Haan},
  {Ishida}, {Iwasawa}, {Kim}, {Lord}, {Murphy}, {Petric}, {Privon}, {Surace},
  \& {Treister}}]{CHU17}
{Chu}, J.~K., {Sanders}, D.~B., {Larson}, K.~L., {et~al.} 2017, ArXiv e-prints,
  arXiv:1702.01756

\bibitem[{{Conley} {et~al.}(2011){Conley}, {Cooray}, {Vieira}, {Gonz{\'a}lez
  Solares}, {Kim}, {Aguirre}, {Amblard}, {Auld}, {Baker}, {Beelen}, {Blain},
  {Blundell}, {Bock}, {Bradford}, {Bridge}, {Brisbin}, {Burgarella},
  {Carpenter}, {Chanial}, {Chapin}, {Christopher}, {Clements}, {Cox},
  {Djorgovski}, {Dowell}, {Eales}, {Earle}, {Ellsworth-Bowers}, {Farrah},
  {Franceschini}, {Frayer}, {Fu}, {Gavazzi}, {Glenn}, {Griffin}, {Gurwell},
  {Halpern}, {Ibar}, {Ivison}, {Jarvis}, {Kamenetzky}, {Krips}, {Levenson},
  {Lupu}, {Mahabal}, {Maloney}, {Maraston}, {Marchetti}, {Marsden},
  {Matsuhara}, {Mortier}, {Murphy}, {Naylor}, {Neri}, {Nguyen}, {Oliver},
  {Omont}, {Page}, {Papageorgiou}, {Pearson}, {P{\'e}rez-Fournon}, {Pohlen},
  {Rangwala}, {Rawlings}, {Raymond}, {Riechers}, {Rodighiero}, {Roseboom},
  {Rowan-Robinson}, {Schulz}, {Scott}, {Scott}, {Serra}, {Seymour}, {Shupe},
  {Smith}, {Symeonidis}, {Tugwell}, {Vaccari}, {Valiante}, {Valtchanov},
  {Verma}, {Viero}, {Vigroux}, {Wang}, {Wiebe}, {Wright}, {Xu}, {Zeimann},
  {Zemcov}, \& {Zmuidzinas}}]{CONLEY11}
{Conley}, A., {Cooray}, A., {Vieira}, J.~D., {et~al.} 2011, \apjl, 732, L35

\bibitem[{{Coppin} {et~al.}(2010){Coppin}, {Pope}, {Men{\'e}ndez-Delmestre},
  {Alexander}, {Dunlop}, {Egami}, {Gabor}, {Ibar}, {Ivison}, {Austermann},
  {Blain}, {Chapman}, {Clements}, {Dunne}, {Dye}, {Farrah}, {Hughes},
  {Mortier}, {Page}, {Rowan-Robinson}, {Scott}, {Simpson}, {Smail}, {Swinbank},
  {Vaccari}, \& {Yun}}]{COPPIN10}
{Coppin}, K., {Pope}, A., {Men{\'e}ndez-Delmestre}, K., {et~al.} 2010, \apj,
  713, 503

\bibitem[{{Coppin} {et~al.}(2015){Coppin}, {Geach}, {Almaini}, {Arumugam},
  {Dunlop}, {Hartley}, {Ivison}, {Simpson}, {Smith}, {Swinbank}, {Blain},
  {Bourne}, {Bremer}, {Conselice}, {Harrison}, {Mortlock}, {Chapman}, {Davies},
  {Farrah}, {Gibb}, {Jenness}, {Karim}, {Knudsen}, {Ibar}, {Micha{\l}owski},
  {Peacock}, {Rigopoulou}, {Robson}, {Scott}, {Stevens}, \& {van der
  Werf}}]{COPPIN15}
{Coppin}, K.~E.~K., {Geach}, J.~E., {Almaini}, O., {et~al.} 2015, \mnras, 446,
  1293

\bibitem[{{Cormier} {et~al.}(2015){Cormier}, {Madden}, {Lebouteiller}, {Abel},
  {Hony}, {Galliano}, {R{\'e}my-Ruyer}, {Bigiel}, {Baes}, {Boselli},
  {Chevance}, {Cooray}, {De Looze}, {Doublier}, {Galametz}, {Hughes},
  {Karczewski}, {Lee}, {Lu}, \& {Spinoglio}}]{CORMIER15}
{Cormier}, D., {Madden}, S.~C., {Lebouteiller}, V., {et~al.} 2015, \aap, 578,
  A53

\bibitem[{{Cortese} {et~al.}(2006){Cortese}, {Boselli}, {Buat}, {Gavazzi},
  {Boissier}, {Gil de Paz}, {Seibert}, {Madore}, \& {Martin}}]{CORTESE06}
{Cortese}, L., {Boselli}, A., {Buat}, V., {et~al.} 2006, \apj, 637, 242

\bibitem[{{Cowie} {et~al.}(2010){Cowie}, {Barger}, \& {Hu}}]{COWIE10}
{Cowie}, L.~L., {Barger}, A.~J., \& {Hu}, E.~M. 2010, \apj, 711, 928

\bibitem[{{Cowie} {et~al.}(2011){Cowie}, {Barger}, \& {Hu}}]{COWIE11}
---. 2011, \apj, 738, 136

\bibitem[{{Curti} {et~al.}(2016){Curti}, {Cresci}, {Mannucci}, {Marconi},
  {Maiolino}, \& {Esposito}}]{CURTI16}
{Curti}, M., {Cresci}, G., {Mannucci}, F., {et~al.} 2016, ArXiv e-prints,
  arXiv:1610.06939

\bibitem[{{da Cunha} {et~al.}(2013){da Cunha}, {Groves}, {Walter}, {Decarli},
  {Weiss}, {Bertoldi}, {Carilli}, {Daddi}, {Elbaz}, {Ivison}, {Maiolino},
  {Riechers}, {Rix}, {Sargent}, \& {Smail}}]{DACUNHA13}
{da Cunha}, E., {Groves}, B., {Walter}, F., {et~al.} 2013, \apj, 766, 13

\bibitem[{{Daddi} {et~al.}(2007){Daddi}, {Dickinson}, {Morrison}, {Chary},
  {Cimatti}, {Elbaz}, {Frayer}, {Renzini}, {Pope}, {Alexander}, {Bauer},
  {Giavalisco}, {Huynh}, {Kurk}, \& {Mignoli}}]{DADDI07a}
{Daddi}, E., {Dickinson}, M., {Morrison}, G., {et~al.} 2007, \apj, 670, 156

\bibitem[{{Dale} {et~al.}(2012){Dale}, {Aniano}, {Engelbracht}, {Hinz},
  {Krause}, {Montiel}, {Roussel}, {Appleton}, {Armus}, {Beir{\~a}o}, {Bolatto},
  {Brandl}, {Calzetti}, {Crocker}, {Croxall}, {Draine}, {Galametz}, {Gordon},
  {Groves}, {Hao}, {Helou}, {Hunt}, {Johnson}, {Kennicutt}, {Koda}, {Leroy},
  {Li}, {Meidt}, {Miller}, {Murphy}, {Rahman}, {Rix}, {Sandstrom}, {Sauvage},
  {Schinnerer}, {Skibba}, {Smith}, {Tabatabaei}, {Walter}, {Wilson}, {Wolfire},
  \& {Zibetti}}]{DALE12}
{Dale}, D.~A., {Aniano}, G., {Engelbracht}, C.~W., {et~al.} 2012, \apj, 745, 95

\bibitem[{{De Barros} {et~al.}(2016){De Barros}, {Reddy}, \&
  {Shivaei}}]{DEBARROS16}
{De Barros}, S., {Reddy}, N., \& {Shivaei}, I. 2016, \apj, 820, 96

\bibitem[{{de Barros} {et~al.}(2014){de Barros}, {Schaerer}, \&
  {Stark}}]{DEBARROS14}
{de Barros}, S., {Schaerer}, D., \& {Stark}, D.~P. 2014, \aap, 563, A81

\bibitem[{{De Looze} {et~al.}(2014){De Looze}, {Cormier}, {Lebouteiller},
  {Madden}, {Baes}, {Bendo}, {Boquien}, {Boselli}, {Clements}, {Cortese},
  {Cooray}, {Galametz}, {Galliano}, {Graci{\'a}-Carpio}, {Isaak}, {Karczewski},
  {Parkin}, {Pellegrini}, {R{\'e}my-Ruyer}, {Spinoglio}, {Smith}, \&
  {Sturm}}]{DELOOZE14}
{De Looze}, I., {Cormier}, D., {Lebouteiller}, V., {et~al.} 2014, \aap, 568,
  A62

\bibitem[{{D{\'{\i}}az-Santos} {et~al.}(2014){D{\'{\i}}az-Santos}, {Armus},
  {Charmandaris}, {Stacey}, {Murphy}, {Haan}, {Stierwalt}, {Malhotra},
  {Appleton}, {Inami}, {Magdis}, {Elbaz}, {Evans}, {Mazzarella}, {Surace}, {van
  der Werf}, {Xu}, {Lu}, {Meijerink}, {Howell}, {Petric}, {Veilleux}, \&
  {Sanders}}]{DIAZSANTOS14}
{D{\'{\i}}az-Santos}, T., {Armus}, L., {Charmandaris}, V., {et~al.} 2014,
  \apjl, 788, L17

\bibitem[{{Draine}(2006)}]{DRAINE06}
{Draine}, B.~T. 2006, \apj, 636, 1114

\bibitem[{{Dunlop}(2016)}]{DUNLOP16}
{Dunlop}, J.~S. 2016, The Messenger, 166, 48

\bibitem[{{Dunne} {et~al.}(2000){Dunne}, {Eales}, {Edmunds}, {Ivison},
  {Alexander}, \& {Clements}}]{DUNNE00}
{Dunne}, L., {Eales}, S., {Edmunds}, M., {et~al.} 2000, \mnras, 315, 115

\bibitem[{{Elbaz} {et~al.}(2007){Elbaz}, {Daddi}, {Le Borgne}, {Dickinson},
  {Alexander}, {Chary}, {Starck}, {Brandt}, {Kitzbichler}, {MacDonald},
  {Nonino}, {Popesso}, {Stern}, \& {Vanzella}}]{ELBAZ07}
{Elbaz}, D., {Daddi}, E., {Le Borgne}, D., {et~al.} 2007, \aap, 468, 33

\bibitem[{{Elmegreen} {et~al.}(2009){Elmegreen}, {Elmegreen}, {Marcus},
  {Shahinyan}, {Yau}, \& {Petersen}}]{ELMEGREEN09}
{Elmegreen}, D.~M., {Elmegreen}, B.~G., {Marcus}, M.~T., {et~al.} 2009, \apj,
  701, 306

\bibitem[{{Erb} {et~al.}(2016){Erb}, {Pettini}, {Steidel}, {Strom}, {Rudie},
  {Trainor}, {Shapley}, \& {Reddy}}]{ERB16}
{Erb}, D.~K., {Pettini}, M., {Steidel}, C.~C., {et~al.} 2016, \apj, 830, 52

\bibitem[{{Faber} {et~al.}(2003){Faber}, {Phillips}, {Kibrick}, {Alcott},
  {Allen}, {Burrous}, {Cantrall}, {Clarke}, {Coil}, {Cowley}, {Davis}, {Deich},
  {Dietsch}, {Gilmore}, {Harper}, {Hilyard}, {Lewis}, {McVeigh}, {Newman},
  {Osborne}, {Schiavon}, {Stover}, {Tucker}, {Wallace}, {Wei}, {Wirth}, \&
  {Wright}}]{FABER03}
{Faber}, S.~M., {Phillips}, A.~C., {Kibrick}, R.~I., {et~al.} 2003, in
  \procspie, Vol. 4841, Instrument Design and Performance for Optical/Infrared
  Ground-based Telescopes, ed. M.~{Iye} \& A.~F.~M. {Moorwood}, 1657--1669

\bibitem[{{Faisst}(2016)}]{FAISST16c}
{Faisst}, A.~L. 2016, \apj, 829, 99

\bibitem[{{Faisst} {et~al.}(2016{\natexlab{a}}){Faisst}, {Capak}, {Hsieh},
  {Laigle}, {Salvato}, {Tasca}, {Cassata}, {Davidzon}, {Ilbert}, {Le
  F{\`e}vre}, {Masters}, {McCracken}, {Steinhardt}, {Silverman}, {de Barros},
  {Hasinger}, \& {Scoville}}]{FAISST16a}
{Faisst}, A.~L., {Capak}, P., {Hsieh}, B.~C., {et~al.} 2016{\natexlab{a}},
  \apj, 821, 122

\bibitem[{{Faisst} {et~al.}(2016{\natexlab{b}}){Faisst}, {Capak}, {Davidzon},
  {Salvato}, {Laigle}, {Ilbert}, {Onodera}, {Hasinger}, {Kakazu}, {Masters},
  {McCracken}, {Mobasher}, {Sanders}, {Silverman}, {Yan}, \&
  {Scoville}}]{FAISST16b}
{Faisst}, A.~L., {Capak}, P.~L., {Davidzon}, I., {et~al.} 2016{\natexlab{b}},
  \apj, 822, 29

\bibitem[{{Feldmann}(2015)}]{FELDMANN15}
{Feldmann}, R. 2015, \mnras, 449, 3274

\bibitem[{{Fitzpatrick}(1986)}]{FITZPATRICK86}
{Fitzpatrick}, E.~L. 1986, \aj, 92, 1068

\bibitem[{{F{\"o}rster Schreiber} {et~al.}(2011){F{\"o}rster Schreiber},
  {Shapley}, {Genzel}, {Bouch{\'e}}, {Cresci}, {Davies}, {Erb}, {Genel},
  {Lutz}, {Newman}, {Shapiro}, {Steidel}, {Sternberg}, \&
  {Tacconi}}]{FORSTERSCHREIBER11}
{F{\"o}rster Schreiber}, N.~M., {Shapley}, A.~E., {Genzel}, R., {et~al.} 2011,
  \apj, 739, 45

\bibitem[{{Fudamoto et al.}(2017)}]{FUDAMOTO17}
{Fudamoto et al.} 2017, \mnras, submitted

\bibitem[{{Genzel} {et~al.}(2015){Genzel}, {Tacconi}, {Lutz}, {Saintonge},
  {Berta}, {Magnelli}, {Combes}, {Garc{\'{\i}}a-Burillo}, {Neri}, {Bolatto},
  {Contini}, {Lilly}, {Boissier}, {Boone}, {Bouch{\'e}}, {Bournaud}, {Burkert},
  {Carollo}, {Colina}, {Cooper}, {Cox}, {Feruglio}, {F{\"o}rster Schreiber},
  {Freundlich}, {Gracia-Carpio}, {Juneau}, {Kovac}, {Lippa}, {Naab}, {Salome},
  {Renzini}, {Sternberg}, {Walter}, {Weiner}, {Weiss}, \& {Wuyts}}]{GENZEL15}
{Genzel}, R., {Tacconi}, L.~J., {Lutz}, D., {et~al.} 2015, \apj, 800, 20

\bibitem[{{Gil de Paz} {et~al.}(2007){Gil de Paz}, {Boissier}, {Madore},
  {Seibert}, {Joe}, {Boselli}, {Wyder}, {Thilker}, {Bianchi}, {Rey}, {Rich},
  {Barlow}, {Conrow}, {Forster}, {Friedman}, {Martin}, {Morrissey}, {Neff},
  {Schiminovich}, {Small}, {Donas}, {Heckman}, {Lee}, {Milliard}, {Szalay}, \&
  {Yi}}]{GILDEPAZ07}
{Gil de Paz}, A., {Boissier}, S., {Madore}, B.~F., {et~al.} 2007, \apjs, 173,
  185

\bibitem[{{Gonz{\'a}lez} {et~al.}(2014){Gonz{\'a}lez}, {Bouwens},
  {Illingworth}, {Labb{\'e}}, {Oesch}, {Franx}, \& {Magee}}]{GONZALEZ14}
{Gonz{\'a}lez}, V., {Bouwens}, R., {Illingworth}, G., {et~al.} 2014, \apj, 781,
  34

\bibitem[{{Gonz{\'a}lez-L{\'o}pez} {et~al.}(2014){Gonz{\'a}lez-L{\'o}pez},
  {Riechers}, {Decarli}, {Walter}, {Vallini}, {Neri}, {Bertoldi}, {Bolatto},
  {Carilli}, {Cox}, {da Cunha}, {Ferrara}, {Gallerani}, \&
  {Infante}}]{GONZALEZLOPEZ14}
{Gonz{\'a}lez-L{\'o}pez}, J., {Riechers}, D.~A., {Decarli}, R., {et~al.} 2014,
  \apj, 784, 99

\bibitem[{{Gordon} {et~al.}(2000){Gordon}, {Witt}, {Rudy}, {Puetter}, {Lynch},
  {Mazuk}, {Misselt}, {Clayton}, \& {Smith}}]{GORDON00}
{Gordon}, K.~D., {Witt}, A.~N., {Rudy}, R.~J., {et~al.} 2000, \apj, 544, 859

\bibitem[{{Green} {et~al.}(2012){Green}, {Froning}, {Osterman}, {Ebbets},
  {Heap}, {Leitherer}, {Linsky}, {Savage}, {Sembach}, {Shull}, {Siegmund},
  {Snow}, {Spencer}, {Stern}, {Stocke}, {Welsh}, {B{\'e}land}, {Burgh},
  {Danforth}, {France}, {Keeney}, {McPhate}, {Penton}, {Andrews},
  {Brownsberger}, {Morse}, \& {Wilkinson}}]{GREEN12}
{Green}, J.~C., {Froning}, C.~S., {Osterman}, S., {et~al.} 2012, \apj, 744, 60

\bibitem[{{Greis} {et~al.}(2016){Greis}, {Stanway}, {Davies}, \&
  {Levan}}]{GREIS16}
{Greis}, S.~M.~L., {Stanway}, E.~R., {Davies}, L.~J.~M., \& {Levan}, A.~J.
  2016, \mnras, 459, 2591

\bibitem[{{Greve} {et~al.}(2012){Greve}, {Vieira}, {Wei{\ss}}, {Aguirre},
  {Aird}, {Ashby}, {Benson}, {Bleem}, {Bradford}, {Brodwin}, {Carlstrom},
  {Chang}, {Chapman}, {Crawford}, {de Breuck}, {de Haan}, {Dobbs}, {Downes},
  {Fassnacht}, {Fazio}, {George}, {Gladders}, {Gonzalez}, {Halverson},
  {Hezaveh}, {High}, {Holder}, {Holzapfel}, {Hoover}, {Hrubes}, {Johnson},
  {Keisler}, {Knox}, {Lee}, {Leitch}, {Lueker}, {Luong-Van}, {Malkan},
  {Marrone}, {McIntyre}, {McMahon}, {Mehl}, {Menten}, {Meyer}, {Montroy},
  {Murphy}, {Natoli}, {Padin}, {Plagge}, {Pryke}, {Reichardt}, {Rest},
  {Rosenman}, {Ruel}, {Ruhl}, {Schaffer}, {Sharon}, {Shaw}, {Shirokoff},
  {Stalder}, {Stanford}, {Staniszewski}, {Stark}, {Story}, {Vanderlinde},
  {Walsh}, {Welikala}, \& {Williamson}}]{GREVE12}
{Greve}, T.~R., {Vieira}, J.~D., {Wei{\ss}}, A., {et~al.} 2012, \apj, 756, 101

\bibitem[{{Heinis} {et~al.}(2013){Heinis}, {Buat}, {B{\'e}thermin}, {Aussel},
  {Bock}, {Boselli}, {Burgarella}, {Conley}, {Cooray}, {Farrah}, {Ibar},
  {Ilbert}, {Ivison}, {Magdis}, {Marsden}, {Oliver}, {Page}, {Rodighiero},
  {Roehlly}, {Schulz}, {Scott}, {Smith}, {Viero}, {Wang}, \&
  {Zemcov}}]{HEINIS13}
{Heinis}, S., {Buat}, V., {B{\'e}thermin}, M., {et~al.} 2013, \mnras, 429, 1113

\bibitem[{{Heinis} {et~al.}(2014){Heinis}, {Buat}, {B{\'e}thermin}, {Bock},
  {Burgarella}, {Conley}, {Cooray}, {Farrah}, {Ilbert}, {Magdis}, {Marsden},
  {Oliver}, {Rigopoulou}, {Roehlly}, {Schulz}, {Symeonidis}, {Viero}, {Xu}, \&
  {Zemcov}}]{HEINIS14}
---. 2014, \mnras, 437, 1268

\bibitem[{{Hemmati} {et~al.}(2014){Hemmati}, {Miller}, {Mobasher}, {Nayyeri},
  {Ferguson}, {Guo}, {Koekemoer}, {Koo}, \& {Papovich}}]{HEMMATI14}
{Hemmati}, S., {Miller}, S.~H., {Mobasher}, B., {et~al.} 2014, \apj, 797, 108

\bibitem[{{Herrera-Camus} {et~al.}(2015){Herrera-Camus}, {Bolatto}, {Wolfire},
  {Smith}, {Croxall}, {Kennicutt}, {Calzetti}, {Helou}, {Walter}, {Leroy},
  {Draine}, {Brandl}, {Armus}, {Sandstrom}, {Dale}, {Aniano}, {Meidt},
  {Boquien}, {Hunt}, {Galametz}, {Tabatabaei}, {Murphy}, {Appleton}, {Roussel},
  {Engelbracht}, \& {Beirao}}]{HERRERACAMUS15}
{Herrera-Camus}, R., {Bolatto}, A.~D., {Wolfire}, M.~G., {et~al.} 2015, \apj,
  800, 1

\bibitem[{{Hodge} {et~al.}(2015){Hodge}, {Riechers}, {Decarli}, {Walter},
  {Carilli}, {Daddi}, \& {Dannerbauer}}]{HODGE15}
{Hodge}, J.~A., {Riechers}, D., {Decarli}, R., {et~al.} 2015, \apjl, 798, L18

\bibitem[{{Hodge} {et~al.}(2016){Hodge}, {Swinbank}, {Simpson}, {Smail},
  {Walter}, {Alexander}, {Bertoldi}, {Biggs}, {Brandt}, {Chapman}, {Chen},
  {Coppin}, {Cox}, {Dannerbauer}, {Edge}, {Greve}, {Ivison}, {Karim},
  {Knudsen}, {Menten}, {Rix}, {Schinnerer}, {Wardlow}, {Weiss}, \& {van der
  Werf}}]{HODGE16}
{Hodge}, J.~A., {Swinbank}, A.~M., {Simpson}, J.~M., {et~al.} 2016, \apj, 833,
  103

\bibitem[{{Hopkins} {et~al.}(2010){Hopkins}, {Younger}, {Hayward}, {Narayanan},
  \& {Hernquist}}]{HOPKINS10}
{Hopkins}, P.~F., {Younger}, J.~D., {Hayward}, C.~C., {Narayanan}, D., \&
  {Hernquist}, L. 2010, \mnras, 402, 1693

\bibitem[{{Howell} {et~al.}(2010){Howell}, {Armus}, {Mazzarella}, {Evans},
  {Surace}, {Sanders}, {Petric}, {Appleton}, {Bothun}, {Bridge}, {Chan},
  {Charmandaris}, {Frayer}, {Haan}, {Inami}, {Kim}, {Lord}, {Madore},
  {Melbourne}, {Schulz}, {U}, {Vavilkin}, {Veilleux}, \& {Xu}}]{HOWELL10}
{Howell}, J.~H., {Armus}, L., {Mazzarella}, J.~M., {et~al.} 2010, \apj, 715,
  572

\bibitem[{{Hu} {et~al.}(2009){Hu}, {Cowie}, {Kakazu}, \& {Barger}}]{HU09}
{Hu}, E.~M., {Cowie}, L.~L., {Kakazu}, Y., \& {Barger}, A.~J. 2009, \apj, 698,
  2014

\bibitem[{{Ilbert} {et~al.}(2015){Ilbert}, {Arnouts}, {Le Floc'h}, {Aussel},
  {Bethermin}, {Capak}, {Hsieh}, {Kajisawa}, {Karim}, {Le F{\`e}vre}, {Lee},
  {Lilly}, {McCracken}, {Michel-Dansac}, {Moutard}, {Renzini}, {Salvato},
  {Sanders}, {Scoville}, {Sheth}, {Silverman}, {Smol{\v c}i{\'c}}, {Taniguchi},
  \& {Tresse}}]{ILBERT15}
{Ilbert}, O., {Arnouts}, S., {Le Floc'h}, E., {et~al.} 2015, \aap, 579, A2

\bibitem[{{Jiang} {et~al.}(2016){Jiang}, {Finlator}, {Cohen}, {Egami},
  {Windhorst}, {Fan}, {Dav{\'e}}, {Kashikawa}, {Mechtley}, {Ouchi},
  {Shimasaku}, \& {Cl{\'e}ment}}]{JIANG16}
{Jiang}, L., {Finlator}, K., {Cohen}, S.~H., {et~al.} 2016, \apj, 816, 16

\bibitem[{{Jones et al.}(2017)}]{JONES17}
{Jones et al.} 2017, \apj, in prep.

\bibitem[{{Kanekar} {et~al.}(2013){Kanekar}, {Wagg}, {Ram Chary}, \&
  {Carilli}}]{KANEKAR13}
{Kanekar}, N., {Wagg}, J., {Ram Chary}, R., \& {Carilli}, C.~L. 2013, \apjl,
  771, L20

\bibitem[{{Karim} {et~al.}(2011){Karim}, {Schinnerer},
  {Mart{\'{\i}}nez-Sansigre}, {Sargent}, {van der Wel}, {Rix}, {Ilbert},
  {Smol{\v c}i{\'c}}, {Carilli}, {Pannella}, {Koekemoer}, {Bell}, \&
  {Salvato}}]{KARIM11}
{Karim}, A., {Schinnerer}, E., {Mart{\'{\i}}nez-Sansigre}, A., {et~al.} 2011,
  \apj, 730, 61

\bibitem[{{Kennicutt} {et~al.}(2011){Kennicutt}, {Calzetti}, {Aniano},
  {Appleton}, {Armus}, {Beir{\~a}o}, {Bolatto}, {Brandl}, {Crocker}, {Croxall},
  {Dale}, {Donovan Meyer}, {Draine}, {Engelbracht}, {Galametz}, {Gordon},
  {Groves}, {Hao}, {Helou}, {Hinz}, {Hunt}, {Johnson}, {Koda}, {Krause},
  {Leroy}, {Li}, {Meidt}, {Montiel}, {Murphy}, {Rahman}, {Rix}, {Roussel},
  {Sandstrom}, {Sauvage}, {Schinnerer}, {Skibba}, {Smith}, {Srinivasan},
  {Vigroux}, {Walter}, {Wilson}, {Wolfire}, \& {Zibetti}}]{KENNICUTT11}
{Kennicutt}, R.~C., {Calzetti}, D., {Aniano}, G., {et~al.} 2011, \pasp, 123,
  1347

\bibitem[{{Kewley} \& {Ellison}(2008)}]{KEWLEY08}
{Kewley}, L.~J., \& {Ellison}, S.~L. 2008, \apj, 681, 1183

\bibitem[{{Kewley} {et~al.}(2013){Kewley}, {Maier}, {Yabe}, {Ohta}, {Akiyama},
  {Dopita}, \& {Yuan}}]{KEWLEY13}
{Kewley}, L.~J., {Maier}, C., {Yabe}, K., {et~al.} 2013, \apjl, 774, L10

\bibitem[{{Knudsen} {et~al.}(2016){Knudsen}, {Richard}, {Kneib}, {Jauzac},
  {Cl{\'e}ment}, {Drouart}, {Egami}, \& {Lindroos}}]{KNUDSEN16}
{Knudsen}, K.~K., {Richard}, J., {Kneib}, J.-P., {et~al.} 2016, \mnras, 462, L6

\bibitem[{{Knudsen} {et~al.}(2017){Knudsen}, {Watson}, {Frayer}, {Christensen},
  {Gallazzi}, {Micha{\l}owski}, {Richard}, \& {Zavala}}]{KNUDSEN17}
{Knudsen}, K.~K., {Watson}, D., {Frayer}, D., {et~al.} 2017, \mnras, 466, 138

\bibitem[{{Kong} {et~al.}(2004){Kong}, {Charlot}, {Brinchmann}, \&
  {Fall}}]{KONG04}
{Kong}, X., {Charlot}, S., {Brinchmann}, J., \& {Fall}, S.~M. 2004, \mnras,
  349, 769

\bibitem[{{Laigle} {et~al.}(2016){Laigle}, {McCracken}, {Ilbert}, {Hsieh},
  {Davidzon}, {Capak}, {Hasinger}, {Silverman}, {Pichon}, {Coupon}, {Aussel},
  {Le Borgne}, {Caputi}, {Cassata}, {Chang}, {Civano}, {Dunlop}, {Fynbo},
  {Kartaltepe}, {Koekemoer}, {Le F{\`e}vre}, {Le Floc'h}, {Leauthaud}, {Lilly},
  {Lin}, {Marchesi}, {Milvang-Jensen}, {Salvato}, {Sanders}, {Scoville},
  {Smolcic}, {Stockmann}, {Taniguchi}, {Tasca}, {Toft}, {Vaccari}, \&
  {Zabl}}]{LAIGLE16}
{Laigle}, C., {McCracken}, H.~J., {Ilbert}, O., {et~al.} 2016, \apjs, 224, 24

\bibitem[{{Larson} {et~al.}(2016){Larson}, {Sanders}, {Barnes}, {Ishida},
  {Evans}, {U}, {Mazzarella}, {Kim}, {Privon}, {Mirabel}, \&
  {Flewelling}}]{LARSON16}
{Larson}, K.~L., {Sanders}, D.~B., {Barnes}, J.~E., {et~al.} 2016, \apj, 825,
  128

\bibitem[{{Lee} {et~al.}(2012){Lee}, {Alberts}, {Atlee}, {Dey}, {Pope},
  {Jannuzi}, {Reddy}, \& {Brown}}]{LEE12}
{Lee}, K.-S., {Alberts}, S., {Atlee}, D., {et~al.} 2012, \apjl, 758, L31

\bibitem[{{Lee} {et~al.}(2015){Lee}, {Sanders}, {Casey}, {Toft}, {Scoville},
  {Hung}, {Le Floc'h}, {Ilbert}, {Zahid}, {Aussel}, {Capak}, {Kartaltepe},
  {Kewley}, {Li}, {Schawinski}, {Sheth}, \& {Xiao}}]{LEE15}
{Lee}, N., {Sanders}, D.~B., {Casey}, C.~M., {et~al.} 2015, \apj, 801, 80

\bibitem[{{Lilly} {et~al.}(2007){Lilly}, {Le F{\`e}vre}, {Renzini}, {Zamorani},
  {Scodeggio}, {Contini}, {Carollo}, {Hasinger}, {Kneib}, {Iovino}, {Le Brun},
  {Maier}, {Mainieri}, {Mignoli}, {Silverman}, {Tasca}, {Bolzonella},
  {Bongiorno}, {Bottini}, {Capak}, {Caputi}, {Cimatti}, {Cucciati}, {Daddi},
  {Feldmann}, {Franzetti}, {Garilli}, {Guzzo}, {Ilbert}, {Kampczyk}, {Kovac},
  {Lamareille}, {Leauthaud}, {Borgne}, {McCracken}, {Marinoni}, {Pello},
  {Ricciardelli}, {Scarlata}, {Vergani}, {Sanders}, {Schinnerer}, {Scoville},
  {Taniguchi}, {Arnouts}, {Aussel}, {Bardelli}, {Brusa}, {Cappi}, {Ciliegi},
  {Finoguenov}, {Foucaud}, {Franceschini}, {Halliday}, {Impey}, {Knobel},
  {Koekemoer}, {Kurk}, {Maccagni}, {Maddox}, {Marano}, {Marconi}, {Meneux},
  {Mobasher}, {Moreau}, {Peacock}, {Porciani}, {Pozzetti}, {Scaramella},
  {Schiminovich}, {Shopbell}, {Smail}, {Thompson}, {Tresse}, {Vettolani},
  {Zanichelli}, \& {Zucca}}]{LILLY07}
{Lilly}, S.~J., {Le F{\`e}vre}, O., {Renzini}, A., {et~al.} 2007, \apjs, 172,
  70

\bibitem[{{Lofthouse} {et~al.}(2017){Lofthouse}, {Houghton}, \&
  {Kaviraj}}]{LOFTHOUSE17}
{Lofthouse}, E.~K., {Houghton}, R.~C.~W., \& {Kaviraj}, S. 2017, ArXiv
  e-prints, arXiv:1701.07015

\bibitem[{{Ma} {et~al.}(2016){Ma}, {Gonzalez}, {Vieira}, {Aravena}, {Ashby},
  {B{\'e}thermin}, {Bothwell}, {Brandt}, {de Breuck}, {Carlstrom}, {Chapman},
  {Gullberg}, {Hezaveh}, {Litke}, {Malkan}, {Marrone}, {McDonald}, {Murphy},
  {Spilker}, {Sreevani}, {Stark}, {Strandet}, \& {Wang}}]{MA16}
{Ma}, J., {Gonzalez}, A.~H., {Vieira}, J.~D., {et~al.} 2016, \apj, 832, 114

\bibitem[{{Madden} {et~al.}(2013){Madden}, {R{\'e}my-Ruyer}, {Galametz},
  {Cormier}, {Lebouteiller}, {Galliano}, {Hony}, {Bendo}, {Smith}, {Pohlen},
  {Roussel}, {Sauvage}, {Wu}, {Sturm}, {Poglitsch}, {Contursi}, {Doublier},
  {Baes}, {Barlow}, {Boselli}, {Boquien}, {Carlson}, {Ciesla}, {Cooray},
  {Cortese}, {de Looze}, {Irwin}, {Isaak}, {Kamenetzky}, {Karczewski}, {Lu},
  {MacHattie}, {O'Halloran}, {Parkin}, {Rangwala}, {Schirm}, {Schulz},
  {Spinoglio}, {Vaccari}, {Wilson}, \& {Wozniak}}]{MADDEN13}
{Madden}, S.~C., {R{\'e}my-Ruyer}, A., {Galametz}, M., {et~al.} 2013, \pasp,
  125, 600

\bibitem[{{Magdis} {et~al.}(2012){Magdis}, {Daddi}, {B{\'e}thermin}, {Sargent},
  {Elbaz}, {Pannella}, {Dickinson}, {Dannerbauer}, {da Cunha}, {Walter},
  {Rigopoulou}, {Charmandaris}, {Hwang}, \& {Kartaltepe}}]{MAGDIS12}
{Magdis}, G.~E., {Daddi}, E., {B{\'e}thermin}, M., {et~al.} 2012, \apj, 760, 6

\bibitem[{{Magnelli} {et~al.}(2014){Magnelli}, {Lutz}, {Saintonge}, {Berta},
  {Santini}, {Symeonidis}, {Altieri}, {Andreani}, {Aussel}, {B{\'e}thermin},
  {Bock}, {Bongiovanni}, {Cepa}, {Cimatti}, {Conley}, {Daddi}, {Elbaz},
  {F{\"o}rster Schreiber}, {Genzel}, {Ivison}, {Le Floc'h}, {Magdis},
  {Maiolino}, {Nordon}, {Oliver}, {Page}, {P{\'e}rez Garc{\'{\i}}a},
  {Poglitsch}, {Popesso}, {Pozzi}, {Riguccini}, {Rodighiero}, {Rosario},
  {Roseboom}, {Sanchez-Portal}, {Scott}, {Sturm}, {Tacconi}, {Valtchanov},
  {Wang}, \& {Wuyts}}]{MAGNELLI14}
{Magnelli}, B., {Lutz}, D., {Saintonge}, A., {et~al.} 2014, \aap, 561, A86

\bibitem[{{Maiolino} {et~al.}(2008){Maiolino}, {Nagao}, {Grazian}, {Cocchia},
  {Marconi}, {Mannucci}, {Cimatti}, {Pipino}, {Ballero}, {Calura}, {Chiappini},
  {Fontana}, {Granato}, {Matteucci}, {Pastorini}, {Pentericci}, {Risaliti},
  {Salvati}, \& {Silva}}]{MAIOLINO08}
{Maiolino}, R., {Nagao}, T., {Grazian}, A., {et~al.} 2008, \aap, 488, 463

\bibitem[{{Maiolino} {et~al.}(2015){Maiolino}, {Carniani}, {Fontana},
  {Vallini}, {Pentericci}, {Ferrara}, {Vanzella}, {Grazian}, {Gallerani},
  {Castellano}, {Cristiani}, {Brammer}, {Santini}, {Wagg}, \&
  {Williams}}]{MAIOLINO15}
{Maiolino}, R., {Carniani}, S., {Fontana}, A., {et~al.} 2015, \mnras, 452, 54

\bibitem[{{Mannucci} {et~al.}(2010){Mannucci}, {Cresci}, {Maiolino}, {Marconi},
  \& {Gnerucci}}]{MANNUCCI10}
{Mannucci}, F., {Cresci}, G., {Maiolino}, R., {Marconi}, A., \& {Gnerucci}, A.
  2010, \mnras, 408, 2115

\bibitem[{{M{\'a}rmol-Queralt{\'o}} {et~al.}(2016){M{\'a}rmol-Queralt{\'o}},
  {McLure}, {Cullen}, {Dunlop}, {Fontana}, \& {McLeod}}]{MARMOLQUERALTO16}
{M{\'a}rmol-Queralt{\'o}}, E., {McLure}, R.~J., {Cullen}, F., {et~al.} 2016,
  \mnras, 460, 3587

\bibitem[{{Martin} {et~al.}(2005){Martin}, {Fanson}, {Schiminovich},
  {Morrissey}, {Friedman}, {Barlow}, {Conrow}, {Grange}, {Jelinsky},
  {Milliard}, {Siegmund}, {Bianchi}, {Byun}, {Donas}, {Forster}, {Heckman},
  {Lee}, {Madore}, {Malina}, {Neff}, {Rich}, {Small}, {Surber}, {Szalay},
  {Welsh}, \& {Wyder}}]{MARTIN05}
{Martin}, D.~C., {Fanson}, J., {Schiminovich}, D., {et~al.} 2005, \apjl, 619,
  L1

\bibitem[{{Mason} {et~al.}(2016){Mason}, {Treu}, {Fontana}, {Jones},
  {Morishita}, {Amorin}, {Bradac}, {Finney}, {Henry}, {Hoag}, {Huang},
  {Schmidt}, {Trenti}, \& {Vulcani}}]{MASON16}
{Mason}, C.~A., {Treu}, T., {Fontana}, A., {et~al.} 2016, ArXiv e-prints,
  arXiv:1610.03075

\bibitem[{{Masters} {et~al.}(2016){Masters}, {Faisst}, \& {Capak}}]{MASTERS16}
{Masters}, D., {Faisst}, A., \& {Capak}, P. 2016, \apj, 828, 18

\bibitem[{{Meurer} {et~al.}(1999){Meurer}, {Heckman}, \& {Calzetti}}]{MEURER99}
{Meurer}, G.~R., {Heckman}, T.~M., \& {Calzetti}, D. 1999, \apj, 521, 64

\bibitem[{{Meurer} {et~al.}(1995){Meurer}, {Heckman}, {Leitherer}, {Kinney},
  {Robert}, \& {Garnett}}]{MEURER95}
{Meurer}, G.~R., {Heckman}, T.~M., {Leitherer}, C., {et~al.} 1995, \aj, 110,
  2665

\bibitem[{{Miller} {et~al.}(2016){Miller}, {Chapman}, {Hayward}, {Behroozi},
  {Bradford}, {Willott}, \& {Wagg}}]{MILLER16}
{Miller}, T.~B., {Chapman}, S.~C., {Hayward}, C.~C., {et~al.} 2016, ArXiv
  e-prints, arXiv:1611.08552

\bibitem[{{Mu{\~n}oz-Mateos} {et~al.}(2009){Mu{\~n}oz-Mateos}, {Gil de Paz},
  {Boissier}, {Zamorano}, {Dale}, {P{\'e}rez-Gonz{\'a}lez}, {Gallego},
  {Madore}, {Bendo}, {Thornley}, {Draine}, {Boselli}, {Buat}, {Calzetti},
  {Moustakas}, \& {Kennicutt}}]{MUNOZMATEOS09}
{Mu{\~n}oz-Mateos}, J.~C., {Gil de Paz}, A., {Boissier}, S., {et~al.} 2009,
  \apj, 701, 1965

\bibitem[{{Narayanan} {et~al.}(2017){Narayanan}, {Dave}, {Johnson}, {Thompson},
  {Conroy}, \& {Geach}}]{NARAYANAN17}
{Narayanan}, D., {Dave}, R., {Johnson}, B., {et~al.} 2017, ArXiv e-prints,
  arXiv:1705.05858

\bibitem[{{Noeske} {et~al.}(2007){Noeske}, {Weiner}, {Faber}, {Papovich},
  {Koo}, {Somerville}, {Bundy}, {Conselice}, {Newman}, {Schiminovich}, {Le
  Floc'h}, {Coil}, {Rieke}, {Lotz}, {Primack}, {Barmby}, {Cooper}, {Davis},
  {Ellis}, {Fazio}, {Guhathakurta}, {Huang}, {Kassin}, {Martin}, {Phillips},
  {Rich}, {Small}, {Willmer}, \& {Wilson}}]{NOESKE07}
{Noeske}, K.~G., {Weiner}, B.~J., {Faber}, S.~M., {et~al.} 2007, \apjl, 660,
  L43

\bibitem[{{Oke}(1974)}]{OKE74}
{Oke}, J.~B. 1974, \apjs, 27, 21

\bibitem[{{Ota} {et~al.}(2014){Ota}, {Walter}, {Ohta}, {Hatsukade}, {Carilli},
  {da Cunha}, {Gonz{\'a}lez-L{\'o}pez}, {Decarli}, {Hodge}, {Nagai}, {Egami},
  {Jiang}, {Iye}, {Kashikawa}, {Riechers}, {Bertoldi}, {Cox}, {Neri}, \&
  {Weiss}}]{OTA14}
{Ota}, K., {Walter}, F., {Ohta}, K., {et~al.} 2014, \apj, 792, 34

\bibitem[{{Ouchi} {et~al.}(2013){Ouchi}, {Ellis}, {Ono}, {Nakanishi}, {Kohno},
  {Momose}, {Kurono}, {Ashby}, {Shimasaku}, {Willner}, {Fazio}, {Tamura}, \&
  {Iono}}]{OUCHI13}
{Ouchi}, M., {Ellis}, R., {Ono}, Y., {et~al.} 2013, \apj, 778, 102

\bibitem[{{Overzier} {et~al.}(2011){Overzier}, {Heckman}, {Wang}, {Armus},
  {Buat}, {Howell}, {Meurer}, {Seibert}, {Siana}, {Basu-Zych}, {Charlot},
  {Gon{\c c}alves}, {Martin}, {Neill}, {Rich}, {Salim}, \&
  {Schiminovich}}]{OVERZIER11}
{Overzier}, R.~A., {Heckman}, T.~M., {Wang}, J., {et~al.} 2011, \apjl, 726, L7

\bibitem[{{Pannella} {et~al.}(2009){Pannella}, {Carilli}, {Daddi}, {McCracken},
  {Owen}, {Renzini}, {Strazzullo}, {Civano}, {Koekemoer}, {Schinnerer},
  {Scoville}, {Smol{\v c}i{\'c}}, {Taniguchi}, {Aussel}, {Kneib}, {Ilbert},
  {Mellier}, {Salvato}, {Thompson}, \& {Willott}}]{PANNELLA09}
{Pannella}, M., {Carilli}, C.~L., {Daddi}, E., {et~al.} 2009, \apjl, 698, L116

\bibitem[{{Pannella} {et~al.}(2015){Pannella}, {Elbaz}, {Daddi}, {Dickinson},
  {Hwang}, {Schreiber}, {Strazzullo}, {Aussel}, {Bethermin}, {Buat},
  {Charmandaris}, {Cibinel}, {Juneau}, {Ivison}, {Le Borgne}, {Le Floc'h},
  {Leiton}, {Lin}, {Magdis}, {Morrison}, {Mullaney}, {Onodera}, {Renzini},
  {Salim}, {Sargent}, {Scott}, {Shu}, \& {Wang}}]{PANNELLA15}
{Pannella}, M., {Elbaz}, D., {Daddi}, E., {et~al.} 2015, \apj, 807, 141

\bibitem[{{Pavesi} {et~al.}(2016){Pavesi}, {Riechers}, {Capak}, {Carilli},
  {Sharon}, {Stacey}, {Karim}, {Scoville}, \& {Smol{\v c}i{\'c}}}]{PAVESI16}
{Pavesi}, R., {Riechers}, D.~A., {Capak}, P.~L., {et~al.} 2016, \apj, 832, 151

\bibitem[{{Peng} {et~al.}(2011){Peng}, {Ho}, {Impey}, \& {Rix}}]{CHIEN11}
{Peng}, C.~Y., {Ho}, L.~C., {Impey}, C.~D., \& {Rix}, H.-W. 2011, {GALFIT:
  Detailed Structural Decomposition of Galaxy Images}, Astrophysics Source Code
  Library, , , ascl:1104.010

\bibitem[{{Pettini} {et~al.}(1998){Pettini}, {Kellogg}, {Steidel}, {Dickinson},
  {Adelberger}, \& {Giavalisco}}]{PETTINI98}
{Pettini}, M., {Kellogg}, M., {Steidel}, C.~C., {et~al.} 1998, \apj, 508, 539

\bibitem[{{Pettini} \& {Pagel}(2004)}]{PETTINI04}
{Pettini}, M., \& {Pagel}, B.~E.~J. 2004, \mnras, 348, L59

\bibitem[{{Pilyugin} \& {Thuan}(2005)}]{PILYUGIN05}
{Pilyugin}, L.~S., \& {Thuan}, T.~X. 2005, \apj, 631, 231

\bibitem[{{Pineda} {et~al.}(2013){Pineda}, {Langer}, {Velusamy}, \&
  {Goldsmith}}]{PINEDA13}
{Pineda}, J.~L., {Langer}, W.~D., {Velusamy}, T., \& {Goldsmith}, P.~F. 2013,
  \aap, 554, A103

\bibitem[{{Pope} {et~al.}(2008){Pope}, {Chary}, {Alexander}, {Armus},
  {Dickinson}, {Elbaz}, {Frayer}, {Scott}, \& {Teplitz}}]{POPE08}
{Pope}, A., {Chary}, R.-R., {Alexander}, D.~M., {et~al.} 2008, \apj, 675, 1171

\bibitem[{{Popping} {et~al.}(2016){Popping}, {Somerville}, \&
  {Galametz}}]{POPPING16}
{Popping}, G., {Somerville}, R.~S., \& {Galametz}, M. 2016, ArXiv e-prints,
  arXiv:1609.08622

\bibitem[{{Prestwich} {et~al.}(2013){Prestwich}, {Tsantaki}, {Zezas},
  {Jackson}, {Roberts}, {Foltz}, {Linden}, \& {Kalogera}}]{PRESTWICH13}
{Prestwich}, A.~H., {Tsantaki}, M., {Zezas}, A., {et~al.} 2013, \apj, 769, 92

\bibitem[{{Prevot} {et~al.}(1984){Prevot}, {Lequeux}, {Prevot}, {Maurice}, \&
  {Rocca-Volmerange}}]{PREVOT84}
{Prevot}, M.~L., {Lequeux}, J., {Prevot}, L., {Maurice}, E., \&
  {Rocca-Volmerange}, B. 1984, \aap, 132, 389

\bibitem[{{Rasappu} {et~al.}(2016){Rasappu}, {Smit}, {Labb{\'e}}, {Bouwens},
  {Stark}, {Ellis}, \& {Oesch}}]{RASAPPU16}
{Rasappu}, N., {Smit}, R., {Labb{\'e}}, I., {et~al.} 2016, \mnras, 461, 3886

\bibitem[{{Reddy} {et~al.}(2012){Reddy}, {Dickinson}, {Elbaz}, {Morrison},
  {Giavalisco}, {Ivison}, {Papovich}, {Scott}, {Buat}, {Burgarella},
  {Charmandaris}, {Daddi}, {Magdis}, {Murphy}, {Altieri}, {Aussel},
  {Dannerbauer}, {Dasyra}, {Hwang}, {Kartaltepe}, {Leiton}, {Magnelli}, \&
  {Popesso}}]{REDDY12}
{Reddy}, N., {Dickinson}, M., {Elbaz}, D., {et~al.} 2012, \apj, 744, 154

\bibitem[{{Reddy} {et~al.}(2010){Reddy}, {Erb}, {Pettini}, {Steidel}, \&
  {Shapley}}]{REDDY10}
{Reddy}, N.~A., {Erb}, D.~K., {Pettini}, M., {Steidel}, C.~C., \& {Shapley},
  A.~E. 2010, \apj, 712, 1070

\bibitem[{{Reddy} {et~al.}(2006){Reddy}, {Steidel}, {Fadda}, {Yan}, {Pettini},
  {Shapley}, {Erb}, \& {Adelberger}}]{REDDY06}
{Reddy}, N.~A., {Steidel}, C.~C., {Fadda}, D., {et~al.} 2006, \apj, 644, 792

\bibitem[{{Reddy} {et~al.}(2015){Reddy}, {Kriek}, {Shapley}, {Freeman},
  {Siana}, {Coil}, {Mobasher}, {Price}, {Sanders}, \& {Shivaei}}]{REDDY15}
{Reddy}, N.~A., {Kriek}, M., {Shapley}, A.~E., {et~al.} 2015, \apj, 806, 259

\bibitem[{{R{\'e}my-Ruyer} {et~al.}(2013){R{\'e}my-Ruyer}, {Madden},
  {Galliano}, {Hony}, {Sauvage}, {Bendo}, {Roussel}, {Pohlen}, {Smith},
  {Galametz}, {Cormier}, {Lebouteiller}, {Wu}, {Baes}, {Barlow}, {Boquien},
  {Boselli}, {Ciesla}, {De Looze}, {Karczewski}, {Panuzzo}, {Spinoglio},
  {Vaccari}, \& {Wilson}}]{REMYRUYER13}
{R{\'e}my-Ruyer}, A., {Madden}, S.~C., {Galliano}, F., {et~al.} 2013, \aap,
  557, A95

\bibitem[{{Rich} {et~al.}(2012){Rich}, {Torrey}, {Kewley}, {Dopita}, \&
  {Rupke}}]{RICH12}
{Rich}, J.~A., {Torrey}, P., {Kewley}, L.~J., {Dopita}, M.~A., \& {Rupke},
  D.~S.~N. 2012, \apj, 753, 5

\bibitem[{{Riechers} {et~al.}(2013){Riechers}, {Bradford}, {Clements},
  {Dowell}, {P{\'e}rez-Fournon}, {Ivison}, {Bridge}, {Conley}, {Fu}, {Vieira},
  {Wardlow}, {Calanog}, {Cooray}, {Hurley}, {Neri}, {Kamenetzky}, {Aguirre},
  {Altieri}, {Arumugam}, {Benford}, {B{\'e}thermin}, {Bock}, {Burgarella},
  {Cabrera-Lavers}, {Chapman}, {Cox}, {Dunlop}, {Earle}, {Farrah}, {Ferrero},
  {Franceschini}, {Gavazzi}, {Glenn}, {Solares}, {Gurwell}, {Halpern},
  {Hatziminaoglou}, {Hyde}, {Ibar}, {Kov{\'a}cs}, {Krips}, {Lupu}, {Maloney},
  {Martinez-Navajas}, {Matsuhara}, {Murphy}, {Naylor}, {Nguyen}, {Oliver},
  {Omont}, {Page}, {Petitpas}, {Rangwala}, {Roseboom}, {Scott}, {Smith},
  {Staguhn}, {Streblyanska}, {Thomson}, {Valtchanov}, {Viero}, {Wang},
  {Zemcov}, \& {Zmuidzinas}}]{RIECHERS13}
{Riechers}, D.~A., {Bradford}, C.~M., {Clements}, D.~L., {et~al.} 2013, \nat,
  496, 329

\bibitem[{{Riechers} {et~al.}(2014){Riechers}, {Carilli}, {Capak}, {Scoville},
  {Smol{\v c}i{\'c}}, {Schinnerer}, {Yun}, {Cox}, {Bertoldi}, {Karim}, \&
  {Yan}}]{RIECHERS14}
{Riechers}, D.~A., {Carilli}, C.~L., {Capak}, P.~L., {et~al.} 2014, \apj, 796,
  84

\bibitem[{{Rupke} {et~al.}(2008){Rupke}, {Veilleux}, \& {Baker}}]{RUPKE08}
{Rupke}, D.~S.~N., {Veilleux}, S., \& {Baker}, A.~J. 2008, \apj, 674, 172

\bibitem[{{Safarzadeh} {et~al.}(2016){Safarzadeh}, {Hayward}, \&
  {Ferguson}}]{SAFARZADEH16}
{Safarzadeh}, M., {Hayward}, C.~C., \& {Ferguson}, H.~C. 2016, ArXiv e-prints,
  arXiv:1604.07402

\bibitem[{{Sanders} {et~al.}(2003){Sanders}, {Mazzarella}, {Kim}, {Surace}, \&
  {Soifer}}]{SANDERS03}
{Sanders}, D.~B., {Mazzarella}, J.~M., {Kim}, D.-C., {Surace}, J.~A., \&
  {Soifer}, B.~T. 2003, \aj, 126, 1607

\bibitem[{{Sanders} {et~al.}(2016){Sanders}, {Shapley}, {Kriek}, {Reddy},
  {Freeman}, {Coil}, {Siana}, {Mobasher}, {Shivaei}, {Price}, \& {de
  Groot}}]{SANDERS16}
{Sanders}, R.~L., {Shapley}, A.~E., {Kriek}, M., {et~al.} 2016, \apjl, 825, L23

\bibitem[{{Schaerer} {et~al.}(2015){Schaerer}, {Boone}, {Zamojski}, {Staguhn},
  {Dessauges-Zavadsky}, {Finkelstein}, \& {Combes}}]{SCHAERER15}
{Schaerer}, D., {Boone}, F., {Zamojski}, M., {et~al.} 2015, \aap, 574, A19

\bibitem[{{Schaerer} {et~al.}(2013){Schaerer}, {de Barros}, \&
  {Sklias}}]{SCHAERER13}
{Schaerer}, D., {de Barros}, S., \& {Sklias}, P. 2013, \aap, 549, A4

\bibitem[{{Schneider} {et~al.}(2016){Schneider}, {Hunt}, \&
  {Valiante}}]{SCHNEIDER16}
{Schneider}, R., {Hunt}, L., \& {Valiante}, R. 2016, \mnras, 457, 1842

\bibitem[{{Schreiber} {et~al.}(2017){Schreiber}, {Pannella}, {Leiton}, {Elbaz},
  {Wang}, {Okumura}, \& {Labb{\'e}}}]{SCHREIBER17}
{Schreiber}, C., {Pannella}, M., {Leiton}, R., {et~al.} 2017, \aap, 599, A134

\bibitem[{{Schreiber} {et~al.}(2015){Schreiber}, {Pannella}, {Elbaz},
  {B{\'e}thermin}, {Inami}, {Dickinson}, {Magnelli}, {Wang}, {Aussel}, {Daddi},
  {Juneau}, {Shu}, {Sargent}, {Buat}, {Faber}, {Ferguson}, {Giavalisco},
  {Koekemoer}, {Magdis}, {Morrison}, {Papovich}, {Santini}, \&
  {Scott}}]{SCHREIBER15}
{Schreiber}, C., {Pannella}, M., {Elbaz}, D., {et~al.} 2015, \aap, 575, A74

\bibitem[{{Scoville} {et~al.}(2015){Scoville}, {Faisst}, {Capak}, {Kakazu},
  {Li}, \& {Steinhardt}}]{SCOVILLE15}
{Scoville}, N., {Faisst}, A., {Capak}, P., {et~al.} 2015, \apj, 800, 108

\bibitem[{{Scoville} {et~al.}(2007){Scoville}, {Aussel}, {Brusa}, {Capak},
  {Carollo}, {Elvis}, {Giavalisco}, {Guzzo}, {Hasinger}, {Impey}, {Kneib},
  {LeFevre}, {Lilly}, {Mobasher}, {Renzini}, {Rich}, {Sanders}, {Schinnerer},
  {Schminovich}, {Shopbell}, {Taniguchi}, \& {Tyson}}]{SCOVILLE07}
{Scoville}, N., {Aussel}, H., {Brusa}, M., {et~al.} 2007, \apjs, 172, 1

\bibitem[{{Scoville} {et~al.}(2016){Scoville}, {Sheth}, {Aussel}, {Vanden
  Bout}, {Capak}, {Bongiorno}, {Casey}, {Murchikova}, {Koda},
  {{\'A}lvarez-M{\'a}rquez}, {Lee}, {Laigle}, {McCracken}, {Ilbert}, {Pope},
  {Sanders}, {Chu}, {Toft}, {Ivison}, \& {Manohar}}]{SCOVILLE16}
{Scoville}, N., {Sheth}, K., {Aussel}, H., {et~al.} 2016, \apj, 820, 83

\bibitem[{{Scoville}(2013)}]{SCOVILLE13}
{Scoville}, N.~Z. 2013, {Evolution of star formation and gas}, ed.
  J.~{Falc{\'o}n-Barroso} \& J.~H. {Knapen}, 491

\bibitem[{{Scoville} \& {Kwan}(1976)}]{SCOVILLE76}
{Scoville}, N.~Z., \& {Kwan}, J. 1976, \apj, 206, 718

\bibitem[{{Seaton}(1979)}]{SEATON79}
{Seaton}, M.~J. 1979, \mnras, 187, 73P

\bibitem[{{Seibert} {et~al.}(2005){Seibert}, {Martin}, {Heckman}, {Buat},
  {Hoopes}, {Barlow}, {Bianchi}, {Byun}, {Donas}, {Forster}, {Friedman},
  {Jelinsky}, {Lee}, {Madore}, {Malina}, {Milliard}, {Morrissey}, {Neff},
  {Rich}, {Schiminovich}, {Siegmund}, {Small}, {Szalay}, {Welsh}, \&
  {Wyder}}]{SEIBERT05}
{Seibert}, M., {Martin}, D.~C., {Heckman}, T.~M., {et~al.} 2005, \apjl, 619,
  L55

\bibitem[{{Shim} {et~al.}(2011){Shim}, {Chary}, {Dickinson}, {Lin}, {Spinrad},
  {Stern}, \& {Yan}}]{SHIM11}
{Shim}, H., {Chary}, R.-R., {Dickinson}, M., {et~al.} 2011, \apj, 738, 69

\bibitem[{{Shivaei} {et~al.}(2015){Shivaei}, {Reddy}, {Shapley}, {Kriek},
  {Siana}, {Mobasher}, {Coil}, {Freeman}, {Sanders}, {Price}, {de Groot}, \&
  {Azadi}}]{SHIVAEI15}
{Shivaei}, I., {Reddy}, N.~A., {Shapley}, A.~E., {et~al.} 2015, \apj, 815, 98

\bibitem[{{Siana} {et~al.}(2009){Siana}, {Smail}, {Swinbank}, {Richard},
  {Teplitz}, {Coppin}, {Ellis}, {Stark}, {Kneib}, \& {Edge}}]{SIANA09}
{Siana}, B., {Smail}, I., {Swinbank}, A.~M., {et~al.} 2009, \apj, 698, 1273

\bibitem[{{Skibba} {et~al.}(2011){Skibba}, {Engelbracht}, {Dale}, {Hinz},
  {Zibetti}, {Crocker}, {Groves}, {Hunt}, {Johnson}, {Meidt}, {Murphy},
  {Appleton}, {Armus}, {Bolatto}, {Brandl}, {Calzetti}, {Croxall}, {Galametz},
  {Gordon}, {Kennicutt}, {Koda}, {Krause}, {Montiel}, {Rix}, {Roussel},
  {Sandstrom}, {Sauvage}, {Schinnerer}, {Smith}, {Walter}, {Wilson}, \&
  {Wolfire}}]{SKIBBA11}
{Skibba}, R.~A., {Engelbracht}, C.~W., {Dale}, D., {et~al.} 2011, \apj, 738, 89

\bibitem[{{Sklias} {et~al.}(2014){Sklias}, {Zamojski}, {Schaerer},
  {Dessauges-Zavadsky}, {Egami}, {Rex}, {Rawle}, {Richard}, {Boone}, {Simpson},
  {Smail}, {van der Werf}, {Altieri}, \& {Kneib}}]{SKLIAS14}
{Sklias}, P., {Zamojski}, M., {Schaerer}, D., {et~al.} 2014, \aap, 561, A149

\bibitem[{{Speagle} {et~al.}(2014){Speagle}, {Steinhardt}, {Capak}, \&
  {Silverman}}]{SPEAGLE14}
{Speagle}, J.~S., {Steinhardt}, C.~L., {Capak}, P.~L., \& {Silverman}, J.~D.
  2014, \apjs, 214, 15

\bibitem[{{Stanway} {et~al.}(2014){Stanway}, {Eldridge}, {Greis}, {Davies},
  {Wilkins}, \& {Bremer}}]{STANWAY14}
{Stanway}, E.~R., {Eldridge}, J.~J., {Greis}, S.~M.~L., {et~al.} 2014, \mnras,
  444, 3466

\bibitem[{{Stark} {et~al.}(2013){Stark}, {Schenker}, {Ellis}, {Robertson},
  {McLure}, \& {Dunlop}}]{STARK13}
{Stark}, D.~P., {Schenker}, M.~A., {Ellis}, R., {et~al.} 2013, \apj, 763, 129

\bibitem[{{Steidel} {et~al.}(2014){Steidel}, {Rudie}, {Strom}, {Pettini},
  {Reddy}, {Shapley}, {Trainor}, {Erb}, {Turner}, {Konidaris}, {Kulas}, {Mace},
  {Matthews}, \& {McLean}}]{STEIDEL14}
{Steidel}, C.~C., {Rudie}, G.~C., {Strom}, A.~L., {et~al.} 2014, \apj, 795, 165

\bibitem[{{Steinhardt} {et~al.}(2014){Steinhardt}, {Speagle}, {Capak},
  {Silverman}, {Carollo}, {Dunlop}, {Hashimoto}, {Hsieh}, {Ilbert}, {Le Fevre},
  {Le Floc'h}, {Lee}, {Lin}, {Lin}, {Masters}, {McCracken}, {Nagao}, {Petric},
  {Salvato}, {Sanders}, {Scoville}, {Sheth}, {Strauss}, \&
  {Taniguchi}}]{STEINHARDT14}
{Steinhardt}, C.~L., {Speagle}, J.~S., {Capak}, P., {et~al.} 2014, \apjl, 791,
  L25

\bibitem[{{Strandet} {et~al.}(2016){Strandet}, {Weiss}, {Vieira}, {de Breuck},
  {Aguirre}, {Aravena}, {Ashby}, {B{\'e}thermin}, {Bradford}, {Carlstrom},
  {Chapman}, {Crawford}, {Everett}, {Fassnacht}, {Furstenau}, {Gonzalez},
  {Greve}, {Gullberg}, {Hezaveh}, {Kamenetzky}, {Litke}, {Ma}, {Malkan},
  {Marrone}, {Menten}, {Murphy}, {Nadolski}, {Rotermund}, {Spilker}, {Stark},
  \& {Welikala}}]{STRANDET16}
{Strandet}, M.~L., {Weiss}, A., {Vieira}, J.~D., {et~al.} 2016, \apj, 822, 80

\bibitem[{{Strom} {et~al.}(2016){Strom}, {Steidel}, {Rudie}, {Trainor},
  {Pettini}, \& {Reddy}}]{STROM16}
{Strom}, A.~L., {Steidel}, C.~C., {Rudie}, G.~C., {et~al.} 2016, ArXiv
  e-prints, arXiv:1608.02587

\bibitem[{{Tacchella} {et~al.}(2015){Tacchella}, {Lang}, {Carollo},
  {F{\"o}rster Schreiber}, {Renzini}, {Shapley}, {Wuyts}, {Cresci}, {Genzel},
  {Lilly}, {Mancini}, {Newman}, {Tacconi}, {Zamorani}, {Davies}, {Kurk}, \&
  {Pozzetti}}]{TACCHELLA15}
{Tacchella}, S., {Lang}, P., {Carollo}, C.~M., {et~al.} 2015, \apj, 802, 101

\bibitem[{{Tacconi} {et~al.}(2010){Tacconi}, {Genzel}, {Neri}, {Cox}, {Cooper},
  {Shapiro}, {Bolatto}, {Bouch{\'e}}, {Bournaud}, {Burkert}, {Combes},
  {Comerford}, {Davis}, {Schreiber}, {Garcia-Burillo}, {Gracia-Carpio}, {Lutz},
  {Naab}, {Omont}, {Shapley}, {Sternberg}, \& {Weiner}}]{TACCONI10}
{Tacconi}, L.~J., {Genzel}, R., {Neri}, R., {et~al.} 2010, \nat, 463, 781

\bibitem[{{Takeuchi} {et~al.}(2010){Takeuchi}, {Buat}, {Heinis}, {Giovannoli},
  {Yuan}, {Iglesias-P{\'a}ramo}, {Murata}, \& {Burgarella}}]{TAKEUCHI10}
{Takeuchi}, T.~T., {Buat}, V., {Heinis}, S., {et~al.} 2010, \aap, 514, A4

\bibitem[{{Takeuchi} {et~al.}(2012){Takeuchi}, {Yuan}, {Ikeyama}, {Murata}, \&
  {Inoue}}]{TAKEUCHI12}
{Takeuchi}, T.~T., {Yuan}, F.-T., {Ikeyama}, A., {Murata}, K.~L., \& {Inoue},
  A.~K. 2012, \apj, 755, 144

\bibitem[{{Tasca} {et~al.}(2015){Tasca}, {Le F{\`e}vre}, {Hathi}, {Schaerer},
  {Ilbert}, {Zamorani}, {Lemaux}, {Cassata}, {Garilli}, {Le Brun}, {Maccagni},
  {Pentericci}, {Thomas}, {Vanzella}, {Zucca}, {Amorin}, {Bardelli},
  {Cassar{\`a}}, {Castellano}, {Cimatti}, {Cucciati}, {Durkalec}, {Fontana},
  {Giavalisco}, {Grazian}, {Paltani}, {Ribeiro}, {Scodeggio}, {Sommariva},
  {Talia}, {Tresse}, {Vergani}, {Capak}, {Charlot}, {Contini}, {de la Torre},
  {Dunlop}, {Fotopoulou}, {Koekemoer}, {L{\'o}pez-Sanjuan}, {Mellier}, {Pforr},
  {Salvato}, {Scoville}, {Taniguchi}, \& {Wang}}]{TASCA15}
{Tasca}, L.~A.~M., {Le F{\`e}vre}, O., {Hathi}, N.~P., {et~al.} 2015, \aap,
  581, A54

\bibitem[{{Temi} {et~al.}(2009){Temi}, {Brighenti}, \& {Mathews}}]{TEMI09}
{Temi}, P., {Brighenti}, F., \& {Mathews}, W.~G. 2009, \apj, 707, 890

\bibitem[{{To} {et~al.}(2014){To}, {Wang}, \& {Owen}}]{TO14}
{To}, C.-H., {Wang}, W.-H., \& {Owen}, F.~N. 2014, \apj, 792, 139

\bibitem[{{U} {et~al.}(2012){U}, {Sanders}, {Mazzarella}, {Evans}, {Howell},
  {Surace}, {Armus}, {Iwasawa}, {Kim}, {Casey}, {Vavilkin}, {Dufault},
  {Larson}, {Barnes}, {Chan}, {Frayer}, {Haan}, {Inami}, {Ishida},
  {Kartaltepe}, {Melbourne}, \& {Petric}}]{U12}
{U}, V., {Sanders}, D.~B., {Mazzarella}, J.~M., {et~al.} 2012, \apjs, 203, 9

\bibitem[{{Vallini} {et~al.}(2017){Vallini}, {Ferrara}, {Pallottini}, \&
  {Gallerani}}]{VALLINI17}
{Vallini}, L., {Ferrara}, A., {Pallottini}, A., \& {Gallerani}, S. 2017,
  \mnras, 467, 1300

\bibitem[{{Vallini} {et~al.}(2013){Vallini}, {Gallerani}, {Ferrara}, \&
  {Baek}}]{VALLINI13}
{Vallini}, L., {Gallerani}, S., {Ferrara}, A., \& {Baek}, S. 2013, \mnras, 433,
  1567

\bibitem[{{Vallini} {et~al.}(2015){Vallini}, {Gallerani}, {Ferrara},
  {Pallottini}, \& {Yue}}]{VALLINI15}
{Vallini}, L., {Gallerani}, S., {Ferrara}, A., {Pallottini}, A., \& {Yue}, B.
  2015, \apj, 813, 36

\bibitem[{{Walter} {et~al.}(2012){Walter}, {Decarli}, {Carilli}, {Riechers},
  {Bertoldi}, {Wei{\ss}}, {Cox}, {Neri}, {Maiolino}, {Ouchi}, {Egami}, \&
  {Nakanishi}}]{WALTER12}
{Walter}, F., {Decarli}, R., {Carilli}, C., {et~al.} 2012, \apj, 752, 93

\bibitem[{{Watson} {et~al.}(2015){Watson}, {Christensen}, {Knudsen}, {Richard},
  {Gallazzi}, \& {Micha{\l}owski}}]{WATSON15}
{Watson}, D., {Christensen}, L., {Knudsen}, K.~K., {et~al.} 2015, \nat, 519,
  327

\bibitem[{{Wei{\ss}} {et~al.}(2013){Wei{\ss}}, {De Breuck}, {Marrone},
  {Vieira}, {Aguirre}, {Aird}, {Aravena}, {Ashby}, {Bayliss}, {Benson},
  {B{\'e}thermin}, {Biggs}, {Bleem}, {Bock}, {Bothwell}, {Bradford}, {Brodwin},
  {Carlstrom}, {Chang}, {Chapman}, {Crawford}, {Crites}, {de Haan}, {Dobbs},
  {Downes}, {Fassnacht}, {George}, {Gladders}, {Gonzalez}, {Greve},
  {Halverson}, {Hezaveh}, {High}, {Holder}, {Holzapfel}, {Hoover}, {Hrubes},
  {Husband}, {Keisler}, {Lee}, {Leitch}, {Lueker}, {Luong-Van}, {Malkan},
  {McIntyre}, {McMahon}, {Mehl}, {Menten}, {Meyer}, {Murphy}, {Padin},
  {Plagge}, {Reichardt}, {Rest}, {Rosenman}, {Ruel}, {Ruhl}, {Schaffer},
  {Shirokoff}, {Spilker}, {Stalder}, {Staniszewski}, {Stark}, {Story},
  {Vanderlinde}, {Welikala}, \& {Williamson}}]{WEISS13}
{Wei{\ss}}, A., {De Breuck}, C., {Marrone}, D.~P., {et~al.} 2013, \apj, 767, 88

\bibitem[{{Willott} {et~al.}(2015){Willott}, {Carilli}, {Wagg}, \&
  {Wang}}]{WILLOTT15}
{Willott}, C.~J., {Carilli}, C.~L., {Wagg}, J., \& {Wang}, R. 2015, \apj, 807,
  180

\end{thebibliography}





\end{document}